\documentclass{elsarticle} %
\usepackage[left=1cm,right=3cm,top=2cm,bottom=2cm]{geometry} 

\usepackage{todonotes}
\usepackage{soul}
\usepackage{amsmath}
\usepackage{amssymb}
\usepackage[linesnumbered,ruled,vlined]{algorithm2e}
\AtBeginEnvironment{algorithm}{\SetArgSty{textrm}}
\usepackage{amsthm}
\usepackage{bm}
\usepackage{proof}
\usepackage{subcaption}
\usepackage{setspace}
\usepackage{rwthcolors}
\colorlet{darkorange100}{orange100!80!black}
\usepackage{rotating}
\usepackage[colorlinks]{hyperref}
\usepackage{varioref}
\usepackage{cleveref}
\usepackage{comment}
\usepackage{wrapfig}
\usepackage{multicol}
\usetikzlibrary{calc}
\usetikzlibrary{decorations.pathreplacing}

\usepackage{pgfplots}

\usepackage{thmtools, thm-restate}

\usepackage[normalem]{ulem}

\usepackage[utf8]{inputenc}
\DeclareUnicodeCharacter{2212}{−}

\usepgfplotslibrary{groupplots,dateplot}
\usetikzlibrary{patterns,shapes.arrows}
\pgfplotsset{compat=newest}

\DeclareMathOperator{\sgnOp}{sgn}
\newcommand\sgn[1]{\sgnOp(#1)}

\newcommand\abs[1]{|#1|}

\newcommand\Undef[0]{\textit{undef}}

\newcommand\true[0]{\textit{true}}

\DeclareMathOperator*{\argmin}{arg\,min}

\DeclareMathOperator{\ordOp}{ord}
\newcommand\ord[2]{\ordOp_{#1}(#2)}

\newcommand\vdeg[2]{\deg_{#1}(#2)}

\DeclareMathOperator{\levelOp}{level}
\newcommand\level[1]{\levelOp(#1)}

\DeclareMathOperator{\ldcfOp}{ldcf}
\newcommand\ldcf[2]{\ldcfOp_{#1}(#2)}
\DeclareMathOperator{\discOp}{disc}
\newcommand\disc[2]{\discOp_{#1}(#2)}
\DeclareMathOperator{\resOp}{res}
\newcommand\res[3]{\resOp_{#1}(#2,#3)}

\DeclareMathOperator{\factorsOp}{factors}
\newcommand\factors[1]{\factorsOp(#1)}

\DeclareMathOperator{\realRootsOp}{realRoots}
\newcommand\realRoots[1]{\realRootsOp(#1)}

\DeclareMathOperator{\irootOp}{root}
\newcommand\iroot[3]{\irootOp_{#1}[#2,#3]}

\newcommand\bools[0]{\mathbb{B}}
\newcommand\rationals[0]{\mathbb{Q}}
\newcommand\reals[0]{\mathbb{R}}

\newcommand\naturals[0]{\mathbb{N}}
\newcommand\integers[0]{\mathbb{Z}}
\newcommand\posints[0]{\mathbb{N}_{>0}}

\DeclareMathOperator{\irexprOp}{irExpr}
\newcommand\irexpr[2]{\irexprOp(#1,#2)}
\newcommand\irexprat[3]{\irexprOp(#1,#2,#3)}

\newcommand\Isymb[0]{\textnormal{\texttt{I}}}

\newcommand\ordinv[1]{\hyperref[def:prop:basic]{\textit{ord\_inv}}(#1)}
\newcommand\sgninv[1]{\hyperref[def:prop:basic]{\textit{sgn\_inv}}(#1)}
\newcommand\nonnull[1]{\hyperref[def:prop:basic]{\textit{non\_null}}(#1)}
\newcommand\del[1]{\hyperref[def:prop:basic]{\textit{an\_del}}(#1)}
\newcommand\pdel[1]{REMOVE}

\newcommand\representation[1]{\hyperref[def:prop:representation]{\textit{repr}}(#1)}
\newcommand\irordering[1]{\hyperref[def:prop:irordering]{\textit{ir\_ord}}(#1)}
\newcommand\submanifold[1]{\hyperref[def:prop:basic]{\textit{an\_sub}}(#1)}
\newcommand\connected[1]{\hyperref[def:prop:basic]{\textit{connected}}(#1)}
\newcommand\irint[1]{\hyperref[def:prop:irint]{\textit{holds}}(#1)}
\newcommand\sample[1]{\hyperref[def:prop:basic]{\textit{sample}}(#1)}

\newcommand\proj[2]{\ensuremath{#1\!\downarrow_{#2}}}
\newcommand\restr[2]{\ensuremath{#1|_{#2}}} %
\newcommand\levels[2]{\ensuremath{#1|_{#2}}}

\DeclareMathOperator{\domOp}{dom}
\newcommand\dom[1]{\domOp(#1)}

\newcommand\propset[0]{\text{Prop}}

\newcommand\liftOp[0]{\ensuremath{\textit{setOf}}}
\newcommand\lift[1]{\liftOp(#1)}

\newtheorem{example}{Example}[section]
\newtheorem{definition}{Definition}[section]
\newtheorem{theorem}{Theorem}[section]
\newtheorem{lemma}{Lemma}[section]

\crefname{claim}{Claim}{Claims}

\Crefname{observation}{Observation}{Observations}

\crefname{mapping}{rule}{rules}

\newtheorem{property}{Property}[section]
\crefname{property}{property}{properties}

\theoremstyle{remark}
\newtheorem*{references}{References}

\newcommand{\realring}{\mathbb{R}}

\usepackage{graphics}

\makeatletter
\g@addto@macro\bfseries{\boldmath}
\makeatother

\begin{document}

\begin{frontmatter}
    \title{Levelwise construction of a single cylindrical algebraic cell}

    \author[rwth,cor1]{Jasper Nalbach}
    \ead{nalbach@cs.rwth-aachen.de}

    \author[rwth]{Erika {\'A}brah{\'a}m}
    \ead{abraham@cs.rwth-aachen.de}

    \author[rwth]{Philippe Specht}
    \ead{philippe.specht@rwth-aachen.de}

    \author[usna]{Christopher W. Brown}
    \ead{wcbrown@usna.edu}

    \author[bath]{James H. Davenport}
    \ead{j.h.davenport@bath.ac.uk}

    \author[cov]{Matthew~England}
    \ead{matthew.england@coventry.ac.uk}

    \cortext[cor1]{Corresponding author}
    \address[rwth]{RWTH Aachen University, 52056 Aachen, Germany}
    \address[usna]{United States Naval Academy, 597 McNair Road, Annapolis, MD 21402-5002, United States}
    \address[bath]{University of Bath, Claverton Down, Bath BA2 7AY, United Kingdom}
    \address[cov]{Coventry University, Coventry CV1 2TL, United Kingdom}

    \begin{abstract}
        \emph{Satisfiability modulo theories (SMT)} solvers check the satisfiability of quantifier-free first-order logic formulae over different theories. We consider the theory of \emph{non-linear real arithmetic} where the formulae are logical combinations of polynomial constraints. Here a commonly used tool is the \emph{cylindrical algebraic decomposition (CAD)} to decompose the real space into cells where the constraints are truth-invariant through the use of \emph{projection polynomials}.
        
        A CAD encodes more information than necessary for checking satisfiability. One approach to address this is to repackage the CAD theory into a search-based algorithm:  one that guesses sample points to satisfy the formula, and generalizes guesses that conflict constraints to cylindrical cells around samples which are avoided in the continuing search.  Such an approach can lead to a satisfying assignment more quickly, or conclude unsatisfiability with far fewer cells.  A notable example of this approach is Jovanovi\'{c} and de Moura's \emph{NLSAT} algorithm.
        Since these cells are being produced locally to a sample there is scope to use fewer projection polynomials than the traditional CAD projection.  The original NLSAT algorithm reduced the set a little; while Brown's \emph{single cell construction} reduced it much further still.  However, it refines a cell polynomial-by-polynomial, meaning the shape and size of the cell produced depends on the order in which the polynomials are considered.  
        
        The present paper proposes a method to construct such cells \emph{levelwise}, i.e. built level-by-level according to a variable ordering instead of polynomial-by-polynomial for all levels.  We still use a reduced number of projection polynomials, but can now consider a variety of different reductions and use \emph{heuristics} to select the projection polynomials in order to optimize the shape of the cell under construction.  The new method can thus improve the performance of the NLSAT algorithm.   
        We formulate all the necessary theory that underpins the algorithm as a \emph{proof system}:  while not a common presentation for work in this field, it is valuable in allowing an elegant decoupling of heuristic decisions from the main algorithm and its proof of correctness.  We expect the symbolic computation community may find uses for it in other areas too. In particular, the proof system could be a step towards formal proofs for non-linear real arithmetic.
        
        This work has been implemented in the SMT-RAT solver and the benefits of the levelwise construction are validated experimentally on the SMT-LIB benchmark library.  We also compare several heuristics for the construction and observe that each heuristic has strengths offering potential for further exploitation of the new approach.
    \end{abstract}

    \begin{keyword}
        Satisfiability modulo theories \sep Cylindrical algebraic decomposition \sep Non-linear real arithmetic \sep Model-constructing satisfiability calculus \sep Formal proofs
    \end{keyword}

\end{frontmatter}

\section{Introduction}
\label{sec:intro}

In this paper we present a new method to construct around a sample point a single cylindrical cell that is truth-invariant for a set of polynomial constraints.  We demonstrate how the new method allows for improved decision procedures to determine the satisfiability of formulae involving such constraints.  We use a proof system presentation for our method, which we consider an important contribution for such algebraic decision procedures.  We take this opportunity to explain the benefit of such a presentation to the symbolic computation community. This introduction continues with a broad overview of context for the contribution, followed by the plan of the paper.

\subsection{Non-linear real arithmetic and CAD}

We are concerned with \emph{non-linear real arithmetic} whose formulae are Boolean combinations of polynomial constraints with rational coefficients.  This is a powerful logic that can express a wide variety of problems.  This logic admits quantifier elimination \cite{Tarski1948}, i.e. any quantified formula in the logic may be replaced by an equivalent quantifier-free one.  In this paper we restrict our attention to the problem of determining the satisfiability of quantifier-free formulae, or equivalently, determining the truth of purely existentially quantified formulae. %

The most commonly used complete methods here are based on the idea of the \emph{cylindrical algebraic decomposition (CAD)} introduced by Collins \cite{Collins1975}.  A CAD is a finite decomposition of $\reals^n$ into cells, traditionally produced relative to a set of polynomials in $n$ variables such that each polynomial has constant sign on each cell.  It thus allows us to use a \emph{finite set} of sample points (one for each cell) to study sign-constraints on those polynomials over the \emph{infinite} space $\reals^n$.  The CAD method offered the first tractable approach to real quantifier elimination and found numerous applications in the years that followed.  However, its practical use is restricted by a doubly exponential worst case complexity in the number of variables \cite{DH88}, that is felt often in practice: the algorithm makes use of iterated resultants \cite{Collins1975} leading to polynomials of doubly-exponential degree.  

It was soon realized that a CAD encoded far more information than needed even for quantifier elimination: a CAD for a set of polynomials can be used to study \emph{any} logical formula built from those polynomials, not just the one of interest.  Some progress was since made in adapting the core CAD algorithm to the logical formula, e.g. \cite{CH91, EBD20}, but these were only partial solutions.

\subsection{NLSAT, MCSAT and single cylindrical cells}
\label{ssec:mcsat}

A novel framework for \emph{satisfiability modulo theories (SMT)} solving
was introduced with the \emph{NLSAT algorithm} of Jovanovi\'{c} and de Moura \cite{jovanovic2012solving} in 2012.  This was since generalized into the \emph{model constructing satisfiability calculus (MCSAT)} framework \cite{dMJ13} and has been applied to other logics such as non-linear integer arithmetic \cite{jovanovic2017integer}.  In MCSAT the search at the Boolean and theory levels are carried out concurrently, mutually guided by each other away from unsatisfiable regions. Partial solution candidates for the Boolean structure and for the corresponding theory constraints are constructed incrementally in parallel, with Boolean conflicts generalized using propositional resolution and, for real algebra, theory conflicts generalized by CAD technology.

For the latter, when a theory model (sample point) is determined not to satisfy all those constraints which should hold according to the current Boolean search, then we seek to guide the future search by an \emph{explanation} which generalizes the sample point to a region containing the point on which the same constraints fail for the same reasons.  This can be achieved by having the polynomials involved in the  combination of constraints which cause the failure all have invariant sign upon this region.  Such regions are constructed as cylindrical algebraic cells, but they are not necessarily cells from the CAD that would be built for the problem.  Instead, they are usually larger since not all constraints are involved in every conflict.  The exclusion of the cell is learned by adding a new clause: the negation of the semi-algebraic description of the cell.

This motivates the optimization of sub-algorithms to produce single cells from a point and a set of polynomial constraints.  Savings can be made not just by building cells with only a subset of the constraints, but also by restricting the combinations of those constraints that we do consider in reference to the current model.  Until now, the state-of-the-art approach is that in \cite{brown2015onecell}.  We continue this research in the present paper by \emph{developing a new method for single cell construction}.

\subsection{First contribution: Proof system presentation}

In contrast to \cite{brown2015onecell}, our new method allows for different choices of how to construct the cell. We will describe and evaluate some of these choices, however, it is important to distinguish that area of work from the broader method to build the cell introduced in the next subsection.   
The choices do not affect the correctness of the cell produced: in all cases the cell meets the essential criteria of containing the sample and being invariant for the truth of the conflicting constraints.  Nor do the choices have an effect on high-level measures of complexity: they achieve similar reductions in algebraic work.  However, it can be observed that the choices do greatly effect the cells produced and thus the performance of the algorithm and so it is worth to try to make an optimal choice.  We do these choices \emph{heuristically}, i.e. using methods not guaranteed to give an optimal answer but hopefully giving a reasonable answer quickly.  To expedite and simplify future research on heuristics it is helpful to clearly separate out these heuristic choices from the broader algorithm and its proof of correctness.

To achieve that, we present our work as a \emph{proof system}.  Such a presentation clearly achieves the separation of heuristic decisions from correctness proof of the method.  Essentially, we must find a chain of proof rules to prove our desired property, and where there is freedom in how the chain can to be built then we can employ a heuristic method.  The system is flexible, extensible, allows for detailed optimizations without changing the fundamental algorithm, and allows for correctness proofs to be portioned nicely. We plan to build on that in future work. We note that this is not just a presentation for the purpose of the paper, but also present in the underlying implementation we report on.

We acknowledge that a proof system presentation is uncommon in symbolic computation.  It is more prevalent in the SAT and SMT communities where there is more intense work on the optimization of such heuristic choices and proof systems are an established presentation method.  However, such a system has not been used before for CAD theory, even when deployed in the SMT context.  We view our proof system presentation as a contribution in its own right, which allowed for greater exploration of heuristic choices in our work.  We also hope the wider symbolic computation community may find it interesting as a potential new tool to use elsewhere. In particular, there is increasing interest in formal proofs. While first SMT solvers \cite{barbosa2022flexible} are able to generate these proofs for a variety of theories, the case of non-linear arithmetic is challenging. Our proof system might pave the way for mechanically-verifiable proofs for non-linear real arithmetic.

\subsection{Second contribution: Levelwise single cells}

The current state-of-the-art for single cylindrical cell construction is that of Brown and Ko{\v{s}}ta \cite{brown2015onecell}.  This constructs the single cell gradually, processing one polynomial at a time, initializing the cell as the entirety of $\reals^n$ and then gradually refining it according to the sign of each polynomial considered.  For each refinement the method needs to consider only the interaction of the next polynomial with the ones currently defining the cell, rather than all those that went before, which allows for savings compared to the original approach used in \cite{jovanovic2012solving}.  However, this approach introduces a sensitivity to the order in which polynomials are considered. A machine learning approach to select the order was considered in \cite{BD20}.

The alternative method contributed in this paper allows to produce the single cell incrementally \emph{by level} (i.e. dimension / variable).  This removed the direct sensitivity to the polynomial ordering, instead introducing at each level decisions about which polynomials to use first.  This approach uncovers a greater range of decisions than the polynomial ordering, and allows for more reasoned heuristics than black-box machine learning.  By allowing for these better heuristic decisions we can produce more optimal cells, in turn improving the performance of algorithms which use them.

\subsection{Plan of the paper}

We continue in \Cref{sec:preliminaries} by introducing the necessary preliminaries and notations used, followed by background material on CAD.  Then in \Cref{sec:singlecell} we present the existing state of the art in single cell construction and an informal motivation for our new levelwise approach to single cell construction. 

In \Cref{sec:proofsystem} we establish the proof system, and in \Cref{sec:levelwise} we present our new algorithm and some heuristics that may be used with it.  
In \Cref{sec:observations} we give some qualitative analysis on our new method and the heuristics. Then an experimental evaluation on the use of the new method for explanation generation in MCSAT is given in \Cref{sec:experiments}. Finally, we conclude in \Cref{sec:conclusion} with an outlook on further research and open questions.

\section{Preliminaries}
\label{sec:preliminaries}
Let $\naturals$ denote the set of all natural numbers including $0$, $\posints=\naturals\setminus\{0\}$, $\rationals$ be the rational numbers, and $\reals$ be the real numbers.  
For $i,j \in \naturals$ with $i<j$, we define the sets of integers $[i..j] = \{ i,\ldots,j \}$ and $[i] = [0..i]$.
For $i,j \in \posints, j \leq i$ and $r \in \reals^i$, we denote by $r_j$ the $j$-th component of $r$ and by $r_{[j]}$ the vector $(r_1,\ldots,r_j)$.
For a tuple $t = (a, b, c, \ldots)$ we denote by $t.a,t.b,t.c,\ldots$ the corresponding tuple entries.

Let $f: D \to E$ be a function, then the \emph{domain} $D$ of $f$ is denoted by $\dom{f}$, and the \emph{restriction of $f$ to $A \subseteq D$} is denoted by $\restr{f}{A}$ (i.e. $\restr{f}{A}: A \to E$ with $\restr{f}{A}(a)=f(a)$ for all $a \in A$). Let $f,g: D \to E$ and let $<$ be a total order on $E$. We write $f<g$ if $f(d)<g(d)$ for all $d \in D$ and $f\leq g$ if $f(d)\leq g(d)$ for all $d \in D$.

\subsection{Variables and polynomials}

We assume that the reader is familiar with the common definitions and terminology related to polynomials. We introduce some notation in this section; for further reading we refer to \cite{Coxetal2006a}.

We work with the \emph{variables} $x_1,\ldots,x_n$ with $n\in\posints$ under a fixed \emph{ordering} $x_1 \prec x_2 \prec ... \prec x_n$.  
A \emph{polynomial} is built from a set of variables and numbers from $\rationals$ using addition and multiplication.   
We use $\rationals[y]$ to denote \emph{univariate} polynomials in some variable $y$ and $\rationals[x_1,\ldots,x_i]$ for \emph{multivariate} polynomials in those variables.
We say that a polynomial $p$ is of level $j$ (denoted as $\level{p}=j$) if $x_j$ is the largest variable appearing in $p$:  i.e. either $j=0$ and $p\in \rationals$; or $j \in [1..n]$ and $p \in \rationals[x_1, \ldots, x_j] \setminus \rationals[x_1, \ldots, x_{j-1}]$. 

Assume in the following some $i \in [n]$ and polynomials $p,q\in \rationals[x_1,\ldots,x_i]$.

We write $p(x_1,\ldots,x_i)$ to indicate $p$'s variable domain.
For $j\in [1..i]$ and $r=(r_1,\ldots,r_j)\in\reals^j$ we write $p(r,x_{j+1},\ldots,x_{i})$ for the polynomial $p$ after substituting $r_1,\ldots,r_j$ for $x_1,\ldots,x_j$ in $p$ and indicating the remaining free variables in $p$.

We use $\realRoots{p} \subseteq \reals^i$ to denote the set of \emph{real roots} of $p$, $\vdeg{x_j}{p}$ to denote the \emph{degree} of $p$ in $x_j$, $\ldcf{x_j}{p}$ the \emph{leading coefficient} of $p$ in $x_j$, $\factors{p}$ to denote the \emph{irreducible factors} of $p$, $\disc{x_j}{p}$ to denote the \emph{discriminant} of $p$ with respect to $x_j$, and $\res{x_j}{p}{q}$ to denote the \emph{resultant} of $p$ and $q$ with respect to $x_j$.

\subsection{Real algebraic numbers, constraints and cells}

\emph{Real algebraic numbers} are real roots of univariate polynomials with rational coefficients. Although we will not distinguish between real and real algebraic numbers in the following for simplicity, the algorithms are complete when restricting all choices of constants to real algebraic numbers.

A \emph{constraint} $p \sim 0$ compares a polynomial $p \in \rationals[x_1, \ldots, x_i],\ \level{p}=i$ to zero using a relation symbol, ${\sim \in \{=,\neq,<,>,\leq,\geq\}}$, and has the \emph{solution set} $\{r\in\reals^i \mid p(r)\sim 0\}$. 

A subset of $\reals^i$ for some $i \in [n]$ is called \emph{semi-algebraic} if it is the solution set of a Boolean combination of polynomial constraints. A \emph{cell} is a non-empty connected subset of $\reals^i$ for some $i \in [n]$. A cell is called \emph{algebraic} if it is a semi-algebraic set.

For simplifying the notation throughout this paper, we define $\reals^0 = \{ () \}$. Given $i,j \in \posints$ with $j<i$, we call $\reals^i$ an \emph{extension} of $\reals^j$ and define the \emph{projection} of a set $R \subseteq \reals^i$ onto $\reals^{j}$ by $\proj{R}{[j]}=\{(r_1,\ldots,r_j) \mid (r_1,\ldots,r_i)\in R\}$.

Given a cell $R \subseteq \reals^i$, $i \in [1..n]$ and continuous functions $f,g: R \to \reals$, we define the sets ${R \times (f,g)} = {\{ (r,r_{i+1}) \mid r\in R , r_{i+1}\in (f(r),g(r))\}}$, analogously ${R \times (-\infty,g)} = {\{ (r,r_{i+1}) \mid r\in R , r_{i+1} < g(r)\}}$, ${R \times (f,\infty)} = {\{ (r,r_{i+1}) \mid r\in R , f(r) < r_{i+1}\}}$, and ${R \times f}={\{ (r,r_{i+1}) \mid r\in R , r_{i+1}=f(r)\}}$. Note that if ${f\leq g}$ (on $R$), then these sets are cells (as a continuous image of a connected set is connected).

The \emph{sign} of $r\in\reals$, denoted $\sgn{r}$, is defined to be $1$ if $r>0$, $-1$ if $r<0$, and $0$ otherwise. A polynomial $p \in \rationals[x_1, \ldots, x_i]$ is \emph{sign-invariant} on a set $R \subseteq \reals^i$ if $\sgn{p(r)} = \sgn{p(r')}$ for all $r,r' \in R$. A set of polynomials $P\subseteq\rationals[x_1,\ldots,x_i]$ is \emph{sign-invariant} on $R\subseteq\reals^i$ if all $p \in P$ are sign-invariant on $R$.

\subsection{CAD definition}

A set $D = \{R_1, \ldots, R_k\}$ of cells of $\reals^n$ such that $\cup_{i=1,\ldots,k} R_i = \reals^n$ and $R_i \cap R_j = \emptyset$ is called a \emph{decomposition} of $\reals^n$.  
A decomposition is called \emph{algebraic} if its cells are algebraic.  A decomposition $D$ of $\reals^n$ is called \emph{cylindrical} over a decomposition $D'$ of $\reals^m,\ m<n$ if all projections of cells $R \in D$ onto $\reals^m$ are themselves cells in $D'$.  I.e. the cells in $\reals^n$ stack up in cylinders over the cells in $\reals^m$.  
A \emph{cylindrical algebraic decomposition (CAD)} is produced relative to a variable ordering: it is an algebraic decomposition $D$ such that there exists a sequence of algebraic decompositions ${(D_1, \ldots, D_n)}, D = D_n$ with each $D_i$ a cylindrical decomposition of $\reals^i$ over $D_{i-1}$ for $i \in [2..n]$.  

A decomposition inherits the invariance properties for polynomials and constraints defined above if they apply to all its cells.  A CAD will usually be computed relative to an input polynomial set to ensure such an invariance property.  
Collins \cite{Collins1975} first introduced the notion of a CAD and an algorithm for computing a sign-invariant CAD in the 1970s.
A central notion for this algorithm is delineability.

\begin{definition}[Delineability \cite{Collins1975}]
    \label{def:delineability}
    Let $i \in \naturals$, $R \subseteq \reals^{i}$ be a cell, and $p \in \rationals[x_1, \ldots, x_{i+1}] \setminus \{ 0 \}$. %
    The polynomial $p$ is called \emph{delineable} on $R$ if and only if there exist finitely many continuous functions $\theta_1, \ldots, \theta_k: R \to \reals$ (for $k\geq 0$) such that
    \begin{itemize}
        \item $\theta_1 < \ldots < \theta_k$;
        \item the set of real roots of the univariate polynomial $p(r,x_{i+1})$ is $\{ \theta_1(r), \ldots, \theta_k(r) \}$ for all $r \in R$; and
        \item there exist constants $m_1,\ldots,m_k \in \posints$ such that for all $r \in R$ and all $j\in[1..k]$, the multiplicity of the root $\theta_j(r)$ of $p(r,x_{i+1})$ is $m_j$.
    \end{itemize}

    The $\theta_j$ are called \emph{real root functions} of $p$ on $R$. The cells $R \times \theta_j,\ j\in[1..k]$ are called $p$-\emph{sections} over $R$. The cells $R \times (-\infty,\theta_1), R \times (\theta_k,\infty),$ and $R \times (\theta_j,\theta_{j+1})$ for $j\in[1..k-1]$, and in case $k=0$ also $R \times (-\infty, \infty)$, are called $p$-\emph{sectors} over $R$.

    These notions are extended to finite sets of polynomials $P \subseteq \rationals[x_1, \ldots, x_{i+1}] \setminus \{ 0 \}$ such that $P$ is delineable on $R$ if the product of the polynomials in $P$ is delineable on $R$. Accordingly, we define the real root functions of $P$, the $P$-sections and $P$-sectors; for the empty polynomial set there are no sections and a single sector $\reals^{i+1}$.
\end{definition}

A \emph{CAD projection operator} is a function $\text{proj}$ that maps a set of polynomials $P$ to a set of lower-level polynomials such that the sign-invariance of $\text{proj}(P)$ on $R$ implies the delineability of $P$ on $R$. The operator $\text{proj}$ induces a \emph{sign-invariant CAD of $P$}, recursively defined as follows. In the case where all polynomials in $P$ are of level $1$, then the CAD is the set of all sections and sectors of $P$. If the polynomials are of higher level, then the CAD contains all sections and sectors of $P$ over each cell of the CAD of $\text{proj}(P)$.

\subsection{McCallum's projection operator}

Although Collin's original projection operator is complete, the projection set is large and thus relatively inefficient.
McCallum presented an improved operator by making the projection set smaller \cite{mccallum1998improved}. To do so, his proof of correctness relies not on sign-invariance but on the stronger property of \emph{order-invariance}.  Although a stronger property, the induced cells in McCallum's CAD are actually bigger than in Collin's CAD. In the following, we present a simplified version of the McCallum CAD projection (simplified as we do not describe the optimization using \emph{delineating polynomials}).

McCallum's theory relies on some notions which we will mention here only on the level of intuition: for more details, we refer to \cite{mccallum1985improved,mccallum1998improved}. An $i$-dimensional \emph{(analytic) submanifold} of $\reals^n$ is a non-empty subset $R \subseteq \reals^n$ that ``looks locally like $\reals^i$''. Given an open subset $U \subseteq \reals^i$, a function $f: U \to \reals$ is called \emph{analytic} if it has a multiple power series representation \cite{mccallum1985improved} around each point of $U$. Given an $i$-dimensional submanifold $R$ of $\reals^n$, a function $f: R \to \reals$ is called \emph{analytic} if for all $r \in R$, $R$ looks locally like $\reals^i$ with respect to a coordinate system about $r$ and $f$ looks locally like an analytic function $\reals^i \to \reals$.
Every open subset of $\reals^i, i \in [1..n]$ is an analytic submanifold. To simplify notation, we say $\reals^0$ is an analytic submanifold. Given an analytic submanifold $R \subseteq \reals^i$ and analytic functions $f,g: R \to \reals$ with $f<g$, the sets $R \times (f,g)$ and $R \times f$ are analytic submanifolds as well \cite[Theorem~2.2.3. and Theorem~2.2.4]{mccallum1985improved}. Note that analytic submanifolds are cells.
Additionally, the notion of delineability is extended to \emph{analytic delineability}, which is only defined on connected analytic submanifolds and the real root functions are required to be analytic. 

Let $p \in \rationals[x_1, \ldots, x_n]$ be a polynomial and $r \in \reals^n$ be a point. Then the \emph{order of $p$ at $r$} is defined as 
\[
\ord{r}{p} = \min(\{ k \in \naturals \mid \text{some partial derivative of total order $k$ of $p$ does not vanish at $r$} \} \cup \{ \infty \}).  
\]
We call $p$ \emph{order-invariant} on $R \subseteq \reals^n$ if $\ord{r}{p} = \ord{r'}{p}$ for all $r,r' \in R$. Note that if $p$ has no root in $R$, then $\ord{r}{p}=0$ for all $r \in R$ and $p$ is trivially order-invariant on $R$. Note also that order-invariance implies sign-invariance.

McCallum's operator requires a smaller projection set than Collins' operator. It does so by maintaining the stronger property of order-invariance instead of sign-invariance; however, order-invariance can only be concluded if no polynomial is \emph{nullified} on a point in the underlying cell.

\begin{definition}[Nullification]
    \label{def:nullification}
    Let $i \in \naturals$, $r \in \reals^i$, and $p \in \rationals[x_1, \ldots, x_{i+1}],\ \level{p}=i+1$. The polynomial $p$ \emph{is nullified} on $r$ if $p(r,x_{i+1}) = 0$.
\end{definition}

The McCallum projection operator is thus incomplete.

We note the recent validation of \emph{Lazard projection} in \cite{MPP19} as an alternative to McCallum projection which is complete and is no bigger than McCallum except for those cases where McCallum cannot be applied \cite{BM20}.  We choose to formalize here with McCallum projection as it is extended to optimizations such as \emph{equational constraints} \cite{mccallum1999equational} \cite{EBD20} which are very powerful in practice (see \Cref{subsec:ec}) and the existing single cell approach \cite{brown2015onecell} is also based on McCallum.  We expect that our work could be reformulated in Lazard projection if so desired.

\subsection{Computing a CAD}

In many applications, a decomposition of a set of polynomials $P$ is not computed explicitly but instead the projection of $P$ is used to generate sample points for every sign-invariant cell that would be formed in such a decomposition. The generation of cells / samples from the projection is called \emph{lifting}. Polynomials can be evaluated at these sample points to check the satisfiability of constraints and formulae which involve them. 

For representing cells explicitly, we need to give witnesses for the real root functions $\theta_j: R \to \reals$ from \Cref{def:delineability} defining the sectors and sections over a cell $R$.

\begin{definition}[Indexed root expression]
    Let $i \in \naturals$, $p \in \rationals[x_1,\ldots,x_{i+1}],\ \level{p}=i+1$, and $j \in \posints$.
    The \emph{indexed root expression} $\iroot{x_{i+1}}{p}{j}: \reals^{i} \to \reals \cup \{ \Undef \}$ is the $j$-th real root of $p$ in $x_{i+1}$ at the given sample if it exists, and $\Undef$ otherwise. That is for each $s \in \reals^i$:
    \[ \iroot{x_{i+1}}{p}{j}(s) = \begin{cases}
        \Undef & \text{if } j > \abs{\realRoots{p(s,x_{i+1})}} \textit{ or } p(s,x_{i+1}) = 0, \textit{ and otherwise }\\
        \xi_j & \text{where } \realRoots{p(s,x_{i+1})} = \{ \xi_1, \ldots, \xi_k \} \textit{ and } \xi_1 < \ldots < \xi_k.
    \end{cases} \]
    Indexed root expressions of this form are also called \emph{indexed root expression of level $i+1$}. Assuming $\xi$ denotes the above indexed root expression, we use $\xi.p$ to refer to $p$ and $\xi.j$ to refer to $j$.
\end{definition}

In the algorithms presented in this paper, we only need to evaluate the indexed root expressions for real algebraic numbers. Note that the existence and index of a real root function depends on the given sample.  Thus, the same indexed root expression may refer to different real root functions at different sample points.

\begin{definition}[Symbolic intervals, single cell and CAD data structures]
    \label{def:datastructure}

    A \emph{symbolic interval $\Isymb$ of level $i$}  is either (i) of the form $\Isymb = (\text{sector}, l, u)$ where $l$ is either an indexed root expression of level $i$ or $-\infty$, and $u$ is either an indexed root expression of level $i$ or $\infty$; or (ii) of the form $\Isymb_i = (\text{section}, b)$ where $b$ is an indexed root expression of level $i$. The polynomials $\Isymb.l$, $\Isymb.u$ respectively $\Isymb.b$ are the \emph{defining polynomials of $\Isymb$} (if they exist).

    A \emph{cell data structure} is a sequence $\texttt{R}=(\Isymb_1,\ldots,\Isymb_n)$ of symbolic intervals $\Isymb_i$ of level $i$.  
    For an empty cell data structure, we define $\lift{()} = \{()\}$. For a cell data structure $(\Isymb_1,\ldots,\Isymb_i)$, $i\geq 1$, we define the corresponding subset of $\reals^i$ as $\lift{\Isymb_1,\ldots,\Isymb_i} = \{ (r,r') \mid r \in \lift{\Isymb_1,\ldots,\Isymb_{i-1}},\ r' \in (\Isymb_i.l(r),\Isymb_i.u(r)) \}$, respectively $\lift{\Isymb_1,\ldots,\Isymb_i} = \{ (r,r') \mid r \in \lift{\Isymb_1,\ldots,\Isymb_{i-1}},\ r' = \Isymb_i.b(r) \}$ if the respective indexed root expressions are not $\Undef$, and $\lift{\Isymb_1,\ldots,\Isymb_i} = \Undef$ otherwise. Similarly, we define $\lift{R,\Isymb_i}$ for arbitrary subsets $R \subseteq \reals^{i-1}$ and $\lift{r,\Isymb_i}$ for points $r \in \reals^{i-1}$.
    
    A \emph{CAD data structure} $\texttt{D}$ is a set of cell data structures. We define $\lift{\texttt{D}} = \{ \lift{\texttt{R}} \mid \texttt{R} \in \texttt{D} \}$.
\end{definition}

In our algorithms, indexed root expressions occurring in a cell data structure will always be defined; the above definition covers the $\Undef$ case just for the sake of completeness.
Furthermore, note that given a cell data structure $\texttt{R}=(\Isymb_1,\ldots,\Isymb_i)$, the restrictions of members $\restr{\Isymb_i.l}{\lift{\Isymb_1,\ldots,\Isymb_{i-1}}}$, $\restr{\Isymb_i.u}{\lift{\Isymb_1,\ldots,\Isymb_{i-1}}}$, $\restr{\Isymb_i.b}{\lift{\Isymb_1,\ldots,\Isymb_{i-1}}}$ are real root functions  of their defining polynomials as $\theta_j$ in \Cref{def:delineability}.

\begin{definition}[Indexed root expressions of polynomials]
    \label{def:irexpr}

    Let $i \in \naturals$, $s \in \reals^{i}$, and $p \in \rationals[x_1,\ldots,x_{i+1}]$ of level $i+1$. The \emph{set of indexed root expressions of $p$ at $s$} is defined as %
    \[ \irexpr{p}{s} = \begin{cases}
        \Undef & p(s,x_{i+1}) = 0,\\
        \{ \iroot{x_{i+1}}{p}{j} \mid j \in [1 .. \abs{\realRoots{p(s,x_{i+1})}}] \} & \text{otherwise}.
    \end{cases} \]
    \noindent Let $P \subseteq \rationals[x_1,\ldots,x_{i+1}]$ be a set of polynomials of level $i+1$. The \emph{set of indexed root expressions of $P$ at $s$} is defined as
    \[ \irexpr{P}{s} = \bigcup_{p \in P} \irexpr{p}{s}. \]
    \noindent Let $\xi \in \reals$. The \emph{set of indexed root expressions of $P$ for $s$ and $\xi$} is defined as \[ \irexprat{P}{s}{\xi} = \{ \iroot{x_{i+1}}{p}{j} \mid p \in P,\ j \in \posints,\ \xi = \iroot{x_{i+1}}{p}{j}(s) \}. \]
\end{definition}

A description of a sign-invariant CAD defined by a projection operator with respect to a set of polynomials can be computed as follows.  
First the projection operator is applied from level $n$ to $1$, called the \emph{projection phase}. This is followed by the \emph{lifting phase} where,  starting from level $1$, all cells from level $i-1$ are extended to the level $i$ such that all sections and sectors in the cylinder above every cell of level $i-1$ are identified as cells. To achieve this, for each cell $\texttt{R}$ of dimension $i-1$, the polynomials on level $i$ are partially evaluated up to their last dimension using a sample $s$ from $\texttt{R}$. This results in univariate polynomials whose roots can be isolated and sorted to give intervals:  point intervals at the roots of the polynomials, the open intervals between them, and the two open intervals below and above all roots (or the whole real line if there are no roots).  Delineability allows the conclusions drawn at the sample to generalize over $\texttt{R}$.  For each interval a symbolic description is determined that together extend $\texttt{R}$ to level $i$.

\section{Single cell computation}
\label{sec:singlecell}
For our problem of study, we are not interested in computing a full sign-invariant CAD, but only a single sign-invariant algebraic cell. That is, given a set of polynomials $P \subseteq \rationals[x_1,\ldots,x_n]$ and a sample $s \in \reals^n$, we need to compute a cell $R \subseteq \reals^n$ such that $s \in R$ and $P$ is sign-invariant on $R$.

In this paper, we focus on the computation of algebraic cells that adhere to \Cref{def:datastructure}. These cells have a triangular algebraic description with respect to the variable ordering (i.e. a condition on $x_1$ alone; then one on $(x_1,x_2)$, and so on).  Such cells are called \emph{(locally) cylindrical} \cite{brown2015onecell, abraham2020covering} as this shape is implied by the cylindricity property of a full CAD.  Although we do not construct entire decompositions in this paper, we make use of CAD theory which results in cells having this property.  Such cells have the advantage of being easy to visualize and compute with.  For example, projection with respect to the variable ordering is trivial, as is checking whether a point is inside such a cell, which is particularly important for our use of such cells. 

\subsection{Previous work on optimizing single cell computations}

As described in Section \ref{ssec:mcsat}, Jovanović and de Moura use a single cell construction for model explanation in their algorithm to decide satisfiability for non-linear arithmetic \cite{jovanovic2012solving}.  This was based on Collins'  CAD projection operator \cite{Collins1975} but did not compute a full CAD projection: it left out coefficients based on the current sample (as we only need to consider subsequent coefficients when the leading coefficient vanishes). 

Brown enhanced this single cell construction, first in the open case \cite{Brown2013} (i.e. building only cells with maximal dimension) and then for the general case \cite{brown2015onecell} in collaboration with Ko{\v{s}}ta.  This work was based on McCallum projection theory, and in addition to coefficients also avoided some computation of resultants and discriminants based on the current sample.  The work led in turn to the consideration of entire decompositions built one cell at a time \cite{Brown2015} and their use for quantifier elimination \cite{Brown2017}.  

The approach in \cite{Brown2015} started with a cell data structure describing the whole of $\reals^n$ which was continually refined by \emph{merging} in new polynomials one at a time.
The process maintains several invariants: (1) the cell contains the sample point $s$; (2) the cell is cylindrical; and (3) all polynomials merged so far have constant sign (more precisely, constant order) in the cell.  
The polynomials are merged one-by-one, and the order in which they are merged affects the cell that gets produced. \Cref{fig:recapproach} illustrates this process for \Cref{ex:motivation} below, showing different results for two different orderings of the polynomials.  From now on we will refer to this approach as \textit{refinement-based single cell construction}.  
The merge operation, unless on the lowest level, projects the current polynomial $p$ and then calls itself iteratively on the projection result. When the call returns to $p$, the polynomial is used to \textit{refine the bounds} of the cell in the dimension that corresponds to the level of $p$. This is done by isolating the roots of the univariate polynomial resulting from partially evaluating $p$ up to its last dimension using the sample $s$. If there is a root closer to $s_i$ than the current bound then it becomes the new bound of the \textit{sector}. In the special case that a root coincides with $s_i$, the sector collapses into a \textit{section} described by said root.  By this recursive refinement, the transitivity of the ordering on real root functions of polynomials induced by the order-invariance of resultants is exploited, so that at most two resultants are calculated per merge-operation and polynomial. Furthermore, as soon as a section is identified, some superfluous discriminants can be detected, and using the current sample irrelevant coefficients can be identified.

\begin{example}
    \label{ex:motivation}

    We consider the
    point $s = (\frac{1}{8},-\frac{3}{4})$ and polynomials
    $p_1 = x_1 - 2 x_2 + 1$,
    $p_2 = x_1^2 + x_2^2 - 1$ and
    $p_3 = x_1 - 2 x_2 - 1$.  
	These are the polynomials studied in \Cref{fig:recapproach} which demonstrates the refinement-based single cell construction process for two different orderings.  We see that one produces a larger cell than the other. It is not obvious how to choose good orders for adding polynomials in this approach (a machine learning heuristic was presented in \cite{BD20}).  Moreover, because the method works by refining the cell by the chosen polynomial, the order in which lower-level polynomials resulting from projections are added is not very flexible.
\end{example}

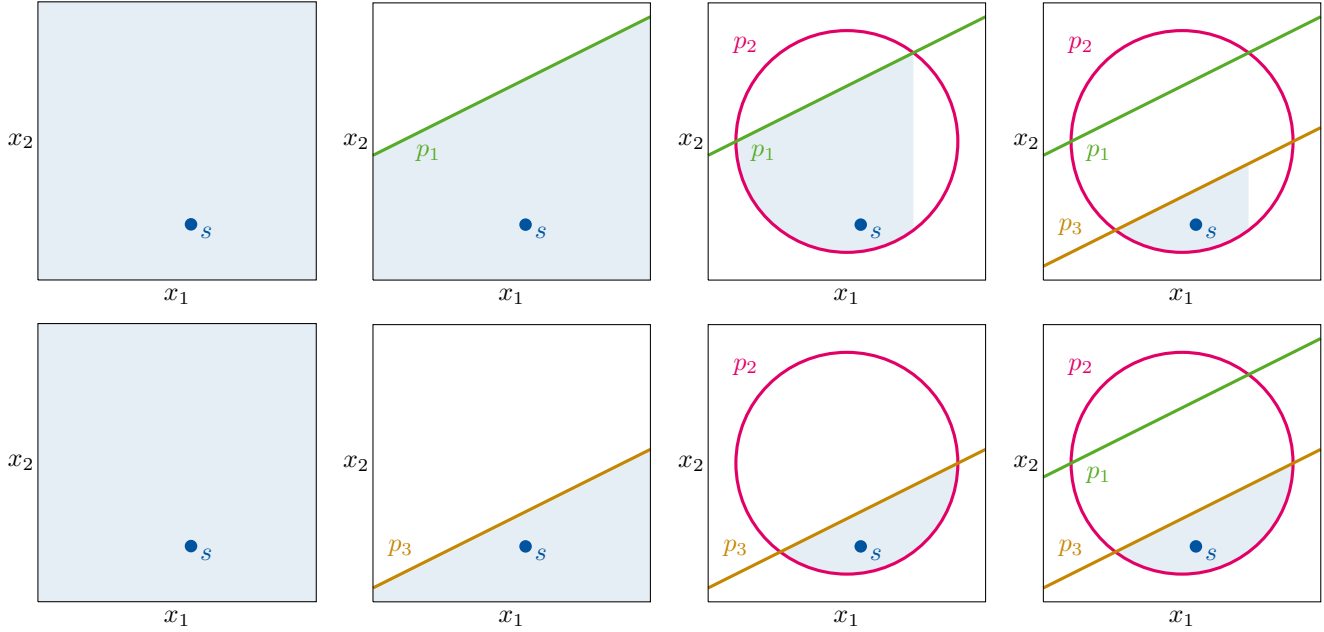
\begin{figure}[t]
  \center
  \begin{subfigure}[b]{0.24\linewidth}
  	\resizebox{\linewidth}{!}{%
  	  \begin{tikzpicture}[scale=1.5]
  	  	\draw[fill, blue100, fill opacity = .1, draw opacity = 0] (-1.25,-1.25) -- (1.25,-1.25) -- (1.25,1.25) -- (-1.25,1.25) -- (-1.25,-1.25);
  	  	\draw (-1.25,-1.25) -- (1.25,-1.25) -- (1.25,1.25) -- (-1.25,1.25) -- (-1.25,-1.25);
		\node at (0,-1.4) {$x_1$};
		\node at (-1.4,0) {$x_2$};
  	  	\node[circle,fill,inner sep=1.7pt,blue100,label={[below right]:\textcolor{blue100}{$s$}}] (dot) at (.125,-.75) {};
  	  \end{tikzpicture}%
    }
  \end{subfigure}
  \hfill
  \begin{subfigure}[b]{0.24\linewidth}
  	\resizebox{\linewidth}{!}{%
  	  \begin{tikzpicture}[scale=1.5]
  	  	\draw[fill, blue100, fill opacity = .1, draw opacity = 0] (-1.25,-1.25) -- (1.25,-1.25) -- (1.25,1.125) -- (-1.25,-.125) -- (-1.25,-1.25);
  	  	\draw[very thick, green100] (1.25,1.125) -- (-1.25,-.125) node[below=0.1,pos=0.8] {\color{green100}$p_1$};
  	  	\draw (-1.25,-1.25) -- (1.25,-1.25) -- (1.25,1.25) -- (-1.25,1.25) -- (-1.25,-1.25);
		\node at (0,-1.4) {$x_1$};
		\node at (-1.4,0) {$x_2$};
  	  	\node[circle,fill,inner sep=1.7pt,blue100,label={[below right]:\textcolor{blue100}{$s$}}] (dot) at (.125,-.75) {};
  	  \end{tikzpicture}%
    }
  \end{subfigure}
  \hfill
  \begin{subfigure}[b]{0.24\linewidth}
  	\resizebox{\linewidth}{!}{%
  	  \begin{tikzpicture}[scale=1.5]
  	  	\begin{scope}
  	  		\clip (-1.25,-1.25) -- (-1,0) -- (.6,.8) -- (.6,-1.25) -- (-1.25,-1.25);
  	  		\draw[fill, blue100, fill opacity = .1, draw opacity = 0] (0,0) ellipse (1cm and 1cm);
  	  	\end{scope}
  	  	\draw[very thick, magenta100] (0,0) ellipse (1cm and 1cm) node[above left=1.5cm] {\color{magenta100}$p_2$};
  	  	\draw[very thick, green100] (1.25,1.125) -- (-1.25,-.125) node[below=0.1,pos=0.8] {\color{green100}$p_1$};
  	  	\draw (-1.25,-1.25) -- (1.25,-1.25) -- (1.25,1.25) -- (-1.25,1.25) -- (-1.25,-1.25);
		\node at (0,-1.4) {$x_1$};
		\node at (-1.4,0) {$x_2$};
  	  	\node[circle,fill,inner sep=1.7pt,blue100,label={[below right]:\textcolor{blue100}{$s$}}] (dot) at (.125,-.75) {};
  	  \end{tikzpicture}%
    }
  \end{subfigure}
  \hfill
  \begin{subfigure}[b]{0.24\linewidth}
  	\resizebox{\linewidth}{!}{%
  	  \begin{tikzpicture}[scale=1.5]
  	  	\begin{scope}
  	  		\clip (-1.25,-1.25) -- (-.6,-.8) -- (.6,-.2) -- (.6,-1.25) -- (-1.25,-1.25);
  	  		\draw[fill, blue100, fill opacity = .1, draw opacity = 0] (0,0) ellipse (1cm and 1cm);
  	  	\end{scope}
  	  	\draw[very thick, magenta100] (0,0) ellipse (1cm and 1cm) node[above left=1.5cm] {\color{magenta100}$p_2$};
  	  	\draw[very thick, green100] (1.25,1.125) -- (-1.25,-.125) node[below=0.1,pos=0.8] {\color{green100}$p_1$};
  	  	\draw[very thick, darkorange100] (1.25,.125) -- (-1.25,-1.125) node[above=0.1,pos=0.9] {\color{darkorange100}$p_3$};
  	  	\draw (-1.25,-1.25) -- (1.25,-1.25) -- (1.25,1.25) -- (-1.25,1.25) -- (-1.25,-1.25);
		\node at (0,-1.4) {$x_1$};
		\node at (-1.4,0) {$x_2$};
  	  	\node[circle,fill,inner sep=1.7pt,blue100,label={[below right]:\textcolor{blue100}{$s$}}] (dot) at (.125,-.75) {};
  	  \end{tikzpicture}%
    }
  \end{subfigure} \\ \vspace{.1 cm}
  \begin{subfigure}[b]{0.24\linewidth}
  	\resizebox{\linewidth}{!}{%
  		\begin{tikzpicture}[scale=1.5]
  			\draw[fill, blue100, fill opacity = .1, draw opacity = 0] (-1.25,-1.25) -- (1.25,-1.25) -- (1.25,1.25) -- (-1.25,1.25) -- (-1.25,-1.25);
  			\draw (-1.25,-1.25) -- (1.25,-1.25) -- (1.25,1.25) -- (-1.25,1.25) -- (-1.25,-1.25);
			  \node at (0,-1.4) {$x_1$};
		\node at (-1.4,0) {$x_2$};
  			\node[circle,fill,inner sep=1.7pt,blue100,label={[below right]:\textcolor{blue100}{$s$}}] (dot) at (.125,-.75) {};
  		\end{tikzpicture}%
  	}
  \end{subfigure}
  \hfill
  \begin{subfigure}[b]{0.24\linewidth}
  	\resizebox{\linewidth}{!}{%
  		\begin{tikzpicture}[scale=1.5]
  			\draw[fill, blue100, fill opacity = .1, draw opacity = 0] (-1.25,-1.25) -- (1.25,-1.25) -- (1.25,.125) -- (-1.25,-1.125) -- (-1.25,-1.25);
  			\draw[very thick, darkorange100] (1.25,.125) -- (-1.25,-1.125) node[above=0.1,pos=0.9] {\color{darkorange100}$p_3$};
  			\draw (-1.25,-1.25) -- (1.25,-1.25) -- (1.25,1.25) -- (-1.25,1.25) -- (-1.25,-1.25);
			  \node at (0,-1.4) {$x_1$};
		\node at (-1.4,0) {$x_2$};
  			\node[circle,fill,inner sep=1.7pt,blue100,label={[below right]:\textcolor{blue100}{$s$}}] (dot) at (.125,-.75) {};
  		\end{tikzpicture}%
  	}
  \end{subfigure}
  \hfill
  \begin{subfigure}[b]{0.24\linewidth}
  	\resizebox{\linewidth}{!}{%
  		\begin{tikzpicture}[scale=1.5]
  			\begin{scope}
  				\clip (-1.25,-1.25) -- (-.6,-.8) -- (1,0) -- (1.25,-1.25) -- (-1.25,-1.25);
  				\draw[fill, blue100, fill opacity = .1, draw opacity = 0] (0,0) ellipse (1cm and 1cm);
  			\end{scope}
  		    \draw[very thick, magenta100] (0,0) ellipse (1cm and 1cm) node[above left=1.5cm] {\color{magenta100}$p_2$};
  			\draw[very thick, darkorange100] (1.25,.125) -- (-1.25,-1.125) node[above=0.1,pos=0.9] {\color{darkorange100}$p_3$};
  			\draw (-1.25,-1.25) -- (1.25,-1.25) -- (1.25,1.25) -- (-1.25,1.25) -- (-1.25,-1.25);
			  \node at (0,-1.4) {$x_1$};
		\node at (-1.4,0) {$x_2$};
  			\node[circle,fill,inner sep=1.7pt,blue100,label={[below right]:\textcolor{blue100}{$s$}}] (dot) at (.125,-.75) {};
  		\end{tikzpicture}%
  	}
  \end{subfigure}
  \hfill
  \begin{subfigure}[b]{0.24\linewidth}
  	\resizebox{\linewidth}{!}{%
  		\begin{tikzpicture}[scale=1.5]
  			\begin{scope}
  				\clip (-1.25,-1.25) -- (-.6,-.8) -- (1,0) -- (1.25,-1.25) -- (-1.25,-1.25);
  				\draw[fill, blue100, fill opacity = .1, draw opacity = 0] (0,0) ellipse (1cm and 1cm);
  			\end{scope}
  			\draw[very thick, magenta100] (0,0) ellipse (1cm and 1cm) node[above left=1.5cm] {\color{magenta100}$p_2$};
  			\draw[very thick, darkorange100] (1.25,.125) -- (-1.25,-1.125) node[above=0.1,pos=0.9] {\color{darkorange100}$p_3$};
  			\draw[very thick, green100] (1.25,1.125) -- (-1.25,-.125) node[below=0.1,pos=0.8] {\color{green100}$p_1$};
  			\draw (-1.25,-1.25) -- (1.25,-1.25) -- (1.25,1.25) -- (-1.25,1.25) -- (-1.25,-1.25);
			  \node at (0,-1.4) {$x_1$};
		\node at (-1.4,0) {$x_2$};
  			\node[circle,fill,inner sep=1.7pt,blue100,label={[below right]:\textcolor{blue100}{$s$}}] (dot) at (.125,-.75) {};
  		\end{tikzpicture}%
  	}
  \end{subfigure}
  
  \caption{Refinement-based construction of a single cell for \Cref{ex:motivation}.  In the upper row the original cell (the entire plane) is refined first by polynomial $p_1$, then by $p_2$, then by $p_3$.
  In the lower row  a different order of refinement is used: $p_3$, $p_2$, $p_1$ giving a larger cell. 
  Note that in the figures of this paper we label the varieties of polynomials with their name, i.e. $p_1$ labels $p_1=0$.
}
\label{fig:recapproach}
\end{figure}

Our approach will build a single cell levelwise, i.e. constructing the cell level-by-level according to a variable ordering instead of polynomial-by-polynomial for all levels.  We note that there exists another \emph{levelwise} single cell construction in an unpublished work on the arXiv, \cite{liSolvingSatisfiabilityPolynomial2020}. In comparison, our formulation using a proof system enables the use of more sophisticated optimizations in the later sections. 

\subsection{Our levelwise approach to single cell computation}

Our new \emph{levelwise} approach is based on using information about the ordering of the roots of polynomials relative to the sample point $s$ and to one another to make decisions during the cell construction process that are likely to lead to larger cells.

We will exploit said information and the transitivity of the ordering on the roots induced by order-invariance of resultants.
Thus, before projection, the roots of the polynomials on the current level are isolated and ordered to see which resultants are required for maintaining sign- or order-invariance of the given polynomials.

\begin{figure}[t]
  \center
  \begin{subfigure}[b]{0.4\linewidth}
  	\centering
  	\begin{tikzpicture}[scale=2]
		\draw (-1.25,-1.25) -- (1.25,-1.25) -- (1.25,1.25) -- (-1.25,1.25) -- (-1.25,-1.25);
		\node at (0,-1.4) {$x_1$};
		\node at (-1.4,0) {$x_2$};
  		
  		\draw[very thick, magenta100, opacity = .25] (0,0) ellipse (1cm and 1cm) node[above left=2cm, opacity = 1] {\color{magenta100}$p_2$};
  		
  		\draw[very thick, green100, opacity = .25] (-1.25,-.125) -- (1.25,1.125) node[below=0.1,pos=0.3, opacity = 1] {\color{green100}$p_1$};
  		
  		\draw[very thick, darkorange100, opacity = .25] (-1.25,-1.125) -- (1.25,.125) node[above=0.15,pos=0.1, opacity = 1] {\color{darkorange100}$p_3$};
  		
  		\draw[thick, dotted, blue100] (.125,-1.25) -- (.125,1.25);
  		
  		\node[circle,fill,inner sep=1.7pt,blue100,label={[below right]:\textcolor{blue100}{$s$}}] (dot) at (.125,-.75) {};
  		
  		\node[circle,fill,inner sep=1.2pt,label={[below right]:\small $\xi_1$}] (dot) at (.125,-.992) {};
  		\node[circle,fill,inner sep=1.2pt,label={[below right]:\small $\xi_2$}] (dot) at (.125,-.4375) {};
  		\node[circle,fill,inner sep=1.2pt,label={[below right]:\small $\xi_3$}] (dot) at (.125,.5625) {};
  		\node[circle,fill,inner sep=1.2pt,label={[above right]:\small $\xi_4$}] (dot) at (.125,.992) {};
	  	
  	\end{tikzpicture}
  \end{subfigure}
  \begin{subfigure}[b]{0.4\linewidth}
  	\centering
  	\begin{tikzpicture}[scale=2]
		  \draw (-1.25,-1.25) -- (1.25,-1.25) -- (1.25,1.25) -- (-1.25,1.25) -- (-1.25,-1.25);
		  \node at (0,-1.5) {$x_1$};
		  \node at (-1.5,0) {$x_2$};

  		\draw[very thick, magenta100, opacity = .25] (0,0) ellipse (1cm and 1cm) node[above left=2cm, opacity = 1] {\color{magenta100}$p_2$};
  		
  		\draw[very thick, green100, opacity = .25] (-1.25,-.125) -- (1.25,1.125) node[below=0.1,pos=0.3,  opacity = 1] {\color{green100}$p_1$};
  		
  		\draw[very thick, darkorange100, opacity = .25] (-1.25,-1.125) -- (1.25,.125) node[above=0.15,pos=0.1, opacity = 1] {\color{darkorange100}$p_3$};
  		
  		\begin{scope}
  			\clip(-.125,1.25) -- (.375,1.25) -- (.375,-1.25) -- (-.125,-1.25) -- (-.125,1.25);
  			\draw[very thick, magenta100] (0,0) ellipse (1cm and 1cm);
  			\draw[very thick, green100] (1.25,1.125) -- (-1.25,-.125);
  			\draw[very thick, darkorange100] (1.25,.125) -- (-1.25,-1.125);
  		\end{scope}
  		
  		\draw[thick, dotted, blue100] (.125,-1.25) -- (.125,1.25);
  		
  		\node[circle,fill,inner sep=1.7pt,blue100,label={[below right]:\textcolor{blue100}{$s$}}] (dot) at (.125,-.75) {};

		\draw [decorate,decoration={brace,amplitude=5pt,mirror,raise=2pt},yshift=0pt, thick] (.35,-.92) -- (.35,-0.33) node [midway,xshift=0.5cm] {$\Isymb_2$};
  	\end{tikzpicture}
  \end{subfigure} \\
  \begin{subfigure}[b]{0.4\linewidth}
  	\centering
  	\begin{tikzpicture}[scale=2]
  		\draw (-1.25,-1.25) -- (1.25,-1.25) -- (1.25,1.25) -- (-1.25,1.25) -- (-1.25,-1.25);
		\node at (0,-1.4) {$x_1$};
		\node at (-1.4,0) {$x_2$};

  		\draw[very thick, magenta100, opacity = .25] (0,0) ellipse (1cm and 1cm) node[above left=2cm, opacity = 1] {\color{magenta100}$p_2$};
  		
  		\draw[very thick, green100, opacity = .25] (-1.25,-.125) -- (1.25,1.125) node[below=0.1,pos=0.3,  opacity = 1] {\color{green100}$p_1$};
  		
  		\draw[very thick, darkorange100, opacity = .25] (-1.25,-1.125) -- (1.25,.125) node[above=0.1,pos=0.15, opacity = 1] {\color{darkorange100}$p_3$};
  		
  		\begin{scope}
  			\clip(-.125,1.25) -- (.375,1.25) -- (.375,-1.25) -- (-.125,-1.25) -- (-.125,1.25);
  			\draw[very thick, magenta100] (0,0) ellipse (1cm and 1cm);
  			\draw[very thick, green100] (1.25,1.125) -- (-1.25,-.125);
  			\draw[very thick, darkorange100] (1.25,.125) -- (-1.25,-1.125);
  		\end{scope}
  		
  		\draw[thick, dotted, blue100] (.125,-1.25) -- (.125,1.25);
  		
  		\node[circle,fill,inner sep=1.7pt,blue100,label={[above right=-0.1cm]:\textcolor{blue100}{$s_{[i-1]}$}}] (dot) at (.125,-1.25) {};
  		
  		\draw[thick, gray, dashed] (-1,-1.25) -- (-1,.5);
  		\draw[thick, gray, dashed] (-.6,-1.25) -- (-.6,-.31);
  		\draw[thick, gray, dashed] (1,-1.25) -- (1,.5);
  		\node[circle,fill,inner sep=1.2pt,label={[above right]:\small $p_4$}] (dot) at (-1,-1.25) {};
  		\node[circle,fill,inner sep=1.2pt,label={[above right]:\small $p_5$}] (dot) at (-.6,-1.25) {};
  		\node[circle,fill,inner sep=1.2pt,label={[above right]:\small $p_4$}] (dot) at (1,-1.25) {};
  		
  	\end{tikzpicture}
  \end{subfigure}
  \begin{subfigure}[b]{0.4\linewidth}
  	\centering
  	\begin{tikzpicture}[scale=2]
		  \draw (-1.25,-1.25) -- (1.25,-1.25) -- (1.25,1.25) -- (-1.25,1.25) -- (-1.25,-1.25);
		  \node at (-.5,-1.5) {$x_1$};
		  \node at (-1.5,0) {$x_2$};

  		\begin{scope}
  			\clip (-1.25,-1.25) -- (-.6,-.8) -- (1,0) -- (1.25,-1.25) -- (-1.25,-1.25);
  			\draw[fill, blue100, fill opacity = .2, draw opacity = 0] (0,0) ellipse (1cm and 1cm);
  		\end{scope}
  		
  		\draw[very thick, magenta100, opacity = .25] (0,0) ellipse (1cm and 1cm) node[above left=2cm, opacity = 1] {\color{magenta100}$p_2$};
  		
  		\draw[very thick, green100, opacity = .25] (-1.25,-.125) -- (1.25,1.125) node[below=0.1,pos=0.3,  opacity = 1] {\color{green100}$p_1$};
  		
  		\draw[very thick, darkorange100, opacity = .25] (-1.25,-1.125) -- (1.25,.125) node[above=0.1,pos=0.15, opacity = 1] {\color{darkorange100}$p_3$};
  		
  		\begin{scope}
  			\clip(-.125,1.25) -- (.375,1.25) -- (.375,-1.25) -- (-.125,-1.25) -- (-.125,1.25);
  			\draw[very thick, magenta100] (0,0) ellipse (1cm and 1cm);
  			\draw[very thick, green100] (1.25,1.125) -- (-1.25,-.125);
  			\draw[very thick, darkorange100] (1.25,.125) -- (-1.25,-1.125);
  		\end{scope}
  		
  		\draw[thick, dotted, blue100] (.125,-1.25) -- (.125,1.25);
  		
  		\node[circle,fill,inner sep=1.7pt,blue100,label={[above right=-0.1cm]:\textcolor{blue100}{$s_{[i-1]}$}}] (dot) at (.125,-1.25) {};
  		
  		\draw[thick, gray, dashed] (-1,-1.25) -- (-1,.5);
  		\draw[thick, gray, dashed] (-.6,-1.25) -- (-.6,-.31);
  		\draw[thick, gray, dashed] (1,-1.25) -- (1,.5);
  		\node[circle,fill,inner sep=1.2pt,label={[above right]:\small $\xi'_1$}] (dot) at (-1,-1.25) {};
  		\node[circle,fill,inner sep=1.2pt,label={[above right]:\small $\xi'_2$}] (dot) at (-.6,-1.25) {};
  		\node[circle,fill,inner sep=1.2pt,label={[above right]:\small $\xi'_3$}] (dot) at (1,-1.25) {};

		\draw [decorate,decoration={brace,amplitude=5pt,mirror,raise=2pt},yshift=0pt, thick] (-.6,-1.25) -- (1,-1.25) node [midway,yshift=-0.5cm] {$\Isymb_1$};
  	\end{tikzpicture}
  \end{subfigure}
  \caption{Construction of a single cell for \Cref{ex:motivation} following the new levelwise approach as described in \Cref{ex:motivation_cont}. 
}
  \label{fig:lwapproach}
\end{figure}
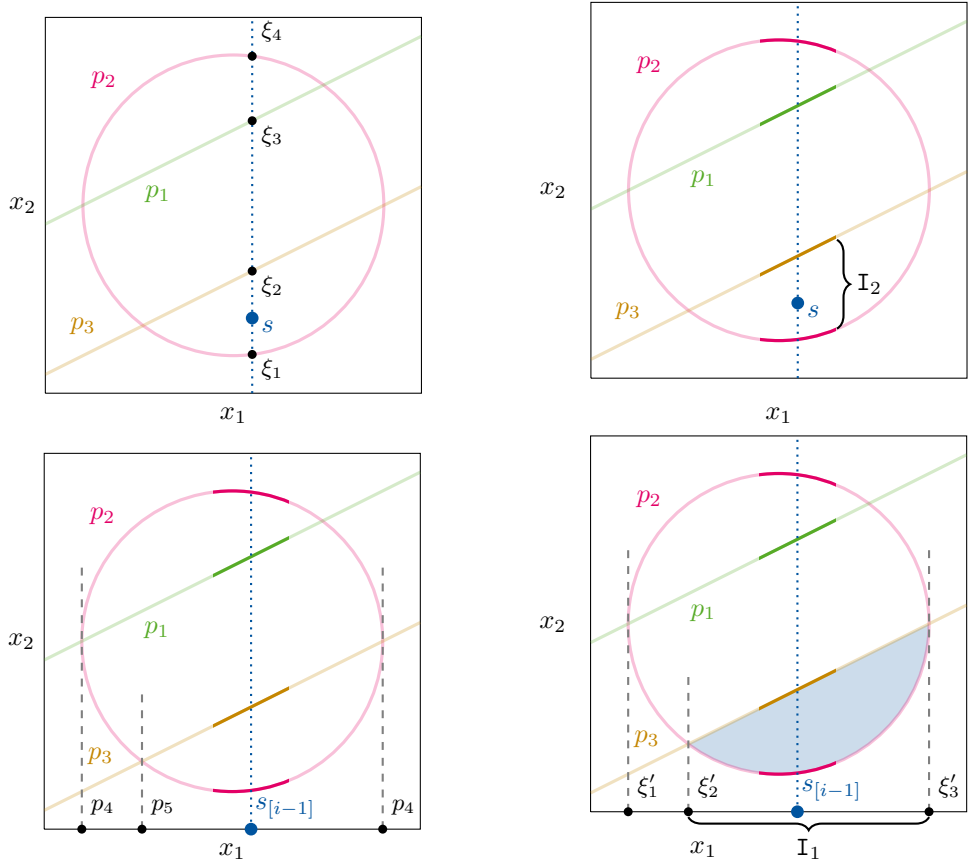

\begin{example}\label{ex:motivation_cont}
	\Cref{fig:lwapproach} shows how our new levelwise approach would operate on the formula from \Cref{ex:motivation}.  We start by considering $x_1 = s_{1}$, the first coordinate of the sample $s$.  We evaluate the polynomials at this coordinate and isolate the real roots to find $\xi_1=\iroot{x_2}{p_2}{1}$, $\xi_2=\iroot{x_2}{p_3}{1}$, $\xi_3=\iroot{x_2}{p_1}{1}$, and $\xi_4=\iroot{x_2}{p_2}{2}$.  We order these, along with the second coordinate of $s$, as in the top-left image.  
	
Our sample lies in the interval for $x_2$ denoted $\Isymb_2=(\text{sector}, \xi_1, \xi_2)$ in the top-right image.  We thus know that, local to $s_{1}$, the cell we want will be bounded from below by $p_2$ and from above by $p_3$.  We also know that the other section of $p_2$ and the section of $p_1$ are above the cell.  
	In this figure the graphs of the polynomials are greyed out because the algorithm is not aware of them: it knows only their roots when evaluated at the sample.  However, this is enough to allow the algorithm to infer behaviour around the sample, as illustrated by the thick partial lines.  

As we generalize from the sample we must ensure than $\xi_1$ and $\xi_2$ remain well-defined and that no root function crosses the symbolic interval described by $\Isymb_2$.  To achieve this we compute a projection consisting of $p_4 = \disc{x_2}{p_2}$ and $p_5 = \res{x_2}{p_3}{p_2}$ (also $\disc{x_2}{p_1}$ and $\disc{x_2}{p_3}$ but they do not have any real roots and so play no further role in this example).  Crucially, because $p_1$ has only one section and that section lies above the cell of interest, the resultant of $p_1$ and the lower-boundary polynomial $p_2$ is not required or computed.  The bottom-left figure shows how the zeros of the projection map onto the geometric features.  

Analogously to $x_2$, we isolate the real roots $\xi'_1 = \iroot{x_1}{p_4}{1}, \xi'_2 = \iroot{x_1}{p_5}{1}, \xi'_3 = \iroot{x_1}{p_4}{2}$ in $x_1$.  We determine the interval $\Isymb_1 = (\text{sector}, \xi'_2, \xi'_3)$ around $s_{1}$ for $x_1$ and thus generate the cell in the bottom-right image.

\end{example}

So we see that for \Cref{ex:motivation_cont} the new levelwise approach produces the second of the two possible outcomes from the refinement-based construction (\Cref{fig:recapproach}).  The example illustrates how the levelwise approach avoids ``mistakes'' that the refinement-based construction can make due to poor choices of orderings for the polynomials.  

The levelwise approach also allows greater flexibility in how polynomials resulting from projection are handled, but that aspect of the algorithm requires more variables than in this simple example in order to be observed.

\section{Proof system for single cells}
\label{sec:proofsystem}

In \Cref{sec:levelwise} we will present our new levelwise procedure to construct a single cell.  We lay the basis for this by first formalizing our work as a proof system in this section.  The system expresses how the properties we want to conclude rely on other ``smaller'' properties.  From this, the procedure in the next section gains maximal freedom to make heuristic choices.  Before we go into the formal presentation, we outline the idea for our running example.

\subsection{Motivating example}

\begin{example}
\label{Ex61}

Recall \Cref{ex:motivation} and \Cref{fig:lwapproach} where we sought to construct a cell $R \subseteq \reals^2$ around $s=(s_1,s_2)$ such that the sign-invariance of $p_1,p_2,p_3$ hold on $R$.  We started by eliminating $x_2$: we isolated real roots of $p_j(s_{1},x_2)$ to define $\xi_1, \dots \xi_4$, sorted them, and observed that $\xi_1(s_1) < s_2 < \xi_2(s_1)$, thus we determined that in the second level the cell was represented by $\Isymb_2 = (sector, \xi_1, \xi_2)$.

This root ordering and the representation should now be generalized such that the underlying cell $R_{1}$ is restricted by specifying certain properties ensuring $\xi_1$ and $\xi_2$ remain as cell boundaries: they must remain well-defined over $R_1$; no further root of $p_1, p_2, p_3$ should appear that intersects $\Isymb_2$ over $R_1$; and the other roots, $\xi_3$ and $\xi_4$, should remain outside the symbolic interval $\Isymb_2$ over $R_1$. For the latter, we extend to a partial ordering $\preceq$ with $\xi_2 \preceq \xi_3$ and $\xi_2 \preceq \xi_4$.

To conclude that $p_1,p_2,p_3$ are sign-invariant in $R$, the proof system shows we need to maintain the properties that: the sample $s$ is included in $R$; $\Isymb_2$ describes the cell's boundaries on the current level; the underlying cell $R_{1}$ is a connected analytic submanifold; and $p_1, p_2, p_3$ are delineable. Furthermore, we maintain that the partial ordering $\preceq$ is maintained over $R_{1}$.  

To maintain these properties the proof system concludes that we must also prove order-invariance and sign-invariance of some coefficients, resultants, and discriminants of the input polynomials.  After simplification, we find that on level $1$ the polynomial $p_4 = \disc{x_2}{p_2}$ and $p_5 = \res{x_2}{p_3}{p_2}$ must be ensured sign-invariant. This leads to a representation $\Isymb_1 = (sector, \xi'_2, \xi'_3)$ and then since all polynomials are univariate and their zeros are algebraic numbers, no further projection or proof steps are necessary. The constructed cell is described by $\texttt{R} = (\Isymb_1, \Isymb_2)$.

The graph of the proof constructed is shown in \Cref{fig:proofex_proof}, in which the sign-invariance properties $\sgninv{p_1}$, $\sgninv{p_2}$, $\sgninv{p_3}$ lead to $R = \lift{\proj{R}{[1]}, \Isymb_2}$ and $\proj{R}{[1]} = \lift{\Isymb_1}$.  The exact proof rules used in that figure will be defined in the rest of this section.
\end{example}

Note how in the example there were multiple choices for $\preceq$.  Choosing $\xi_2 \preceq \xi_3$ and $\xi_3 \preceq \xi_4$ would have been a valid choice but that leads to a smaller cell (the one depicted in the top right of \Cref{fig:recapproach}). We will discuss ordering heuristics to make this choice in \Cref{sec:orderingheuristics}.

\begin{figure}[t]
    \begin{tikzpicture}
        \node (a) {
            \begin{minipage}{1\linewidth}
            \begin{align*}
                \sgninv{p_1}(R) \dashv & \sample{s}(R), \representation{\Isymb_2, s_1}(R), \irordering{\preceq,s_1}(R), \del{p_1}(R), \\ &\quad  \submanifold{1}(R), \connected{1}(R) \\
                \sgninv{p_2}(R) \dashv & \sample{s}(R), \representation{\Isymb_2, s_1}(R), \irordering{\preceq,s_1}(R), \del{p_2}(R),\\ &\quad \submanifold{1}(R), \connected{1}(R) \\
                \sgninv{p_3}(R) \dashv & \sample{s}(R), \representation{\Isymb_2, s_1}(R), \irordering{\preceq,s_1}(R), \del{p_3}(R), \\ &\quad  \submanifold{1}(R), \connected{1}(R) \\
                \sample{s}(R) \dashv & \representation{\Isymb_2, s_1}(R), \sample{s_1}(R) \\
                \representation{\Isymb_2, s_1}(R) \dashv & R = \lift{\proj{R}{[1]}, \Isymb_2}, \del{p_2}(R), \del{p_3}(R), \sample{s_1}(R) \\
                \irordering{\preceq,s_1}(R) \dashv & \del{p_1}(R), \del{p_2}(R), \del{p_3}(R), \ordinv{\res{x_2}{p_3}{p_1}}(R), \\ &\quad \ordinv{\res{x_2}{p_3}{p_2}}(R),  \submanifold{1}(R), \connected{1}(R), \sample{s_1}(R) \\ %
                \del{p_1}(R)\dashv & \nonnull{p_1}(R), \ordinv{\disc{x_2}{p_1}}(R), \sgninv{\ldcf{x_2}{p_1}}(R), \\ &\quad \submanifold{1}(R), \connected{1}(R)\\
                \del{p_2}(R)\dashv & \nonnull{p_2}(R), \ordinv{\disc{x_2}{p_2}}(R), \sgninv{\ldcf{x_2}{p_2}}(R), \\ &\quad \submanifold{1}(R), \connected{1}(R)\\
                \del{p_3}(R)\dashv & \nonnull{p_3}(R), \ordinv{\disc{x_2}{p_3}}(R), \sgninv{\ldcf{x_2}{p_3}}(R), \\ &\quad \submanifold{1}(R), \connected{1}(R)\\
                \nonnull{p_1}(R)\dashv & \true \\
                \nonnull{p_2}(R)\dashv & \true \\
                \nonnull{p_3}(R)\dashv & \true \\
                \ordinv{\disc{x_2}{p_1}}(R) \dashv & \true \\
                \ordinv{\disc{x_2}{p_2}}(R)\dashv & \sgninv{p_4}(R), \sample{s_1}(R) & \\
                \ordinv{\disc{x_2}{p_3}}(R)\dashv & \true \\
                \sgninv{p_4}(R) \dashv & \representation{\Isymb_1}(R) \\
                \ordinv{\res{x_2}{p_3}{p_1}}(R) \dashv & \true \\
                \ordinv{\res{x_2}{p_3}{p_2}}(R) \dashv & \sgninv{p_5}(R), \sample{s_1}(R) \\
            \end{align*}
        \end{minipage}
        };
        \node[anchor=south east] at (a.south east) {
            \begin{minipage}{0.5\linewidth}
            \begin{align*}
                \sgninv{p_5}(R) \dashv & \representation{\Isymb_1}(R) \\
                \sgninv{\ldcf{x_2}{p_1}}(R) \dashv & \true \\
                \sgninv{\ldcf{x_2}{p_2}}(R) \dashv & \true \\
                \sgninv{\ldcf{x_2}{p_3}}(R) \dashv & \true \\
                \submanifold{1}(R) \dashv & \representation{\Isymb_1}(R)\\
                \connected{1}(R) \dashv & \true \\
                \sample{s_1}(R) \dashv & \representation{\Isymb_1}(R) \\
                \representation{\Isymb_1}(R)\dashv & \proj{R}{[1]} = \lift{\Isymb_1} \\
            \end{align*}
        \end{minipage}
        };
    \end{tikzpicture}
    
    \caption{Proof for \Cref{Ex61} (slightly truncated, read in columns). $Q \dashv P_1, \ldots, P_k$ means that $Q$ is derived from $P_1, \ldots, P_k$.}
    \label{fig:proofex_proof}
\end{figure}
\subsection{Proof system: Properties and rules}

Our system is for constructing cells.  We will define \emph{properties} of the cell that is being constructed and \emph{rules of inference} to infer that a property holds for a cell given some other properties and conditions.  As indicated in the previous example, there may be different rules that can be used to prove a property.

\begin{definition}[Property]
    Let $i \in \naturals$.
    A \emph{property of level $i$} is a function $q: \{ R \mid  R \subseteq \reals^i \} \to \bools$. A property $q$ of level $i$ \emph{holds} on a set $R  \subseteq \reals^i$ if $q(R) = 1$. The set $\propset$ denotes the set of all properties of any level. For a set $Q \subseteq \propset$, we denote by $\levels{Q}{i}$ the set of properties of level $i$ and by $\levels{Q}{[i]}$ the set of properties of level at most $i$.
\end{definition}

Note that the concept of levels of polynomials has been extended to properties in the sense that the satisfaction of a property of level $i$ cannot be influenced by values of the variables $x_{i+1}, \ldots, x_n$.

We denote a \emph{rule of inference} in the form $P_1, \ldots, P_k \vdash Q$, where $P_1, \ldots, P_k$ are the antecedents and $Q$ is the consequent.
 
For convenience, we allow lifting properties from lower levels to higher levels as follows.
\begin{definition}
    We extend every property $q$ of level $i$ to $q: \{ R \mid R \subseteq \reals^j,\ j \geq i \} \to \bools$ such that $q(R) = q(\proj{R}{[i]})$ for any $R \subseteq \reals^j,\ j \geq i$. Accordingly, we introduce the following rule of inference:
    \begin{align*}
        & R \subseteq \reals^{i+1} ,\ q \in \levels{\propset}{i} ,\ q(\proj{R}{[i]}) && \vdash q(R)
    \end{align*}
\end{definition}
So, for example, this rule allows us to say that because the property of sign-invariance holds for $x^2 -2$ in the one-level cell $-1 < x < 1$, sign-invariance holds for $x^2 - 2$ in the two-level cell which is the interior of the unit circle, since this projects down to $-1 < x < 1$.

The set of inference rules will be defined in such a way that the property of sign-invariance of a polynomial is reduced to the property of sign-invariance of polynomials at a lower level: thus the chain of rules is finite. On each level, we will additionally determine a \emph{representation} describing the symbolic boundaries of the cell on this level as well as an ordering of the indexed roots of some polynomials assuring their sign-invariance which will be used later for heuristic decisions.

\subsection{Basic cell properties}

We will first define the basic properties on polynomials and cells that follow from the original work on McCallum CAD projection.  Throughout, $i$ will refer to the level of the given property and $s$ refers to \emph{algebraic} points. 
It is important to recall that an $i$-level property is a function that maps subsets of $\reals^i$ to Boolean values.  Consider for example the first element of \Cref{def:prop:basic}: ``$\sample{s}$'' for $s\in\reals^i$ is a property, meaning  that $\textit{sample}$ is a function that maps a point (and a level $i$ that is given implicitly by the point) to a function mapping subsets of $\reals^i$ to Booleans.  
Thus, $\sample{s}$ is not a true/false value.  Rather, it is the function $\sample{s}$ applied to a given subset of $\reals^i$ that yields a true/false value.  Note that for the fifth and sixth items in \Cref{def:prop:basic} the level $i$ must be given explicitly, while with the others the level is implicit in other arguments.

\begin{property}
    \label{def:prop:basic}

    Let $i \in \naturals$, and $R \subseteq \reals^i$. %

    \begin{enumerate}
        \item Let $s \in \reals^i$ be a sample point. The property $\sample{s}$ holds on $R$ if and only if $s \in R$.
        \item Let $p \in \rationals[x_1,\ldots,x_i]$ be a polynomial of level $i$. The property $\ordinv{p}$ holds on $R$ if and only if $p$ is order-invariant on $R$.
        \item Let $p \in \rationals[x_1,\ldots,x_i]$ be a polynomial of level $i$. The property $\sgninv{p}$ holds on $R$ if and only if $p$ is sign-invariant on $R$.
        \item Let $p \in \rationals[x_1,\ldots,x_{i+1}]$ be a polynomial of level $i+1$. The property $\nonnull{p}$ holds on $R$ if and only if $p$ is not nullified on any point in $R$.
        \item The property $\submanifold{i}$ holds on $R$ if and only if $R$ is an analytic submanifold.
        \item The property $\connected{i}$ holds on $R$ if and only if $R$ is connected.
        \item Let $p \in \rationals[x_1,\ldots,x_{i+1}]$ be a polynomial of level $i+1$. The property $\del{p}$ holds on $R$ if and only if $p$ is analytically delineable on some connected superset $R' \supseteq R$ and $p$ is order-invariant in all $p$-sections over $R'$.
    \end{enumerate}
\end{property}

Note that $\del{p}$ is defined such that it holds on a subset $R \subseteq \reals^i$ if and only if there is a connected superset of $R$ on which $p$ is analytically delineable. This is due to the fact that in general, we cannot assume that $R$ is a connected set, but delineability is only defined on connected sets. We will use this trick in the following for other properties as well.

Further, note that for some properties such as $\sgninv{p}$, if they hold on $R \subseteq \reals^i$, then they also hold on all subsets of $R$. For others such as $\submanifold{i}$ or $\connected{i}$, this is not the case: a subset of $R$ is not necessarily an analytic submanifold nor connected! This is one of the reasons why the proof rules below cannot cover all subsets where a given property holds, but they do cover sufficiently many subsets for our purposes.

For the remainder of this subsection we will present proof rules for these properties (with proofs of correctness available in \ref{sec:proofsystemcorrectness}).  For the ease of navigation we give references to the rules which apply to prove each property above (and provide similar references after introducing further properties in later subsections).

In the following rule definitions we list assumptions before the actual inference rule. These assumptions are formally part of the inference rule, but we keep them separated to improve readability. 

We start by defining rules that relate to McCallum's projection. 

\begin{restatable}{mapping}{defmapdel}
    \label{def:map:del}

    Let $i \in \naturals$, $R \subseteq \reals^i$, and $p \in \rationals[x_1, \ldots, x_{i+1}],\ \level{p}=i+1$. Assume that $p$ is irreducible.
    \begin{align*}
        & \submanifold{i}(R),\  \connected{i}(R),\  \nonnull{p}(R),\  \ordinv{\disc{x_{i+1}}{p}}(R),\  \sgninv{\ldcf{x_{i+1}}{p}}(R) & & \vdash  \del{p}(R)
    \end{align*}
\end{restatable}

\begin{restatable}{mapping}{defmapnonnull}
    \label{def:map:nonnull}

    Let $i \in \naturals$, $R \subseteq \reals^i$, $s \in \reals^i$, and $p \in \rationals[x_1, \ldots, x_{i+1}],\ \level{p}=i+1$. Assume that $p$ is irreducible, and $p=c_m \cdot x_{i+1}^m + \ldots + c_1 \cdot x_{i+1} + c_0$ such that $c_m, \ldots, c_0 \in \rationals[x_1, \ldots, x_{i}]$.

    \begin{align*}
        & \sample{s}(R) ,\  \vdeg{x_{i+1}}{p}>1,\  \disc{x_{i+1}}{p}(s) \neq 0,\  \sgninv{\disc{x_{i+1}}{p}}(R) & & \vdash \nonnull{p}(R) \\
        & \sample{s}(R) ,\  \exists j \in [m].\; (c_j(s) \neq 0 \wedge \sgninv{c_j}(R)) & & \vdash \nonnull{p}(R)
    \end{align*}
\end{restatable}

Note that the condition $\vdeg{x_{i+1}}{p}>1 \wedge \disc{x_{i+1}}{p}(s) \neq 0$ already implies that $p(s,x_{i+1}) \neq 0$, as does $c_j(s) \neq 0$. In practice, it is good to delay choosing which rule to use because we may observe that one of the non-zero $c_j$'s is already required to be sign-invariant from some other part of the projection process (e.g. leading coefficients are often added to the projection). Moreover, the second case is trivial if the $c_j$ is constant.

\begin{restatable}{mapping}{defmapinvtrivial}
    \label{def:map:invtrivial}

    Let $p \in \rationals$.
    \begin{align*}
        & & \vdash \ordinv{p}(\reals^0) \\
        & & \vdash \sgninv{p}(\reals^0) \\
    \end{align*}
\end{restatable}

\begin{restatable}{mapping}{defmapinvreducible}
    \label{def:map:invreducible}

    Let $i \in \posints$, $R \subseteq \reals^i$, and $p \in \rationals[x_1, \ldots, x_{i}],\ \level{p}=i$. Assume that $p$ is reducible, and $\factors{p} = \{ q_1, \ldots, q_k \}$.
    \begin{align*}
        \ordinv{q_1}(R),\ \ldots,\ \ordinv{q_k}(R) & & \vdash \ordinv{p}(R)\\
        \sgninv{q_1}(R),\ \ldots,\ \sgninv{q_k}(R) & & \vdash \sgninv{p}(R)
    \end{align*}
\end{restatable}

\begin{restatable}{mapping}{defmapordinv}
    \label{def:map:ordinv}

    Let $i \in \posints$, $R \subseteq \reals^i$, $s \in \reals^i$, and $p \in \rationals[x_1, \ldots, x_{i}],\ \level{p}=i$. Assume that $p$ is irreducible. 
    \begin{align*}
        & p(s) \neq 0,\  \sample{s}(R),\  \sgninv{p}(R) & & \vdash \ordinv{p}(R) \\
        & p(s) = 0,\  \sample{s}(R),\   \submanifold{i-1}(R),\   \connected{i}(R),\   \sgninv{p}(R),\   \del{p}(R)  & & \vdash \ordinv{p}(R)
    \end{align*}
\end{restatable}

\subsection{Sign invariance for polynomials without roots}

\begin{figure}
    \begin{subfigure}{0.49\textwidth}
        \center
        \begin{tikzpicture}

            \draw (-.2,-.2) -- (4.2,-.2) -- (4.2,4.2) -- (-.2,4.2) -- (-.2,-.2);
		    \node at (2,-.5) {$x_1$};
		    \node at (-.5,2) {$x_2$};

            \draw[very thick, magenta100] (3,2) ellipse (1cm and 1cm);
            \node[circle,fill,inner sep=2pt,blue100,label={[right]:\textcolor{blue100}{$s$}}] (dot) at (1,2) {};

            \draw[dashed, thick, gray] (2,4.2) -- (2,-.2);
        \end{tikzpicture}

        \caption{Without singularities.}
    \end{subfigure}
    \begin{subfigure}{0.49\textwidth}
        \center
        \begin{tikzpicture}

            \draw (-.2,-.2) -- (4.2,-.2) -- (4.2,4.2) -- (-.2,4.2) -- (-.2,-.2);
		    \node at (2,-.5) {$x_1$};
		    \node at (-.5,2) {$x_2$};

            \draw[very thick, color=magenta100,domain=-0.2:1.55, variable=\x] plot ({\x},{(2*\x-3)/(\x-2)});
            \draw[very thick, color=magenta100,domain=2.45:4.2, variable=\x] plot ({\x},{(2*\x-3)/(\x-2)});
            \node[circle,fill,inner sep=2pt,blue100,label={[right]:\textcolor{blue100}{$s$}}] (dot) at (2,2) {};

            \draw[dashed, thick, gray] (2,4.2) -- (2,-.2);
        \end{tikzpicture}
        
        \caption{With singularity.}
    \end{subfigure}

    \caption{Examples of the two cases of polynomials without roots at the projection $s_1$ of the sample $s=(s_1,s_2)$. Dashed lines denote cylinder boundaries over the projected cell.}
    \label{fig:map:nozero}
\end{figure}
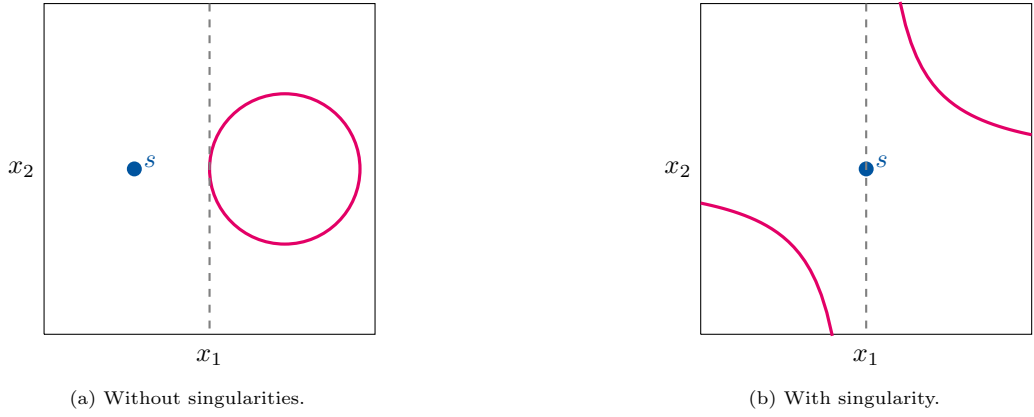

For polynomials without roots over the current sample, we must ensure that no further root appears over the underlying cell. In this case, it is sufficient to make the polynomial delineable. \Cref{fig:map:nozero} illustrates some cases.

\begin{restatable}{mapping}{defmapnozero}
    \label{def:map:nozero}

    Let $i \in \naturals$, $R \subseteq \reals^i$, $s \in \reals^{i-1}$, and $p \in \rationals[x_1, \ldots, x_{i}],\ \level{p}=i$. Assume that $p$ is irreducible, and $\realRoots{p(s,x_{i})} = \emptyset$.
    \begin{align*}
        & \sample{s}(R),\   \del{p}(R) & & \vdash \sgninv{p}(R) 
    \end{align*}
\end{restatable}

\subsection{Cell boundary representations}

We are aiming to generate the description of a sign-invariant cell for the input polynomials. In the language of our proof system: at every level we identify a symbolic interval upon which we prove sign-invariance of the polynomials of that level. To do this, we take their real roots over the current sample into account.

We determine whether we are in a section or sector (i.e. whether a polynomial has a root exactly at the current sample point or not) and pick indexed root expressions (as given by \Cref{def:irexpr}) to be symbolic descriptions of the cell boundaries that generalize a concrete interval around the sample point:  either the root at the current sample or the largest root below and the lowest root above. Note that the choice of such boundaries might not be unique in the cases where two root functions cross over the underlying sample. We will discuss more general choices of the symbolic intervals in \Cref{sec:heuristics}.

Our next property is designed to state that a designated representation describes the cell boundaries on the current level.

To do so, we define how we use indexed root expressions to refer to real root functions of a polynomial. 
An indexed root expression has incontinuities where the number of the roots of its defining polynomial changes. Therefore a root function of a polynomial is only uniquely determined by an indexed root expression in combination with a given sample. In the following, we define the unique real root function determined this way:

\begin{definition}[Root functions]
    \label{def:rrfats}
    Let $i \in \posints$, $s \in \reals^{i-1}$, and $\xi \in \irexpr{\xi.p}{s}$ be an indexed root expression of level $i$.

    Let $R \subseteq \reals^i$ maximal such that $\xi.p$ is delineable on $R$, $R$ is connected, and $s \in R$. Then $\theta_{\xi,s}$ denotes the real root function of $p$ on $R$ that witnesses analytic delineability of $p$ and equals $\restr{\xi}{R}$.
\end{definition}

\begin{property}
    \label{def:prop:representation}
    Let $i \in \posints$, $R \subseteq \reals^i$, $s \in \reals^{i-1}$, and $\Isymb$ be a symbolic interval of level $i$.
    
    The property $\representation{\Isymb,s}$ holds on $R$ if and only if $\Isymb.l \in \irexpr{\Isymb.l.p}{s}$ (if $\Isymb.l \neq -\infty$), $\Isymb.u \in \irexpr{\Isymb.u.p}{s}$ (if $\Isymb.u \neq \infty$) respectively $\Isymb.b \in \irexpr{\Isymb.b.p}{s}$ and one of the following holds:
    \begin{itemize}
        \item $\Isymb = (\text{sector},l,u)$, $\dom{\theta_{l,s}} \cap \dom{\theta_{u,s} }\supseteq \proj{R}{[i-1]}$  and $R = \{ (r,r') \mid r \in \proj{R}{[i-1]}, r' \in (\theta_{l,s}(r), \theta_{u,s}(r)) \}$; or %
        \item $\Isymb = (\text{sector},-\infty,u)$, $\dom{\theta_{u,s} } \supseteq \proj{R}{[i-1]}$ and $R = \{ (r,r') \mid r \in \proj{R}{[i-1]}, r' \in (-\infty, \theta_{u,s}(r)) \}$; or %
        \item $\Isymb = (\text{sector},l,\infty)$, $\dom{\theta_{l,s}}  \supseteq \proj{R}{[i-1]}$ and $R = \{ (r,r') \mid r \in \proj{R}{[i-1]}, r' \in (\theta_{l,s}(r), \infty) \}$; or %
        \item $\Isymb = (\text{sector},-\infty,\infty)$ and $R = \{ (r,r') \mid r \in \proj{R}{[i-1]}, r' \in \reals \}$; or
        \item $\Isymb = (section,b)$, $\dom{\theta_{b,s}} \supseteq \proj{R}{[i-1]}$ and $R = \{ (r,r') \mid r \in \proj{R}{[i-1]}, r' = \theta_{b,s}(r) \}$, %
    \end{itemize}
    (for the real root functions $\theta_{l,s}$, $\theta_{u,s}$ respectively $\theta_{u,b}$ according to \Cref{def:rrfats}).
\end{property}

Since the cell's boundaries are described by real root functions, we can prove that it is an analytic submanifold.

\begin{restatable}{mapping}{defmapsubmanifold}
    \label{def:map:submanifold}

    Let $i \in \posints$, $R \subseteq \reals^i$, $s \in \reals^{i-1}$, and $\Isymb$ be a symbolic interval of level $i$.
    \begin{align*}
        & & & \vdash  \submanifold{0}(\reals^0) \\
        & \representation{\Isymb,s}(R),\  \submanifold{i-1}(R) & & \vdash  \submanifold{i}(R)
    \end{align*}
\end{restatable}

\subsection{Equational constraint projection}
\label{subsec:ec}

An optimization to CAD which tailors it to the logical structure of the problem is the theory of equational constraints.  A polynomial equation is an \emph{equational constraint} if it is logically implied by the truth of the input formula.  If the input to CAD has an equational constraint then we may perform fewer projection and lifting operations to achieve truth-invariance.  This optimization dates back to \cite{mccallum1999equational} with the recent paper \cite{EBD20} summarizing the state of the art.  The examples there demonstrate the drastic savings that are achievable in these cases (the analysis in \cite{EBD20} shows the double exponent in the complexity bound decreases for each equational constraint).

In the context of building a single cell we have even greater scope to use equational constraint savings.  Here, any time we are generalizing in the section case, we can apply the reduced projection and lifting.  I.e., if a polynomial is zero at our sample then it must also be zero throughout the cell we are generalizing the sample to, and so for the purposes of this cell, it is an equational constraint.  In this case we need only make the section defining polynomial delineable.  All other constraints can be made sign-invariant simply by including resultants with the section defining polynomial (but not other projection polynomials).

\begin{restatable}{mapping}{defmapeqproj}
    \label{def:map:eqproj}

    Let $i \in \posints$, $R \subseteq \reals^i$, $s \in \reals^{i-1}$, $p \in \rationals[x_1, \ldots, x_{i}],\ \level{p}=i$, and $\Isymb$ be a symbolic interval of level $i$. Assume that $p$ is irreducible, and $\Isymb = (\text{section},b)$.

    Let $Q := \submanifold{i-1}(R) \wedge \connected{i-1}(R) \wedge \representation{\Isymb,s}(R) \wedge \del{b.p}(R)$.
    \begin{align*}
        & Q,\  b.p = p && \vdash \sgninv{p}(R) \\
        & Q,\  b.p \neq p,\  \ordinv{\res{x_i}{b.p}{p}}(R) && \vdash \sgninv{p}(R)
    \end{align*}
\end{restatable}

\subsection{Root orderings}

We define a partial ordering on the indexed root expressions of the given set of polynomials, which we will use to ensure that none of their roots crosses the cell's boundary on the current level.

We first give a general definition of indexed root orderings.

\begin{definition}[Indexed root ordering]
    \label{def:irordering}

    Let $i \in \naturals$, and $\Xi$ be a set of indexed root expressions of level $i+1$. An \emph{indexed root ordering on $\Xi$} is a relation $\preceq \subseteq \Xi \times \Xi$ such that its reflexive and transitive closure $\preceq^t$ is a partial order on $\Xi$. Indexed root orderings of this form are also called \emph{indexed root ordering of level $i+1$}, and we define $\dom{\preceq} = \{ \xi, \xi' \mid (\xi,\xi') \in \preceq \}$.

    Let $s \in \reals^i$. An indexed root ordering $\preceq$ of level $i+1$ \emph{matches $s$} if and only if $\xi \in \irexpr{\xi.p}{s}$ for all $\xi \in \dom{\preceq}$ and $\xi(s) \leq \xi'(s)$ for all $(\xi,\xi') \in \preceq$.
\end{definition}

In our algorithm we will pick an indexed root ordering such that, in the sector case, roots lower than the interval's lower bound remain lower, and roots greater than the interval's upper bound remain greater. In the section case, the lower and upper bounds in that condition both refer to a single bound. We do not give an explicit definition here yet, as it will be part of the rules defined below.

The motivation for introducing the general concept of indexed root orderings is to allow for choices between different root orderings. We will discuss them in \Cref{sec:heuristics}.

\begin{example}
    Given indexed root expressions $\Xi = \{ \xi_1,\ldots,\xi_5 \}$ of level $i+1$ and a sample $s \in \reals^i$ such that $\xi_1(s)<\ldots<\xi_5(s)$ and $\Isymb = (\text{sector}, \xi_1, \xi_2)$; then $\preceq \, = \{ (\xi_2,\xi_3), (\xi_2,\xi_4), (\xi_2,\xi_5) \}$, $\preceq' \,= \{ (\xi_2,\xi_3), (\xi_3,\xi_4), (\xi_2,\xi_5) \}$, and $\preceq'' \,= \{ (\xi_2,\xi_3), (\xi_3,\xi_4), (\xi_4,\xi_5) \}$ are all indexed root orderings on $\Xi$ that allow us to derive sign-invariance of the corresponding polynomials above the sample $s$ and the interval $\Isymb$.
\end{example}

\begin{figure}
    \begin{subfigure}[t]{0.49\textwidth}
        \center
        \begin{tikzpicture}[scale=1]

            \draw (0,0) -- (4,0) -- (4,4) -- (0,4) -- (0,0);
            \node at (2,-.5) {$x_1$};
            \node at (-.5,2) {$x_2$};

            \draw[very thick,color=magenta100,domain=0:4, variable=\x] plot ({\x},{0.125*\x*(\x-1.5)*(\x-3))*(\x-4)+1});
            \draw[very thick,color=green100,domain=0:4, variable=\x] plot ({\x},{-0.5*(\x-2)*(\x-2)+2.5});
            \draw[very thick,color=darkorange100,domain=0.25:3.32, variable=\x] plot ({\x},{-0.5*(\x-1.5)*(\x-1.5)*(\x-1.5)+3});
            \node at (1,1) {\textcolor{magenta100}{$\xi_1$}};
            \node at (1,2.5) {\textcolor{green100}{$\xi_2$}};
            \node at (1,3.5) {\textcolor{darkorange100}{$\xi_3$}};

            \draw[blue100, thick, dotted] (2,4) -- (2,0) node [above right]{$s$};

            \draw[dashed, thick, gray] (0.15,4) -- (0.15,0);
            \draw[dashed, thick, gray] (2.6,4) -- (2.6,0);
        \end{tikzpicture}

        \caption{Assume $\preceq = \{ (\xi_1,\xi_2), (\xi_2,\xi_3) \}$. All three real root functions are defined over the underlying cell. Only two resultants are necessary to maintain the ordering of the root functions, e.g. $\res{x_2}{\xi_1.p}{\xi_2.p}$ and $\res{x_2}{\xi_2.p}{\xi_3.p}$.  We conclude the ordering between $\xi_1$ and $\xi_3$ by transitivity.}
        \label{fig:map:irordering:1}
    \end{subfigure}\hfill
    \begin{subfigure}[t]{0.49\textwidth}
        \center
        \begin{tikzpicture}[scale=1]

            \draw (0,0) -- (4,0) -- (4,4) -- (0,4) -- (0,0);
            \node at (2,-.5) {$x_1$};
            \node at (-.5,2) {$x_2$};

            \draw[very thick,color=magenta100] (2,1.5) ellipse (1.8cm and 0.8cm);
            \draw[very thick,color=darkorange100,domain=0.25:3.32, variable=\x] plot ({\x},{-0.5*(\x-1.5)*(\x-1.5)*(\x-1.5)+3});
            \node at (1,0.6) {\textcolor{magenta100}{$\xi_1$}};
            \node at (1,1.8) {\textcolor{magenta100}{$\xi_2$}};
            \node at (1,3.5) {\textcolor{darkorange100}{$\xi_3$}};

            \draw[blue100, thick, dotted] (2,4) -- (2,0) node [above right]{$s$};

            \draw[dashed, thick, gray] (0.2,4) -- (0.2,0);
            \draw[dashed, thick, gray] (2.65,4) -- (2.65,0);
        \end{tikzpicture} 

        \caption{Assume $\preceq = \{ (\xi_1,\xi_2), (\xi_1,\xi_3) \}$. The lower bound on $x_1$ is due to the delineability of the common defining polynomial of $\xi_1$ and $\xi_2$. The ordering $\xi_1 \preceq \xi_2$ is maintained by $\disc{x_2}{\xi_1.p}$. The ordering $\xi_1 \preceq \xi_3$ by the resultant $\res{x_2}{\xi_1.p}{\xi_3.p}$. Note that latter causes the upper bound on $x_1$ to be smaller than necessary, as $\xi_3.p$ intersects $\theta_{\xi_2,s}$ ``before'' it intersects $\theta_{\xi_1,s}$. More sophisticated rules circumventing this are future work.}
        \label{fig:map:irordering:2}
    \end{subfigure}

    \caption{Two examples for the property $\irordering{\preceq,s}$.}
    \label{fig:map:irordering}
\end{figure}

We need to maintain that an indexed root ordering holds over the underlying cell. First, we will define this property formally.

\begin{property}
    \label{def:prop:irordering}

    Let $i \in \naturals$, $R \subseteq \reals^i$, $s \in \reals^i$, $\preceq$ be an indexed root ordering of level $i+1$, and $\preceq^t$ be the reflexive and transitive closure of $\preceq$.

    The property $\irordering{\preceq,s}$ holds on $R$ if and only if $\preceq$ matches $s$, for all $\xi \in \dom{\preceq}$ it holds $R \subseteq \dom{\theta_{\xi,s}}$, and for all $\xi,\xi' \in \dom{\preceq}$ it holds that
    \[ \xi \preceq^t \xi' \wedge \xi(s) < \xi'(s) \implies \theta_{\xi,s}(r) < \theta_{\xi',s}(r) \text{ for all } r \in R \]
    and
    \[ \xi \preceq^t \xi' \wedge \xi(s) = \xi'(s)  \implies \theta_{\xi,s}(r) = \theta_{\xi',s}(r) \text{ for all } r \in R \]
    (for the real root functions $\theta_{\xi,s}$, $\theta_{\xi',s}$ according to \Cref{def:rrfats}).
\end{property}

Note that the semantics of an indexed root expression $\xi$ is only defined in combination with the sample $s$, as this uniquely determines the real root function $\theta_{\xi,s}$; thus, we add $s$ to the signature of the property. To clarify this further: the sample $s$ is only used to identify the referred real root functions and is not necessarily contained in the constructed cell by definition of the above property (although in the following proof rules, this will be required).

The property is maintained by delineability and adding resultants for the defining polynomials of a pair of indexed root expressions to the projection. We essentially prove that the ordering of root functions ensured by the order-invariance of resultants is transitive. \Cref{fig:map:irordering} shows an example where transitivity is maintained for three root functions.

\begin{restatable}{mapping}{defmapirordering}
    \label{def:map:irordering}

    Let $i \in \naturals$, $R \subseteq \reals^i$, $s \in \reals^i$, and $\preceq$ be an indexed root ordering of level $i+1$. Assume that $\xi.p$ is irreducible for all $\xi \in \dom{\preceq}$, and that $\preceq$ matches $s$.
    \begin{align*}
        \sample{s}(R),\ \submanifold{i}(R),\ \connected{i}(R),\  \forall \xi \in \dom{\preceq}.\; \del{\xi.p}(R),\ \\ \forall (\xi, \xi') \in \preceq.\; \ordinv{\res{x_{i+1}}{\xi.p}{\xi'.p}}(R) && \vdash \irordering{\preceq,s}(R)
    \end{align*}
\end{restatable}

\begin{figure}
    \begin{subfigure}[t]{0.49\textwidth}
        \center
        \begin{tikzpicture}[scale=0.75]
            \draw (0,0) -- (4,0) -- (4,6) -- (0,6) -- (0,0);
		    \node at (2,-.5) {$x_1$};
		    \node at (-.5,2) {$x_2$};

            \node[circle,fill,inner sep=2pt,blue100,label={[right]:\textcolor{blue100}{$s$}}] (dot) at (1,3) {};

            \draw[very thick,color=magenta100,domain=0:4, variable=\x] plot ({\x},{0.5*(\x-2)*(\x-2)+2});
            \node[magenta100] at (0.4,4) {$p_1$};

            \draw[very thick,color=darkorange100,domain=0:4, variable=\x] plot ({\x},{-0.5*(\x-2)*(\x-2)+4});
            \node[darkorange100] at (3.6,2) {$p_2$};

            \draw[very thick,color=green100,domain=0:4, variable=\x] plot ({\x},{-0.005*(\x-4)*(\x-4)+5});
            \draw[very thick,color=green100,domain=0:4, variable=\x] plot ({\x},{0.1*(\x+1)*(\x+1)+1.5});
            \node[green100] at (0.3,1.4) {$p_3$};
            \node[green100] at (0.3,5.3) {$p_3$};

            \draw[dashed, thick, gray] (0.6,6) -- (0.6,0);
            \draw[dashed, thick, gray] (1.58,6) -- (1.58,0);
        \end{tikzpicture}

        \caption{The indexed root ordering ensures that real root functions remain outside the cell. As $p_3$ is delineable, this is sufficient to prove sign invariance.}
        \label{fig:map:sgninvord:1}
    \end{subfigure}\hfill
    \begin{subfigure}[t]{0.49\textwidth}
        \center
        \begin{tikzpicture}[scale=0.75]
            \draw (0,0) -- (4,0) -- (4,6) -- (0,6) -- (0,0);
		    \node at (2,-.5) {$x_1$};
		    \node at (-.5,2) {$x_2$};

            \node[circle,fill,inner sep=2pt,blue100,label={[right]:\textcolor{blue100}{$s$}}] (dot) at (1,3) {};

            \draw[very thick,color=magenta100,domain=0:4, variable=\x] plot ({\x},{0.5*(\x-2)*(\x-2)+2});
            \node[magenta100] at (0.4,4) {$p_1$};

            \draw[very thick,color=darkorange100,domain=0:4, variable=\x] plot ({\x},{-0.5*(\x-2)*(\x-2)+4});
            \node[darkorange100] at (3.6,2) {$p_2$};

            \draw[very thick,color=green100,domain=0:1.143, variable=\x] plot ({\x},{52*\x*\x-169*\x+134)/(16*\x*\x-52*\x+40});
            \draw[very thick,color=green100,domain=1.375:1.875, variable=\x] plot ({\x},{52*\x*\x-169*\x+134)/(16*\x*\x-52*\x+40});
            \draw[very thick,color=green100,domain=2.105:4, variable=\x] plot ({\x},{52*\x*\x-169*\x+134)/(16*\x*\x-52*\x+40});
            \node[green100] at (1.5,5.3) {$p_3$};
            \node[green100] at (2.2,0.5) {$p_3$};
            \node[green100] at (2.5,5.3) {$p_3$};

            \draw[dashed, thick, gray] (0.6,6) -- (0.6,0);
            \draw[dashed, thick, gray] (1.2,6) -- (1.2,0);
        \end{tikzpicture}

        \caption{We note that if $p_3$ has singularities, the cell might be made smaller than required for ensuring sign-invariance due to the strong notion of delineability. Relaxing this notion is part of future research.}
        \label{fig:map:sgninvord:2}
    \end{subfigure}\hfill
    \caption{Two examples for maintaining the sign-invariance of a polynomial $p_3$ in a cell (here defined by the enclosed region between the graphs of $p_1$ and $p_2$).}
    \label{fig:map:sgninvord}
\end{figure}
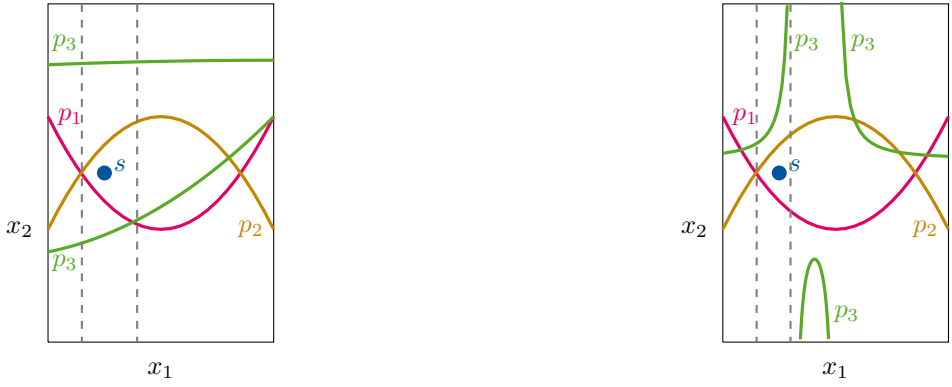

Now we are ready to define the rule for ensuring the sign-invariance of a polynomial $p$, given a representation $\Isymb$ and an indexed root ordering $\preceq$. The latter will be used to ensure that no real root function of $p$ crosses the cell's boundaries. The cases are depicted in \Cref{fig:map:sgninvord}.

\begin{restatable}{mapping}{defmapsgninvord}
    \label{def:map:sgninvord}
    Let $i \in \posints$, $R \subseteq \reals^i$, $s \in \reals^{i-1}$, $p \in \rationals[x_1, \ldots, x_{i}],\ \level{p}=i$, $\Isymb$ be a symbolic interval of level $i$, $\preceq$ be an indexed root ordering of level $i$, and $\preceq^t$ be the reflexive and transitive closure of $\preceq$.

    We choose $l,u$ such that either $\Isymb=(\text{sector},l,u)$ or ($\Isymb=(\text{section},b)$ for $b=l=u$).
    
    Assume that $p$ is irreducible, $\irexpr{p}{s} \neq \emptyset$, $\xi.p$ is irreducible for all $\xi \in \dom{\preceq}$, $\preceq$ matches $s$, and for all $\xi \in \irexpr{p}{s}$ it holds either $\xi \preceq^t l$ or $u \preceq^t \xi$.

    \begin{align*}
        \sample{s}(R),\ \representation{\Isymb,s}(R),\ \irordering{\preceq,s}(R),\ \del{p}(R)&& \vdash \sgninv{p}(R)
    \end{align*}
\end{restatable}

\subsection{Connectedness}

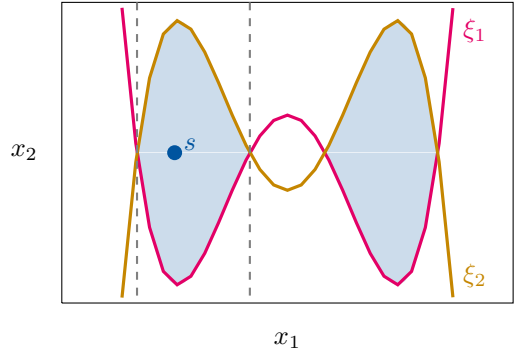
\begin{figure}%
    \begin{center}
        \begin{tikzpicture}

            \draw (-3,0) -- (3,0) -- (3,4) -- (-3,4) -- (-3,0);
		    \node at (0,-.5) {$x_1$};
		    \node at (-3.5,2) {$x_2$};

            \fill[color=blue100,opacity=0.2,domain=-2:-0.5, variable=\x] plot ({\x},{0.5*(\x-2)*(\x-0.5)*(\x+0.5)*(\x+2)+2}) -- plot ({\x},{-0.5*(\x-2)*(\x-0.5)*(\x+0.5)*(\x+2)+2});
            \fill[color=blue100,opacity=0.2,domain=0.5:2, variable=\x] plot ({\x},{0.5*(\x-2)*(\x-0.5)*(\x+0.5)*(\x+2)+2}) -- plot ({\x},{-0.5*(\x-2)*(\x-0.5)*(\x+0.5)*(\x+2)+2});
            
            \draw[very thick, color=magenta100,domain=-2.2:2.2, variable=\x] plot ({\x},{0.5*(\x-2)*(\x-0.5)*(\x+0.5)*(\x+2)+2}) node[below right] {$\xi_1$};
            \draw[very thick, color=darkorange100,domain=-2.2:2.2, variable=\x] plot ({\x},{-0.5*(\x-2)*(\x-0.5)*(\x+0.5)*(\x+2)+2}) node[above right] {$\xi_2$};

            \node[circle,fill,inner sep=2pt,blue100,label={[right]:\textcolor{blue100}{$s$}}] (dot) at (-1.5,2) {};

            \draw[dashed, thick, gray] (-2,4) -- (-2,0);
            \draw[dashed, thick, gray] (-0.5,4) -- (-0.5,0);
        \end{tikzpicture}
    \end{center}
    \caption{Let $\Isymb = (sector,\xi_1,\xi_2)$, then $\representation{\Isymb,s_1}$ holds in the whole blue area, thus forming a non-connected subset of $\reals^2$. By adding $\res{x_2}{\xi_1.p}{\xi_2.p}$ to the projection, the subset is made connected by restricting the underlying cell as indicated by the dashed vertical lines.}
    \label{fig:map:connectedness}
\end{figure}

For maintaining properties (in particular the order-invariance of some polynomials) on higher levels, we need to maintain connectedness. Besides some preconditions, i.e. that the cell's boundaries are described by two well-defined real root functions, we need to ensure that the lower and upper bound of a sector do not cross as illustrated in \Cref{fig:map:connectedness}.

\begin{restatable}{mapping}{defmapconnectedness}
    \label{def:map:connectedness}

    Let $i \in \posints$, $R \subseteq \reals^i$, $s \in \reals^{i-1}$, $\Isymb$ be a symbolic interval of level $i$, $\preceq$ be an indexed root ordering of level $i$, and $\preceq^t$ be the reflexive and transitive closure of $\preceq$. Assume that $\preceq$ matches $s$. 

    \ \newline
    Let $Q := \connected{i-1}(R) \wedge \representation{\Isymb,s}(R)$.
    \begin{align*}
        & && \vdash \connected{0}(\reals^0) \\
        & Q,\ \Isymb=(\textit{sector},l,u),\  l \neq -\infty,\  u \neq \infty,\  l \preceq^t u,\ \irordering{\preceq,s} && \vdash \connected{i}(R) \\
        & Q,\ \Isymb=(\textit{sector},l,u),\ l = -\infty \vee u = \infty && \vdash \connected{i}(R) \\
        & Q,\ \Isymb=(\textit{section},b) && \vdash \connected{i}(R) 
    \end{align*}
\end{restatable}

\subsection{Generalization of the current sample}

We define a mapping for generalizing the sample to the cell on the current level.

\begin{restatable}{mapping}{defmapsample}
    \label{def:map:sample}

    Let $i \in \posints$, $R \subseteq \reals^i$, $s \in \reals^i$, and $\Isymb$ be a symbolic interval of level $i$. Assume that $s_i \in \lift{s_{[i-1]},\Isymb}$.
    \begin{align*}
        & && \vdash \sample{()}(\reals^0) \\
        & \sample{s_{[i-1]}}(R),\   \representation{\Isymb,s_{[i-1]}}(R) && \vdash \sample{s}(R) 
    \end{align*}
\end{restatable}

\subsection{Cell descriptions}

Now, we define a property that states that the cell is described by its representation.

\begin{property}
    \label{def:prop:irint}

    Let $i \in \posints$, $R \subseteq \reals^i$, and $\Isymb$ be a symbolic interval of level $i$. 

    The property $\irint{\Isymb}$ holds on $R$ if and only if $R=\lift{\proj{R}{[i-1]},\Isymb}$.
\end{property}

This property is the only one that cannot be mapped to a set of other properties, i.e. properties of this kind are the assumptions in our proof system.

Given $\irint{\Isymb}$, we can maintain $\representation{\Isymb,s}$. To do so, we need to ensure that the indexed root expressions describing the cell's boundaries always refers to the same root on the underlying cell by making their defining polynomials delineable (such that the referred real root function is defined and the number of roots below the referred root function is constant over the underlying cell).

\begin{restatable}{mapping}{defmaprepr}
    \label{def:map:repr}

    Let $i \in \posints$, $R \subseteq \reals^{i-1}$, $s \in \reals^{i-1}$, and $\Isymb$ be a symbolic interval of level $i$. Assume that $\Isymb.l \in \irexpr{\Isymb.l.p}{s}$ (if $\Isymb.l \neq -\infty$), $\Isymb.u \in \irexpr{\Isymb.u.p}{s}$ (if $\Isymb.u \neq \infty$) respectively $\Isymb.b \in \irexpr{\Isymb.b.p}{s}$.
    \begin{align*}
        & \sample{s}(R) ,\  \irint{\Isymb}(R) ,\  \Isymb = (\text{section}, b),\  \del{b.p}(R)  && \vdash \representation{\Isymb,s}(R) \\
        & \sample{s}(R) ,\  \irint{\Isymb}(R) ,\  \Isymb = (\text{sector}, l, u),\  l = -\infty \vee \del{l.p}(R) ,\  u = \infty \vee \del{u.p}(R)  && \vdash \representation{\Isymb,s}(R)
    \end{align*}
\end{restatable}

\subsection{Ordering of properties}

We observe that the rules of inference defined in \Cref{def:map:del,def:map:nonnull,def:map:invtrivial,def:map:invreducible,def:map:ordinv,def:map:nozero,def:map:submanifold,def:map:eqproj,def:map:irordering,def:map:sgninvord,def:map:connectedness,def:map:sample,def:map:repr} are cycle-free in the sense that all properties in the antecedents are smaller than the consequent property according to some ordering $\vartriangleleft$:

\begin{definition}
    \label{def:cyclefree}

    The ordering $\vartriangleleft$ is defined such that properties of level $i$ are greater than any property of level $i-1$, and such that it satisfies the following partial order of properties of each level $i$ (starting with the greatest element):

    \begin{enumerate}
        \item $\irordering{\preceq,s}$ for all $\preceq$ for roots of level $i$ and $s \in \reals^{i-1}$
        \item $\del{p}$ for all $p$ of level $i$
        \item $\nonnull{p}$ for all $p$ of level $i$
        \item $\ordinv{p}$ for all reducible $p$ of level $i$
        \item $\ordinv{p}$ for all irreducible $p$ of level $i$
        \item $\sgninv{p}$ for all reducible $p$ of level $i$
        \item $\sgninv{p}$ for all irreducible $p$ of level $i$
        \item $\connected{i}$
        \item $\submanifold{i}$
        \item $\sample{s}$ for all $s \in \reals^i$ of level $i$
        \item $\representation{\Isymb,s}$ for all $\Isymb$ of level $i$ and $s \in \reals^{i-1}$
        \item $\irint{\Isymb}$ for all $\Isymb$ of level $i$
    \end{enumerate}
\end{definition}

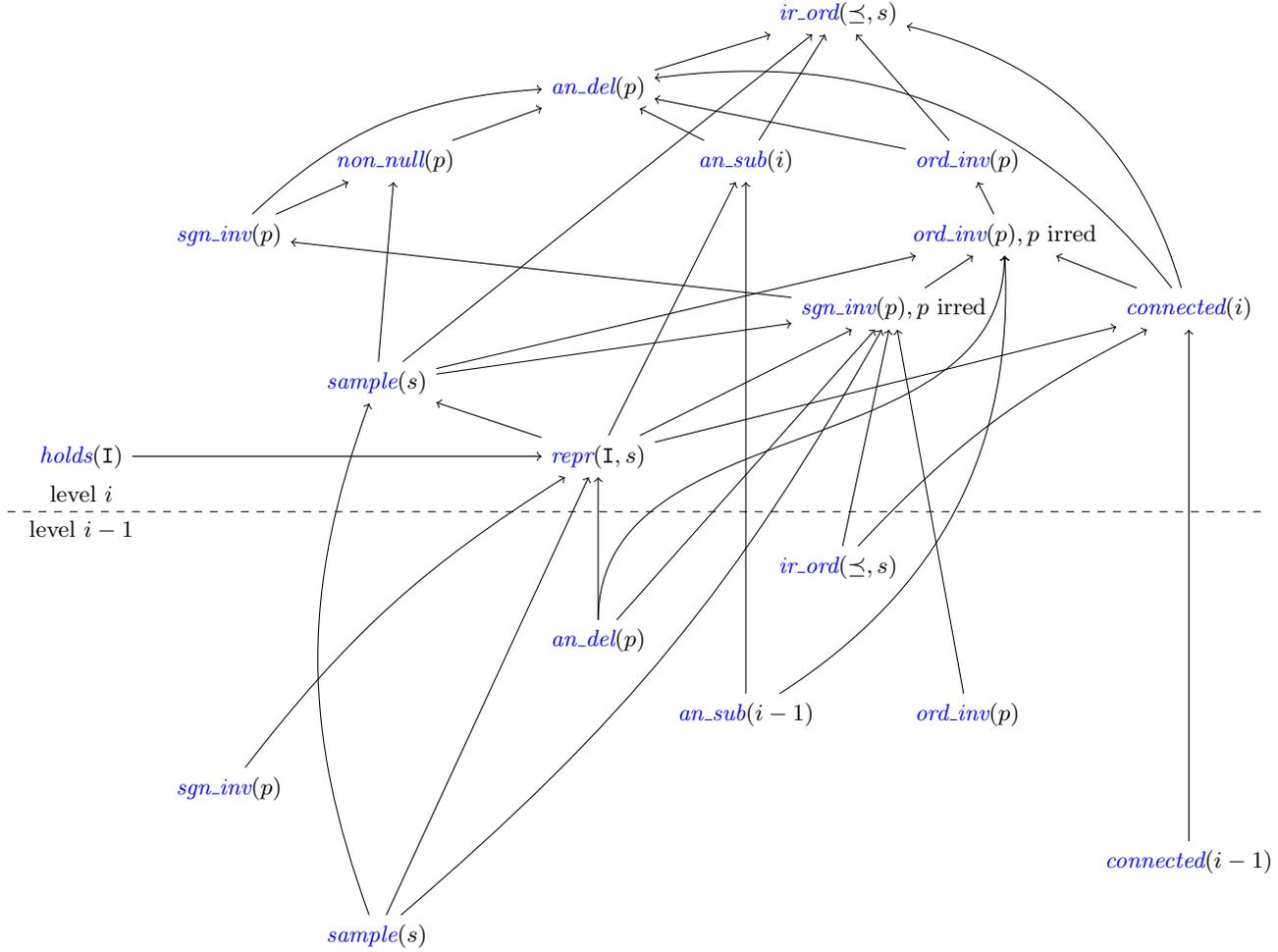
\begin{figure}
    \centering
    \begin{tikzpicture}
    \tikzstyle{every node}=[font=\small]

    \def\levelheight{7.5}

    \foreach \pos in {0} {
        \node (irordering_\pos) at (8.25,0-\pos*\levelheight) {$\irordering{\preceq,s}$};
        \node (pdel_\pos) at (5,-1-\pos*\levelheight) {$\del{p}$};
        \node (nonnull_\pos) at (2.25,-2-\pos*\levelheight) {$\nonnull{p}$};
        \node (submanifold_\pos) at (7,-2-\pos*\levelheight) {$\submanifold{i}$};
        \node (ordinv_\pos) at (10,-2-\pos*\levelheight) {$\ordinv{p}$};
        \node (sgninv_\pos) at (0,-3-\pos*\levelheight) {$\sgninv{p}$};
        \node (ordinv_irred_\pos) at (10.5,-3-\pos*\levelheight) {$\ordinv{p}, p$ irred};
        \node (sgninv_irred_\pos) at (9,-4-\pos*\levelheight) {$\sgninv{p}, p$ irred};
        \node (connected_\pos) at (13,-4-\pos*\levelheight) {$\connected{i}$};
        \node (sample_\pos) at (2,-5-\pos*\levelheight) {$\sample{s}$};
        \node (representation_\pos) at (5,-6-\pos*\levelheight) {$\representation{\Isymb,s}$};
        \node (irint_\pos) at (-2,-6-\pos*\levelheight) {$\irint{\Isymb}$};
    }
    \foreach \pos in {1} {
        \node (irordering_\pos) at (8.25,0-\pos*\levelheight) {$\irordering{\preceq,s}$};
        \node (pdel_\pos) at (5,-1-\pos*\levelheight) {$\del{p}$};
        \node (submanifold_\pos) at (7,-2-\pos*\levelheight) {$\submanifold{i-1}$};
        \node (ordinv_\pos) at (10,-2-\pos*\levelheight) {$\ordinv{p}$};
        \node (sgninv_\pos) at (0,-3-\pos*\levelheight) {$\sgninv{p}$};
        \node (connected_\pos) at (13,-4-\pos*\levelheight) {$\connected{i-1}$};
        \node (sample_\pos) at (2,-5-\pos*\levelheight) {$\sample{s}$};
    }

    \node at (-2,-6.5) {level $i$};
    \draw[dashed] (-3,-6.75) -- (14,-6.75);
    \node at (-2,-7) {level $i-1$};

    \foreach \pos [evaluate=\pos as \posb using int(\pos+1)] in {0} {
        \draw[->] (submanifold_\pos) -- (pdel_\pos);
        \draw[->] (connected_\pos) edge[bend right] (pdel_\pos);
        \draw[->] (nonnull_\pos) -- (pdel_\pos);
        \draw[->] (ordinv_\pos) -- (pdel_\pos);
        \draw[->] (sgninv_\pos) edge[bend left=20] (pdel_\pos);

        \draw[->] (sample_\pos) -- (nonnull_\pos);
        \draw[->] (sgninv_\pos) -- (nonnull_\pos);

        \draw[->] (ordinv_irred_\pos) -- (ordinv_\pos);

        \draw[->] (sgninv_irred_\pos) -- (sgninv_\pos);

        \draw[->] (sgninv_irred_\pos) -- (ordinv_irred_\pos);
        \draw[->] (sample_\pos) -- (ordinv_irred_\pos);
        \draw[->] (connected_\pos) -- (ordinv_irred_\pos);
        \draw[->] (submanifold_\posb) edge[bend right=30] (ordinv_irred_\pos);
        \draw[->] (pdel_\posb) edge[out=90, in=-90] (ordinv_irred_\pos);

        \draw[->] (sample_\posb) edge[bend right=10] (sgninv_irred_\pos);

        \draw[->] (submanifold_\posb) -- (submanifold_\pos);
        \draw[->] (representation_\pos) -- (submanifold_\pos);

        \draw[->] (sample_\pos) -- (irordering_\pos);
        \draw[->] (submanifold_\pos) -- (irordering_\pos);
        \draw[->] (connected_\pos) edge[bend right] (irordering_\pos);
        \draw[->] (pdel_\pos) -- (irordering_\pos);
        \draw[->] (ordinv_\pos) -- (irordering_\pos);

        \draw[->] (sample_\pos) -- (sgninv_irred_\pos);
        \draw[->] (representation_\pos) -- (sgninv_irred_\pos);
        \draw[->] (irordering_\posb) -- (sgninv_irred_\pos);
        \draw[->] (pdel_\posb) -- (sgninv_irred_\pos);
        \draw[->] (ordinv_\posb) -- (sgninv_irred_\pos);

        \draw[->] (connected_\posb) -- (connected_\pos);
        \draw[->] (representation_\pos) -- (connected_\pos);
        \draw[->] (irordering_\posb) edge[bend left=10] (connected_\pos);

        \draw[->] (representation_\pos) -- (sample_\pos);
        \draw[->] (sample_\posb) edge[bend left=20] (sample_\pos);

        \draw[->] (pdel_\posb) -- (representation_\pos);
        \draw[->] (sgninv_\posb) edge[bend left=10] (representation_\pos);

        \draw[->] (irint_\pos) -- (representation_\pos);
        \draw[->] (sample_\posb) -- (representation_\pos);
    }
\end{tikzpicture}

    \caption{Proof rule relationships; \emph{irred} stands for \emph{irreducible}.}
    \label{fig:rules_overview}
\end{figure}

The proof rules introduced in this section are visualized in \Cref{fig:rules_overview} which shows the relationships between rules of different levels.

\section{Levelwise construction of a single cell}
\label{sec:levelwise}

So far, we presented an abstract proof system, i.e. a set of rules which allow us to construct a proof for sign-invariance of some polynomials in a cell. In the following, we will give an algorithm for constructing a single cell in a levelwise manner by specifying how the proof rules are applied. We first give the algorithmic framework and then define heuristics needed to instantiate the algorithm.

\subsection{Algorithm to construct a single cell}

\Cref{alg:single_cell} constructs a single cell. The input to the algorithm is a set of polynomials for which a cell is to be computed on which these polynomials are sign-invariant. This initializes the set of properties $Q$ on \Cref{line:Qinit} which is maintained to contain  properties that need to be fulfilled by the yet to be chosen representations on the lower levels. I.e. at the beginning of iteration $i$, the data structure $(\Isymb_1, \ldots, \Isymb_i)$ needs to maintain all properties in $Q$. To do so, \Cref{alg:single_cell} iteratively calls \Cref{alg:construct_interval} to process the properties and construct the interval on level $i$, taking note of any failure cases due to nullifications that might occur.

\Cref{alg:apply_pre} is a sub-algorithm that replaces a property in $Q$ by a set of properties $Q'$ induced by a proof rule if no such set is already contained in $Q$, and otherwise simply removes the property from $Q$. \textit{FAIL} is returned here if and only if no proof rule is applicable; which is the case only when a polynomial is nullified (i.e. \Cref{def:map:nonnull}) and equational constraint projection does not allow us to ignore that fact.
The sets $Q'$ induced by a proof rule are determined in \Cref{alg:apply_pre:choices}
as follows. Assume we want to derive the property $\nonnull{p}$ of $R \subseteq \reals^i$ for some polynomial $p$ of level $i+1$ with respect to the current sample $s \in \reals^i$. The algorithm would check if the conditions of the proof rules in \Cref{def:map:nonnull} are satisfied, and then determine the corresponding properties of $R$ which form the set $Q'$. In the first case, we check whether $p$ is irreducible, $\vdeg{x_{i+1}}{p}>1$, and $\disc{x_{i+1}}{p}(s) \neq 0$ hold; if yes, we add the set $Q' = \{ \sample{s}, \sgninv{\disc{x_{i+1}}{p}} \}$ to the set of choices. In the second case, we check whether $p$ is irreducible and there exists a coefficient $c_j$ of $p$ such that $c_j(s) \neq 0$ holds; if yes, we add the set $Q' = \{ \sample{s}, \sgninv{c_j} \}$ to the set of choices.

\Cref{alg:construct_interval} is used to extend the cell construction by a level.  It applies the proof rules according to the ordering $\vartriangleleft$ on the properties from \Cref{def:cyclefree} until the only properties of level $i$ remaining are the sign-invariance of some irreducible polynomials (note that all smaller properties w.r.t. $\vartriangleleft$ are provable with only the sample $s$ present except if a nullification occurs). Then, a \emph{representation} consisting of a symbolic interval $\Isymb$, a set of polynomials $E$ and an indexed root ordering $\preceq$ is chosen. The roles of $\Isymb$ and $\preceq$ are already discussed above; the polynomials in $E$ are meant to be excluded from the indexed root ordering, and thus the application of the equational constraint projection rule (\Cref{def:map:eqproj}) is enforced. The formal requirements on the representation will be defined in \Cref{def:representation}. Finally, the remaining properties on level $i$ are eliminated by application of the proof rules, until only a property of the form $\irint{\Isymb}$ is left on the level (which is the greatest property of the current level w.r.t. $\vartriangleleft$).

\begin{algorithm}[p]
    \caption{$\texttt{single\_cell}(i,P,s)$}
    \label{alg:single_cell}
    \SetKwInOut{Input}{Input}
    \SetKwInOut{Output}{Output}
    \DontPrintSemicolon
    \SetKwFunction{FApply}{apply}
    \SetKwFunction{FConstructInterval}{construct\_interval} 
    \SetKwProg{Fn}{Function}{:}{}

    \Input{finite $P \subseteq \rationals[x_1,\ldots,x_n]$, $s \in \reals^n$}
    \Output{cell data structure $\texttt{R}$ such that $s \in \lift{\texttt{R}}$ and all polynomials in $P$ are sign-invariant on $\lift{\texttt{R}}$; or \textit{FAIL}}

    $Q := \{ \sgninv{p} \mid p \in P \}$ \label{line:Qinit}\;
    \For{$i=n,\ldots,1$}{
        $\textit{result} := $ \FConstructInterval{$i$, $Q$, $s$} \tcp{\Cref{alg:construct_interval}}
        \If{\textit{result} $ \neq $ \textit{FAIL}}{
            $\Isymb_i, Q :=  \textit{result}$ \;
        }
        \Else{
            \Return{\textit{FAIL}} \;
        }
    }
    \Return{$(\Isymb_1, \ldots, \Isymb_n)$}
\end{algorithm}

\begin{algorithm}[p]
    \caption{$\texttt{construct\_interval}(i,Q,s)$}
    \label{alg:construct_interval}
    \SetKwInOut{Input}{Input}
    \SetKwInOut{Output}{Output}
    \DontPrintSemicolon
    \SetKwFunction{FApply}{apply\_pre}
    \SetKwProg{Fn}{Function}{:}{}

    \Input{$i \in \posints$, finite $Q \subseteq \levels{\propset}{[i]}$, $s \in \reals^i$}
    \Output{an interval data structure $\Isymb$ such that $s \in \lift{s_{[i-1]}, \Isymb}$, and a set $Q' \subseteq \levels{\propset}{[i-1]}$ such that for any $R \subseteq \reals^{i-1}$ it holds $(\forall q' \in Q'.\; q'(R)) \implies (\forall q \in Q.\; q(\lift{R, \Isymb}))$; or \textit{FAIL}}

    \ForEach{$q \in \levels{Q}{i}$ where $q$ is the greatest element with respect to $\vartriangleleft$ (from \Cref{def:cyclefree}) and $q \neq \sgninv{p}$ for an irreducible $p$}{
        $Q := $ \FApply{$i$,$Q$,$q$,$(s)$} \tcp{\Cref{alg:apply_pre}}
        \If{$Q = $ \textit{FAIL}}{
            \Return{\textit{FAIL}} \; 
        }
    }
    
    $P_{nonnull} := \{ p \mid \sgninv{p} \in \levels{Q}{i} \text{ s.t. } p(s_{[i-1]},x_i) \neq 0 \}$ \; %

    $\Xi := \irexpr{P_{nonnull}}{s_{[i-1]}}$ \;

    \textbf{choose} representation $(\Isymb,E,\preceq)$ of $\Xi$ with respect to $s$ \label{algline:choice-rep}\;

    \If{$i>1$}{
        \ForEach{$q \in \levels{Q}{i}$ where $q$ is the greatest element with respect to $\vartriangleleft$ (from \Cref{def:cyclefree}) and $q \neq \irint{\Isymb}$}{
            $Q := $ \FApply{$i$,$Q$,$q$,$(s,\text{\Isymb},\preceq)$} \tcp*{\Cref{alg:apply_pre}}
            \If{$Q = $ \textit{FAIL}}{
                \Return{\textit{FAIL}} \; 
            }
        }
    }
    \tcc{$\levels{Q}{i}$ contains now only $\irint{\Isymb}$} 
    \Return{$\Isymb$,$(Q \setminus \levels{Q}{i})$}
\end{algorithm}

\begin{algorithm}[p]
    \caption{$\texttt{apply\_pre}(i,Q,q,\textit{parameters})$}
    \label{alg:apply_pre}
    \SetKwInOut{Input}{Input}
    \SetKwInOut{Output}{Output}
    \DontPrintSemicolon
    \SetKwFunction{FApply}{apply}
    \SetKwProg{Fn}{Function}{:}{}

    \Input{$i \in \posints$, finite $Q \subseteq \levels{\propset}{[i]}$, $q \in \levels{\propset}{i} \cap Q$ and either $\textit{parameters} = (s)$ or $\textit{parameters} = (s,\Isymb, \preceq)$ with $s \in \reals^i$, symbolic interval $\Isymb$ on level $i$, indexed root ordering $\preceq$ on level $i$}
    \Output{a set $(Q \setminus \{ q \}) \cup Q' \subseteq \levels{\propset}{[i]} \setminus \{ q \}$ such that for all $R \subseteq \reals^i$ it holds $(\forall q' \in Q'.\; q'(R)) \implies q(R)$; or \textit{FAIL}}

    \textbf{let} $\textit{choices} \subseteq 2^\propset$ such that each $Q' \in \textit{choices}$ is a set of properties from a proof rule which can be applied according to \textit{parameters} to derive $q$ (see explanatory text above for details) \;\label{alg:apply_pre:choices}

    \If{$\textit{choices} = \emptyset$}{
        \Return{\textit{FAIL}} \;
    }

    \If{$\nexists Q' \in \textit{choices}.\; Q' \subseteq Q$}{
        \textbf{choose} $Q' \in \textit{choices}$ \label{algline:choice-pre}\;
        \Return{$(Q \setminus \{ q \}) \cup Q'$} \;
    }
    \Else {
        \Return{$Q \setminus \{ q \}$} \;
    }
\end{algorithm}

\begin{theorem}
    Let $P \subseteq \rationals[x_1,\ldots,x_n]$, and $s \in \reals^n$. The method $\texttt{single\_cell}$ (\Cref{alg:single_cell}) either returns a cell data structure $\texttt{R}$ such that all polynomials in $P$ are sign-invariant on $\lift{\texttt{R}}$ and $s \in \lift{\texttt{R}}$, or returns $\textit{FAIL}$.
\end{theorem}

\begin{proof}
    Note that the algorithm will terminate as \Cref{alg:apply_pre} replaces properties by smaller properties according to the ordering as defined in \Cref{def:cyclefree}, there does exist a smallest property, and the CAD projection of a set of polynomials is finite.
    The correctness follows from the correctness proofs of the rules of inference. %
\end{proof}

\subsection{Heuristic choices}
\label{sec:heuristics}

Our algorithm has two points where a choice must be made: (1) when choosing a \emph{rule of inference} to apply in \Cref{alg:apply_pre} \Cref{algline:choice-pre}; and (2) when choosing the \emph{representation} in \Cref{alg:construct_interval} \Cref{algline:choice-rep}.

For the first choice, the rules of inference may define multiple possible property sets for a property that each imply that property and so we have the freedom to choose which to use.  We aim to pick the ``best'' such set. One could compute each of these sets individually and compare them, or, to avoid superfluous heavy computations of resultants or discriminants, prefer sets which seem easier to compute.  The latter approach, making the choice heuristically (with respect to e.g. estimated computational effort, or degrees of polynomials, or number of polynomials / real roots) is the approach taken by our implementation. For instance, in \Cref{def:map:nonnull}, we always pick a coefficient rather than computing a discriminant. Note that even if sets involving resultants or discriminants are avoided where possible, the algorithm should first check whether one such set is already maintained before computing any new set.

For the second choice, we formalize what we mean by a representation. As already discussed, there might be multiple choices for the cell description $\Isymb$ on the current level and the indexed root ordering $\preceq$ for fulfilling the requirements for \Cref{def:map:sgninvord}. Additionally, if $\Isymb$ is a section, we can make use of the equational constraints projection in \Cref{def:map:eqproj}. A representation determines all these parameters.

\begin{definition}[Representation for $\Xi$]
    \label{def:representation}

    Let $i \in \posints$, $s \in \reals^i$, and $\Xi$ be a set of indexed root expressions of level $i$ such that $\xi \in \irexpr{\xi.p}{s_{[i-1]}}$ for every $\xi \in \Xi$.
    
    A \emph{representation} for $\Xi$ with respect to $s$ is a tuple $(\Isymb, E, \preceq)$ where $\Isymb$ is a symbolic interval of level $i$, $E$ is a set of polynomials of level $i$, and $\preceq$ is an indexed root ordering of level $i$ such that
    \begin{itemize}
        \item $s \in \lift{s_{[i-1]},\Isymb}$,
        \item either $p \in E$ or $\irexpr{p}{s_{[i-1]}} \subseteq \dom{\preceq}$ for all $p \in \{ \xi.p \mid \xi \in \Xi \}$,
        \item if $E \neq \emptyset$ then $\Isymb = (\text{section}, b)$ for some indexed root expression $b$,
        \item $\preceq$ matches $s_{[i-1]}$, and
        \item $\xi \preceq^t \Isymb.l$ or $\Isymb.u \preceq^t \xi$ for all $\xi \in \Xi$ if $\Isymb = (\text{sector}, l,u)$ respectively $\xi \preceq^t \Isymb.b$ or $\Isymb.b \preceq^t \xi$ for all $\xi \in \Xi$ if $\Isymb = (\text{section}, b)$.
    \end{itemize}
\end{definition}

This definition is quite general by allowing \emph{additional} indexed root expressions in $\Xi'$ compared to the set $\Xi$ describing the roots of the present polynomials. This for example enables heuristics to under-approximate the constructed cell to reduce heavy computations in future work (see \Cref{sec:futurework}).

In the following however, we will assume that $\Xi' = \Xi$, that is, $\Isymb$ and $\preceq$ will only consider roots from $\Xi$ corresponding to the polynomials for which we need to derive sign-invariance.

Note that we omit the requirements of \Cref{def:map:connectedness} here yet for readability reasons and as the adaption is trivial. If required, we only need to add the pair $(\Isymb.l,\Isymb.u)$ to the indexed root ordering.

For the symbolic interval $\Isymb$, we minimize the degrees in the main variable of the defining polynomials.

\begin{definition}[Choice of the symbolic interval]
    \label{def:heuristics_representation}

    Let $i \in \posints$, $P \subseteq \rationals[x_1, \ldots, x_{i}]$ be a set of irreducible polynomials of level $i$, $s \in \reals^{i}$ be a sample such that no $p \in P$ is nullified on $s_{[i-1]}$, and $\Xi \subseteq \irexpr{P}{s_{[i-1]}}$.

    We define the sets $\Xi_{lo}$ and $\Xi_{up}$ of the closest lower respectively upper bounds of $s_i$ as follows:
    \begin{align*}
        \Xi_{lo} & = \{ \xi \in \Xi \mid \xi(s_{[i-1]}) \leq s_i \wedge \forall \xi' \in \Xi.\; \big( \xi'(s_{[i-1]}) \leq s_i \implies \xi'(s_{[i-1]}) \leq \xi(s_{[i-1]}) \big) \}, \\
        \Xi_{up} & = \{ \xi \in \Xi \mid s_i \leq \xi(s_{[i-1]}) \wedge \forall \xi' \in \Xi.\; \big( s_i \leq \xi'(s_{[i-1]}) \implies \xi(s_{[i-1]}) \leq \xi'(s_{[i-1]}) \big) \}.
    \end{align*}
    \noindent We pick $\xi_\textit{lo} \in \Xi_{lo}$ such that $\vdeg{x_i}{\xi_\textit{lo}.p} \text{ is minimal}$ and $\xi_\textit{up} \in \Xi_{up}$ such that $\vdeg{x_i}{\xi_\textit{up}.p} \text{ is minimal}$.  Additionally, if $\Xi_{lo} = \Xi_{up}$, then we require $\xi_\textit{lo} = \xi_\textit{up}$.

    We define the \emph{lowest degree interval} $\Isymb_{\text{ldeg}}$ of $s$ with respect to $\Xi$ as
    \begin{align*}
        & \Isymb_{\text{ldeg}}  = (section, \xi_\textit{lo}) & \text{if } \xi_\textit{lo} = \xi_\textit{up}, \\
        & \Isymb_{\text{ldeg}} = (sector, \xi_\textit{lo}, \xi_\textit{up}) & \text{otherwise}.
    \end{align*}
\end{definition}

For the indexed root ordering $\preceq$, there are two possibilities: aim to make the underlying cell as big as possible; or aim to avoid heavy resultant computations. In theory, we could compute the results for all (or all promising) possible indexed root orderings and pick the best one. However, as this is infeasible in practise, we define below several alternative heuristics with different rationales. 

To achieve this we employ a somewhat idealistic view on the problem. Recall that an indexed root ordering is ensured by making resultants order-invariant, which is often a stronger ordering on the roots than required by the picked indexed root ordering. We ignore this fact and base our heuristics on the set of indexed roots without considering common defining polynomials between them (for now).  This is further discussed in \Cref{sec:observations}.

We start with the following observation: as the given derivation rules always require delineability for all polynomials whose sign-invariance is proven using an indexed root ordering, a fixed ordering of all real root functions defined at $s_{[i-1]}$ of such polynomials is guaranteed anyway. Thus, we can restrict the heuristics for the choice of $\preceq$ by computing the ordering on the set $\tilde{\Xi}$ containing for each polynomial only the closest lower and upper roots to $s_i$, and extending an ordering $\tilde{\preceq}$ of $\tilde{\Xi}$ to an ordering $\preceq$ on $\Xi$. This is formalized in the following definition.

\begin{definition}[Choice of the indexed root ordering: reduction]
    Let $i, P, s, \Xi$ be as in \Cref{def:heuristics_representation}. Then
    \begin{align*}
        \tilde{\Xi} = \quad & \{ \xi \in \Xi \mid \xi(s_{[i-1]}) \leq s_i \wedge \forall \xi' \in \Xi \setminus \{ \xi \}.\; (\xi'.p = \xi.p \implies \neg (\xi(s_{[i-1]}) < \xi'(s_{[i-1]}) \leq s_i)) \} \\
        \cup \, & \{ \xi \in \Xi \mid s_i \leq \xi(s_{[i-1]}) \wedge \forall \xi' \in \Xi \setminus \{ \xi \}.\; (\xi'.p = \xi.p \implies \neg(s_i \leq \xi'(s_{[i-1]})  < \xi(s_{[i-1]}) )) \}.
    \end{align*}

    \noindent For any indexed root ordering $\tilde{\preceq}_X$ on $\tilde{\Xi}$ matching $s_{[i-1]}$, we define the ordering $\preceq_X$ on $\Xi$ matching $s_{[i-1]}$ as
    \begin{align*}
        \preceq_X \, = \, & \tilde{\preceq}_X \cup \{ (\xi, \xi') \mid \xi, \xi' \in \Xi \text{ such that } \xi.p = \xi'.p \wedge \xi(s_{[i-1]}) < \xi'(s_{[i-1]})) \}.
    \end{align*}
\end{definition}

In the following definitions we give different heuristics to choose an indeed root ordering $\tilde{\preceq}_X$ which in each case can be extended to a corresponding ordering $\preceq_X$.

First, in the section case, the below \textsc{equational constraint} heuristic enforces simply the application of the equational constraint rule to all polynomials.

\begin{definition}[\textsc{Equational constraint} representation]
    \label{def:heuristics_preceq_eq}

    Let $i,\ P,\ s,\ \Xi,\ \xi_\textit{lo},\ \xi_\textit{up},\ \Isymb_{\text{ldeg}}$ be as in \Cref{def:heuristics_representation}.
    If $\xi_\textit{lo} = \xi_\textit{up}$, we define the \emph{\textsc{Equational constraint} representation} as the tuple $(\Isymb_\text{ldeg}, P, \emptyset)$. 
\end{definition}

\begin{figure}
    \centering
    \begin{subfigure}[b]{0.49\textwidth}
        \centering
        \begin{tikzpicture}
            \draw[thick,->] (0,0) -- (8,0) node[right]{$\mathbb{R}$};
                                        
            \node[circle,fill,inner sep=1pt,label={[below=0.25]:\small $\xi_1(s_{[i-1]})$}] (xi1) at (0.2,0) {};
            \node[circle,fill,inner sep=1pt,label={[below=0.75]:\small $\xi_2(s_{[i-1]})$}] (xi2) at (1.2,0) {};
            \node[circle,fill,inner sep=1pt,label={[below=0.25]:\small $\xi_3(s_{[i-1]})$}] (xi3) at (2.2,0) {};
            \node[circle,fill,inner sep=1pt,label={[below=0.75]:\small $\xi_4(s_{[i-1]})$}] (xi4) at (3.2,0) {};
            \node[circle,fill,inner sep=1pt,label={[below=0.25]:\small $\xi_5(s_{[i-1]})$}] (xi5) at (4.2,0) {};
            \node[circle,fill,inner sep=1pt,red100,label={[above=0.25]:\small\textcolor{red100}{$s_i$}}] (dot) at (4.7,0) {};
            \node[circle,fill,inner sep=1pt,label={[below=0.75]:\small $\xi_6(s_{[i-1]})$}] (xi6) at (5.2,0) {};
            \node[circle,fill,inner sep=1pt,label={[below=0.25]:\small $\xi_7(s_{[i-1]})$}] (xi7) at (6.2,0) {};
            \node[circle,fill,inner sep=1pt,label={[below=0.75]:\small $\xi_8(s_{[i-1]})$}] (xi8) at (7.2,0) {};
        
            \draw[-,gray,thick] (xi1) to [bend left=40] (xi5);
            \draw[-,gray,thick] (xi2) to [bend left=40] (xi5);
            \draw[-,gray,thick] (xi3) to [bend left=40] (xi5);
            \draw[-,gray,thick] (xi4) to [bend left=40] (xi5);
            \draw[-,gray,thick] (xi5) to [bend left=40] (xi6);
            \draw[-,gray,thick] (xi6) to [bend left=40] (xi7);
            \draw[-,gray,thick] (xi6) to [bend left=40] (xi8);
        \end{tikzpicture}
        \caption{\textsc{Biggest cell}\label{fig:heuristicsVis_bc}}
    \end{subfigure}
    \hfill
    \begin{subfigure}[b]{0.49\textwidth}
        \centering
        \begin{tikzpicture}
            \draw[thick,->] (0,0) -- (8,0) node[right]{$\mathbb{R}$};
                                        
            \node[circle,fill,inner sep=1pt,label={[below=0.25]:\small $\xi_1(s_{[i-1]})$},label={[above=0.5]:$7$}] (xi1) at (0.2,0) {};
            \node[circle,fill,inner sep=1pt,label={[below=0.75]:\small $\xi_2(s_{[i-1]})$},label={[above=0.5]:$4$}] (xi2) at (1.2,0) {};
            \node[circle,fill,inner sep=1pt,label={[below=0.25]:\small $\xi_3(s_{[i-1]})$},label={[above=0.5]:$3$}] (xi3) at (2.2,0) {};
            \node[circle,fill,inner sep=1pt,label={[below=0.75]:\small $\xi_4(s_{[i-1]})$},label={[above=0.5]:$10$}] (xi4) at (3.2,0) {};
            \node[circle,fill,inner sep=1pt,label={[below=0.25]:\small $\xi_5(s_{[i-1]})$},label={[above=0.5]:$5$}] (xi5) at (4.2,0) {};
            \node[circle,fill,inner sep=1pt,red100,label={[above=0.25]:\small\textcolor{red100}{$s_i$}}] (dot) at (4.7,0) {};
            \node[circle,fill,inner sep=1pt,label={[below=0.75]:\small $\xi_6(s_{[i-1]})$},label={[above=0.5]:$5$}] (xi6) at (5.2,0) {};
            \node[circle,fill,inner sep=1pt,label={[below=0.25]:\small $\xi_7(s_{[i-1]})$},label={[above=0.5]:$3$}] (xi7) at (6.2,0) {};
            \node[circle,fill,inner sep=1pt,label={[below=0.75]:\small $\xi_8(s_{[i-1]})$},label={[above=0.5]:$10$}] (xi8) at (7.2,0) {};
        
            \draw[-,gray,thick] (xi1) to [bend left=40] (xi3);
            \draw[-,gray,thick] (xi2) to [bend left=40] (xi3);
            \draw[-,gray,thick] (xi3) to [bend left=40] (xi5);
            \draw[-,gray,thick] (xi4) to [bend left=40] (xi5);
            \draw[-,gray,thick] (xi5) to [bend left=40] (xi6);
            \draw[-,gray,thick] (xi6) to [bend left=40] (xi7);
            \draw[-,gray,thick] (xi7) to [bend left=40] (xi8);
        \end{tikzpicture}
        \caption{\textsc{Lowest degree barriers}\label{fig:heuristicsVis_ldb}}
    \end{subfigure}
    \hfill
    \begin{subfigure}[b]{0.49\textwidth}
        \centering
        \begin{tikzpicture}
            \draw[thick,->] (0,0) -- (8,0) node[right]{$\mathbb{R}$};
                                        
            \node[circle,fill,inner sep=1pt,label={[below=0.25]:\small $\xi_1(s_{[i-1]})$}] (xi1) at (0.2,0) {};
            \node[circle,fill,inner sep=1pt,label={[below=0.75]:\small $\xi_2(s_{[i-1]})$}] (xi2) at (1.2,0) {};
            \node[circle,fill,inner sep=1pt,label={[below=0.25]:\small $\xi_3(s_{[i-1]})$}] (xi3) at (2.2,0) {};
            \node[circle,fill,inner sep=1pt,label={[below=0.75]:\small $\xi_4(s_{[i-1]})$}] (xi4) at (3.2,0) {};
            \node[circle,fill,inner sep=1pt,label={[below=0.25]:\small $\xi_5(s_{[i-1]})$}] (xi5) at (4.2,0) {};
            \node[circle,fill,inner sep=1pt,red100,label={[above=0.25]:\small\textcolor{red100}{$s_i$}}] (dot) at (4.7,0) {};
            \node[circle,fill,inner sep=1pt,label={[below=0.75]:\small $\xi_6(s_{[i-1]})$}] (xi6) at (5.2,0) {};
            \node[circle,fill,inner sep=1pt,label={[below=0.25]:\small $\xi_7(s_{[i-1]})$}] (xi7) at (6.2,0) {};
            \node[circle,fill,inner sep=1pt,label={[below=0.75]:\small $\xi_8(s_{[i-1]})$}] (xi8) at (7.2,0) {};
        
            \draw[-,gray,thick] (xi1) to [bend left=40] (xi2);
            \draw[-,gray,thick] (xi2) to [bend left=40] (xi3);
            \draw[-,gray,thick] (xi3) to [bend left=40] (xi4);
            \draw[-,gray,thick] (xi4) to [bend left=40] (xi5);
            \draw[-,gray,thick] (xi5) to [bend left=40] (xi6);
            \draw[-,gray,thick] (xi6) to [bend left=40] (xi7);
            \draw[-,gray,thick] (xi7) to [bend left=40] (xi8);
        \end{tikzpicture}
        \caption{\textsc{Chain}\label{fig:heuristicsVis_chain}}
    \end{subfigure}
    \hfill
    \begin{subfigure}[b]{0.49\textwidth}
        \centering
        \begin{tikzpicture}
            \draw[thick,->] (0,0) -- (8,0) node[right]{$\mathbb{R}$};
                                        
            \node[circle,fill,inner sep=1pt,label={[below=0.25]:\small $\xi_1(s_{[i-1]})$}] (xi1) at (0.2,0) {};
            \node[circle,fill,inner sep=1pt,label={[below=0.75]:\small $\xi_2(s_{[i-1]})$}] (xi2) at (1.2,0) {};
            \node[circle,fill,inner sep=1pt,label={[below=0.25]:\small $\xi_3(s_{[i-1]})$}] (xi3) at (2.2,0) {};
            \node[circle,fill,inner sep=1pt,label={[below=0.75]:\small $\xi_4(s_{[i-1]})$}] (xi4) at (3.2,0) {};
            \node[circle,fill,inner sep=1pt,label={[below=0.25]:\small $\xi_5(s_{[i-1]})$}] (xi5) at (4.2,0) {};
            \node[circle,fill,inner sep=1pt,red100,label={[above=0.25]:\small\textcolor{red100}{$s_i$}}] (dot) at (4.7,0) {};
            \node[circle,fill,inner sep=1pt,label={[below=0.75]:\small $\xi_6(s_{[i-1]})$}] (xi6) at (5.2,0) {};
            \node[circle,fill,inner sep=1pt,label={[below=0.25]:\small $\xi_7(s_{[i-1]})$}] (xi7) at (6.2,0) {};
            \node[circle,fill,inner sep=1pt,label={[below=0.75]:\small $\xi_8(s_{[i-1]})$}] (xi8) at (7.2,0) {};
        
            \draw[-,gray,thick] (xi1) to [bend left=40] (xi2);
            \draw[-,gray,thick] (xi1) to [bend left=40] (xi3);
            \draw[-,gray,thick] (xi1) to [bend left=40] (xi4);
            \draw[-,gray,thick] (xi1) to [bend left=40] (xi5);
            \draw[-,gray,thick] (xi1) to [bend left=40] (xi6);
            \draw[-,gray,thick] (xi1) to [bend left=40] (xi7);
            \draw[-,gray,thick] (xi1) to [bend left=40] (xi8);
            \draw[-,gray,thick] (xi2) to [bend left=40] (xi3);
            \draw[-,gray,thick] (xi2) to [bend left=40] (xi4);
            \draw[-,gray,thick] (xi2) to [bend left=40] (xi5);
            \draw[-,gray,thick] (xi2) to [bend left=40] (xi6);
            \draw[-,gray,thick] (xi2) to [bend left=40] (xi7);
            \draw[-,gray,thick] (xi2) to [bend left=40] (xi8);
            \draw[-,gray,thick] (xi3) to [bend left=40] (xi4);
            \draw[-,gray,thick] (xi3) to [bend left=40] (xi5);
            \draw[-,gray,thick] (xi3) to [bend left=40] (xi6);
            \draw[-,gray,thick] (xi3) to [bend left=40] (xi7);
            \draw[-,gray,thick] (xi3) to [bend left=40] (xi8);
            \draw[-,gray,thick] (xi4) to [bend left=40] (xi5);
            \draw[-,gray,thick] (xi4) to [bend left=40] (xi6);
            \draw[-,gray,thick] (xi4) to [bend left=40] (xi7);
            \draw[-,gray,thick] (xi4) to [bend left=40] (xi8);
            \draw[-,gray,thick] (xi5) to [bend left=40] (xi6);
            \draw[-,gray,thick] (xi5) to [bend left=40] (xi7);
            \draw[-,gray,thick] (xi5) to [bend left=40] (xi8);
            \draw[-,gray,thick] (xi6) to [bend left=40] (xi7);
            \draw[-,gray,thick] (xi6) to [bend left=40] (xi8);
            \draw[-,gray,thick] (xi7) to [bend left=40] (xi8);
        \end{tikzpicture}
        \caption{\textsc{Full}\label{fig:heuristicsVis_full}}
    \end{subfigure}
    \caption{Visualization of indexed root ordering heuristics. \label{fig:heuristicsVis}}
\end{figure}
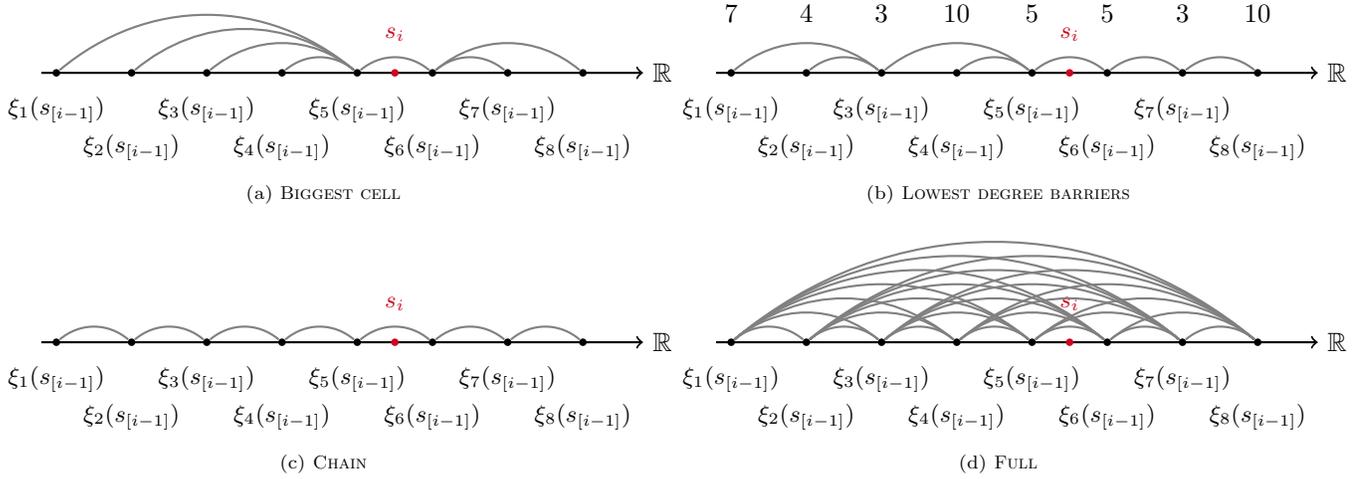

Next, the \textsc{biggest cell} heuristic defines the weakest ordering on the indexed roots according to \Cref{def:representation} and thus defines the biggest possible underlying cell (under the assumption that this ordering can be realized perfectly, i.e. the resultants have roots only below the crossing of a polynomial's root with a cell boundary) as visualized in \Cref{fig:heuristicsVis_bc}. Note that for the section case, the \textsc{biggest cell} heuristic is the \textsc{equational constraint} heuristic plus some discriminants and coefficients; thus, the application of \textsc{biggest cell} only makes sense in the sector case. 

\begin{definition}[\textsc{Biggest cell} representation]
    \label{def:heuristics_preceq_biggestcell}

    Let $i,\ P,\ s,\ \Xi,\ \xi_\textit{lo},\ \xi_\textit{up},\ \Isymb_{\text{ldeg}}$ be as in \Cref{def:heuristics_representation}. 

    We define the indexed root ordering $\preceq_{\text{biggest}}$ on $\Xi$ according to
    \begin{align*}
        \tilde{\preceq}_{biggest} = \quad & \{ (\xi, \xi_\textit{lo}) \mid \xi \in \tilde{\Xi} \setminus \{ \xi_\textit{lo} \} \text{ s.t. } \xi(s_{[i-1]}) \leq \xi_\textit{lo}(s_{[i-1]}) \} \\
        \cup \, & \{ (\xi_\textit{up}, \xi) \mid \xi \in \tilde{\Xi} \setminus \{ \xi_\textit{up} \} \text{ s.t. } \xi_\textit{up}(s_{[i-1]}) \leq \xi(s_{[i-1]}) \}
    \end{align*}
    and the \emph{\textsc{Biggest cell} representation} as the tuple $(\Isymb_\text{ldeg}, \emptyset, \preceq_{\text{biggest}})$.
\end{definition}

The below \textsc{lowest degree barriers} heuristic minimizes the degrees of the defining polynomials (locally per level), and thus also the degree of the computed resultants (under the above assumption) as visualized in \Cref{fig:heuristicsVis_ldb}. Furthermore, it enforces the equational constraints rule whenever possible.

This heuristic has two motivations.  First, that the polynomial degrees grow doubly exponentially during the CAD projection, see e.g. \cite[Table 1]{BDEMW16}. Second, that the running time of the resultant computation depends quadratically on the degree in the main variable of the input polynomials, see \cite{ducos2000optimizations}.

In the following definition, we use the \emph{lexicographical ordering} on tuples, that is $t = (t_1,\ldots,t_k) < (t'_1,\ldots,t'_k) = t'$ holds if $t_1 < t'_1 \vee (t_1 = t'_1 \wedge t_2 < t'_2) \vee (t_1 = t'_1 \wedge t_2 = t'_2 \wedge t_3 < t'_3) \vee \ldots$.

\begin{definition}[\textsc{Lowest degree barriers} representation]
    \label{def:heuristics_preceq_ldb}

    Let $i,\ P,\ s,\ \Xi,\ \xi_\textit{lo},\ \xi_\textit{up},\ \Isymb_{\text{ldeg}}$ be as in \Cref{def:heuristics_representation}. 

    We assume an injection $o: \Xi \to \naturals$ that orders the elements of $\Xi$ such that $o(\Isymb_{\text{ldeg}}.b)=0$ respectively $o(\Isymb_{\text{ldeg}}.l)=0$, $o(\Isymb_{\text{ldeg}}.u)=1$ (if both are indexed root expressions), or $o(\Isymb_{\text{ldeg}}.l)=0$ (if $\Isymb_{\text{ldeg}}.u=\infty$), or $o(\Isymb_{\text{ldeg}}.u)=0$ (if $\Isymb_{\text{ldeg}}.l=-\infty$).
    
    Let $\Xi' \subseteq \Xi$. For $\xi \in \Xi'$, we define the \emph{barrier} of $\xi$ w.r.t. $\Xi'$, that is a root in $\Xi'$ between $\xi(s_{[i-1]})$ and $s_i$ with minimal degree in the main variable:
    \begin{align*}
        & \text{if } \xi(s_{[i-1]}) \leq s_i & \text{ then } & \text{barrier}_{\Xi'}(\xi) = \argmin_{\{ \xi' \in \Xi' \mid \xi(s_{[i-1]}) \leq \xi'(s_{[i-1]}) \leq s_i \}}  (\vdeg{x_i}{\xi'.p},  s_i-\xi'(s_{[i-1]}), o(\xi')) ; \\
        & \text{if } s_i \leq \xi(s_{[i-1]}) & \text{ then } & \text{barrier}_{\Xi'}(\xi) = \argmin_{\{ \xi' \in \Xi' \mid s_i \leq \xi'(s_{[i-1]}) \leq \xi(s_{[i-1]})  \}}  (\vdeg{x_i}{\xi'.p},  \xi'(s_{[i-1]})-s_i, o(\xi')) . \\
    \end{align*}
    If $\xi_\textit{lo} \neq \xi_\textit{up}$, we define the indexed root ordering $\preceq_{\text{barriers}}$ on $\Xi$ according to
    \begin{align*}
        \tilde{\preceq}_{barriers} = \quad & \{ (\xi, \text{barrier}_{\tilde{\Xi}}(\xi)) \mid \xi \in \tilde{\Xi}  \text{ s.t. } \xi(s_{[i-1]}) < s_i \text{ and } \xi \neq \text{barrier}_{\tilde{\Xi}}(\xi) \}, \\
        \cup \, & \{ ( \text{barrier}_{\tilde{\Xi}}(\xi) , \xi) \mid \xi \in \tilde{\Xi} \text{ s.t. }  s_i < \xi(s_{[i-1]})  \text{ and } \xi \neq \text{barrier}_{\tilde{\Xi}}(\xi) \}
    \end{align*}
    and the \emph{\textsc{Lowest degree barriers} representation} as the tuple $(\Isymb_\text{ldeg}, \emptyset, \preceq_{\text{barriers}})$.

    If $\xi_\textit{lo} = \xi_\textit{up}$, we exclude the polynomials with roots around $s_i$ which do not qualify as a barrier for some other roots from the indexed root ordering (thus enforcing the application of the equational constraint rule). To do so, for a set of polynomials $P' \subseteq P$, we define the set $\restr{\tilde{\Xi}}{P'} = \{ \xi \mid \xi \in \tilde{\Xi} \text{ s.t. } \xi.p \in P' \}$ and let $P_\text{eq} \subseteq P$ be the result of the following fixed point computation (i.e. $P_\text{eq} = P_j = P_{j+1}$ for some $j$):
    \begin{align*}
        P_0 =\ & \emptyset \\
        P_{j+1} =\ & P_j \cup \{ p \in P \mid p \neq \Isymb_{\text{ldeg}}.b.p \\
        & \quad\quad \wedge \exists \xi \in \tilde{\Xi} \setminus \restr{\tilde{\Xi}}{P_j}.\; (\xi.p = p \wedge \text{barrier}_{\tilde{\Xi} \setminus \restr{\tilde{\Xi}}{P_j}}(\xi) = \Isymb_{\text{ldeg}}.b \ \wedge \nexists \xi' \in \tilde{\Xi} \setminus \restr{\tilde{\Xi}}{P_j}.\; \text{barrier}_{\tilde{\Xi} \setminus \restr{\tilde{\Xi}}{P_j}}(\xi') = \xi ) \}.
    \end{align*}
    \noindent We define the indexed root ordering $\preceq_{\text{barriers}}$ on $\Xi \setminus \restr{\Xi}{P_\text{eq}}$ according the definition of $\tilde{\preceq}_{\text{barriers}}$ as above but only considering the roots $\tilde{\Xi} \setminus \restr{\tilde{\Xi}}{P_\text{eq}}$ and the \emph{\textsc{Lowest degree barriers} representation} as the tuple $(\Isymb_\text{ldeg}, P_\text{eq}, \preceq_{\text{barriers}})$.
\end{definition}

The below \textsc{chain} heuristic fixes the total ordering on the roots as visualized in \Cref{fig:heuristicsVis_chain}.

\begin{definition}[\textsc{Chain} representation]
    \label{def:heuristics_preceq_chain}

    Let $i,\ P,\ s,\ \Xi,\ \xi_\textit{lo},\ \xi_\textit{up},\ \Isymb_{\text{ldeg}}$ be as in \Cref{def:heuristics_representation}. 

    Let $\{ \xi_1, \ldots, \xi_{k} \} = \tilde{\Xi}$ s.t. $\xi_j(s_{[i-1]}) \leq \xi_{j+1}(s_{[i-1]})$ for all $j \in [1..k-1]$.
    
    We define the indexed root ordering $\preceq_{\text{chain}}$ on $\Xi$ according to
    \[ \tilde{\preceq}_{chain} = \{ (\xi_{j}, \xi_{j+1}) \mid j \in [1..k-1] \} \]
    and the \emph{\textsc{Chain} representation} as the tuple $(\Isymb_\text{ldeg}, \emptyset, \preceq_{\text{chain}})$.
\end{definition}

Finally, the \textsc{full} heuristic fixes the same total ordering as it is the unique transitive closure of the \textsc{chain} heuristic as visualized in \Cref{fig:heuristicsVis_full}. Note that we include this heuristic only for illustrative purposes: it resembles the cells constructed naively by making a full projection without adapting to the behaviour at the sample.

\begin{definition}[\textsc{Full} representation]
    \label{def:heuristics_preceq_full}

    Let $i,\ P,\ s,\ \Xi,\ \xi_\textit{lo},\ \xi_\textit{up},\ \Isymb_{\text{ldeg}}$ as in \Cref{def:heuristics_representation}. 

    Let $\{ \xi_1, \ldots, \xi_{k} \} = \tilde{\Xi}$ s.t. $\xi_j(s_{[i-1]}) \leq \xi_{j+1}(s_{[i-1]})$ for all $j \in [1..k-1]$.
    
    We define the indexed root ordering $\preceq_{\text{full}}$ on $\Xi$ according to
    \[ \tilde{\preceq}_{full} = \{ (\xi_j, \xi_{j'}) \mid j,j' \in [1..k-1],\ j < j' \} \]
    and the \emph{\textsc{Full} representation} as the tuple $(\Isymb_\text{ldeg}, \emptyset, \preceq_{\text{full}})$.
\end{definition}

\section{Qualitative Observations}
\label{sec:observations}

\subsection{Comparing ordering heuristics}
\label{sec:orderingheuristics}

In this section we will explain the intuition behind the various ordering heuristics.  They are designed to optimize desirable characteristics.  We acknowledge their heuristic nature, and in particular that the intuition is made under the idealistic view that any indexed root ordering $\preceq$ can be perfectly realized: that is, the resultants calculated only have roots which indicate a crossing of two root functions considered in $\preceq$ when in reality they also have ``spurious'' roots which do not have relevance for the problem at hand.

\begin{figure}
    \begin{subfigure}[t]{0.3\textwidth}
        \begin{center}
            \begin{tikzpicture}

                \draw (0,0) -- (4.5,0) -- (4.5,6.5) -- (0,6.5) -- (0,0);
		        \node at (2.5,-.5) {$x_1$};
		        \node at (-.5,3) {$x_2$};
                
                \draw[very thick, magenta100] (2,2) ellipse (1cm and 1cm);
                \draw[very thick, green100] (0,6) -- (4.5,1.5);
                \draw[very thick, color=blue75,domain=0.41:3.35] plot (\x,{\x*\x*\x - 6*\x*\x + 12*\x - 4});
                \draw[very thick, color=darkorange100,domain=1.09:3.91] plot (\x,{-12.75 + 15*\x - 3*\x*\x});
                \node[circle,fill,inner sep=2pt,blue100,label={[right]:\textcolor{blue100}{$s$}}] (dot) at (2.5,2) {};

                \node[color=magenta100] at (2.75,1) {$p_1$};
                \node[color=green100] at (0.75,4.75) {$p_2$};
                \node[color=blue100] at (3.5,5.5) {$p_3$};
                \node[color=darkorange100] at (1.75,5.5) {$p_4$};
    
                \draw[dashed, thick, gray] (2,6.5) -- (2,0);
                \draw[dashed, thick, gray] (3,6.5) -- (3,0);
    
                \draw[dotted, thick, black] (2.5,2.85) -- (2.5,3.5);
                \draw[dotted, thick, black] (2.4,3.6) -- (2.4,4.1);
                \draw[dotted, thick, black] (2.6,4.3) -- (2.6,6);
            \end{tikzpicture}
        \end{center}
        \caption{\textsc{Chain} heuristic}
        \label{fig:obs_chain}
    \end{subfigure}
    \hfill
    \begin{subfigure}[t]{0.3\textwidth}
        \begin{center}
            \begin{tikzpicture}
                \draw (0,0) -- (4.5,0) -- (4.5,6.5) -- (0,6.5) -- (0,0);
		        \node at (2.5,-.5) {$x_1$};
		        \node at (-.5,3) {$x_2$};
                
                \draw[very thick, magenta100] (2,2) ellipse (1cm and 1cm);
                \draw[very thick, green100] (0,6) -- (4.5,1.5);
                \draw[very thick, color=blue75,domain=0.41:3.35] plot (\x,{\x*\x*\x - 6*\x*\x + 12*\x - 4});
                \draw[very thick, color=darkorange100,domain=1.09:3.91] plot (\x,{-12.75 + 15*\x - 3*\x*\x});
                \node[circle,fill,inner sep=2pt,blue100,label={[right]:\textcolor{blue100}{$s$}}] (dot) at (2.5,2) {};
    
                \draw[dashed, thick, gray] (1.5,6.5) -- (1.5,0);
                \draw[dashed, thick, gray] (3,6.5) -- (3,0);
    
                \draw[dotted, thick, black] (2.5,2.85) -- (2.5,3.5);
                \draw[dotted, thick, black] (2.65,2.8) -- (2.65,4.3);
                \draw[dotted, thick, black] (2.35,2.95) -- (2.35,5.9);
            \end{tikzpicture}
        \end{center}
        \caption{\textsc{Biggest cell} heuristic}
        \label{fig:obs_biggestcell}
    \end{subfigure}
    \hfill
    \begin{subfigure}[t]{0.3\textwidth}
        \begin{center}
            \begin{tikzpicture}
                \draw (0,0) -- (4.5,0) -- (4.5,6.5) -- (0,6.5) -- (0,0);
		        \node at (2.5,-.5) {$x_1$};
		        \node at (-.5,3) {$x_2$};
                
                \draw[very thick, magenta100] (2,2) ellipse (1cm and 1cm);
                \draw[very thick, green100] (0,6) -- (4.5,1.5);
                \draw[very thick, color=blue75,domain=0.41:3.35] plot (\x,{\x*\x*\x - 6*\x*\x + 12*\x - 4});
                \draw[very thick, color=darkorange100,domain=1.09:3.91] plot (\x,{-12.75 + 15*\x - 3*\x*\x});
                \node[circle,fill,inner sep=2pt,blue100,label={[right]:\textcolor{blue100}{$s$}}] (dot) at (2.5,2) {};

                \draw[dashed, thick, gray] (2,6.5) -- (2,0);
                \draw[dashed, thick, gray] (3,6.5) -- (3,0);

                \draw[dotted, thick, black] (2.5,2.85) -- (2.5,3.5);
                \draw[dotted, thick, black] (2.4,3.6) -- (2.4,4.1);
                \draw[dotted, thick, black] (2.6,3.5) -- (2.6,5.95);
            \end{tikzpicture}
        \end{center}
        \caption{\textsc{Lowest degree barriers} heuristic}
        \label{fig:obs_barriers}
    \end{subfigure}

    \caption{Root ordering heuristics; the dotted lines indicate for which pairs of polynomials the resultant is added to the projection.}
\end{figure}
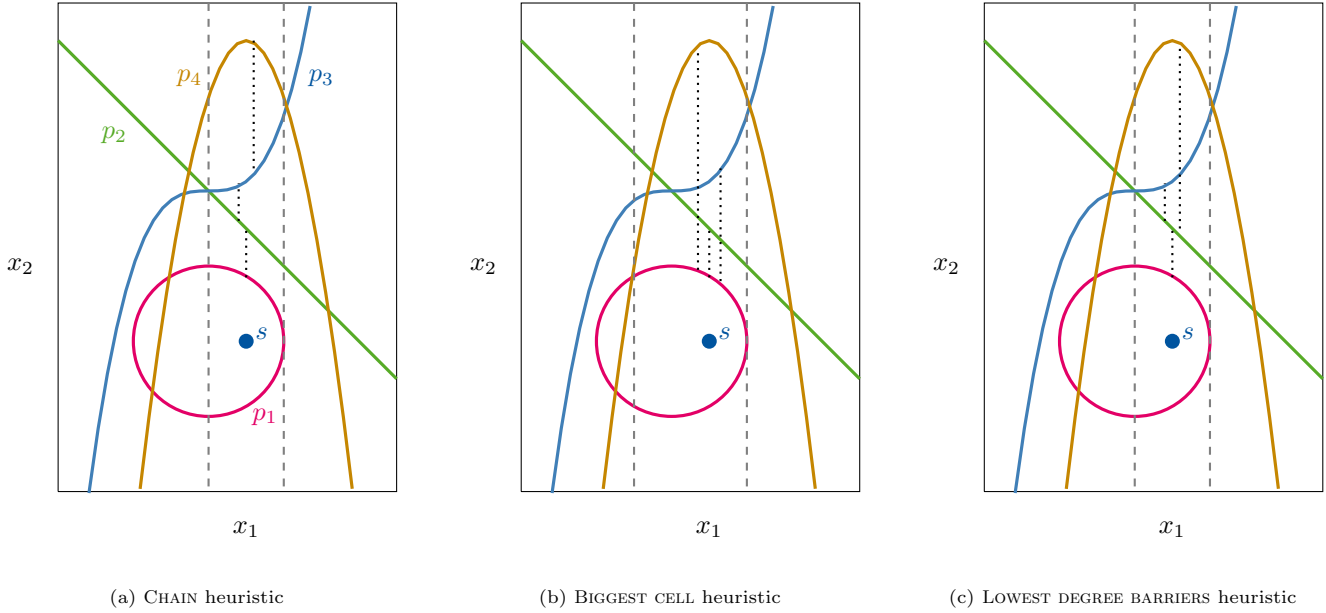

\paragraph{From the \textsc{full} to the \textsc{chain} heuristic}

As mentioned above, both the \textsc{full} and the \textsc{chain} heuristic fix the same ordering on the real root functions over any cell containing the current sample as depicted in \Cref{fig:obs_chain}.

Obviously, the \textsc{chain} heuristic is more efficient here as it uses a strict subset of the work done by the \textsc{full} heuristic.  Unlike the \textsc{full} heuristic, the \textsc{chain} heuristic takes the underlying sample into account, and thus the resulting projection is only valid for a single cell, i.e. \emph{locally delineable}. The \textsc{full} heuristic is independent from the sample and makes the set of polynomials fully delineable.

\paragraph{From the \textsc{chain} to the \textsc{biggest cell} heuristic}

The \textsc{chain} heuristic still fixes a stronger ordering on the root functions than necessary. As defined above, the \textsc{biggest cell} heuristic is the minimal requirement on the ordering to maintain sign-invariance on the constructed cell. This way, we hope that the size of the underlying cell is maximized, as in \Cref{fig:obs_biggestcell}. Note that while the \textsc{chain} heuristic only takes the sample in one dimension less into account, the \textsc{biggest cell} heuristic also considers the highest dimension of the sample.

\paragraph{From the \textsc{biggest cell} heuristic to the \textsc{lowest degree barriers} heuristic}

The \textsc{lowest degree barriers} heuristic minimizes the degrees of the resultants, as illustrated in \Cref{fig:obs_barriers}.  The rationale here is that resultant computations are heavy and their complexity depends on the degrees of the input polynomials, and that polynomials with lower degrees have fewer roots.  Thus reducing degrees may reduce the case of resultants having real roots that do not actually correspond to relevant points.  Note that despite the naming, the \textsc{lowest degree barriers} heuristic could in theory lead to bigger cells than the \textsc{biggest cell} heuristic in certain cases.

\paragraph{Equational constraint projection}

The equational constraint projection allows us to leave out discriminants and coefficients of all polynomials except the section-defining polynomial by only adding resultants of all polynomials with the defining polynomial. Thus, we expect this rule to be more efficient than the presented heuristics in most cases. However, we added the possibility for the application of heuristics for cases where the section-defining polynomial has high degree and thus computing resultants with that polynomial is desirable to be avoided.

\paragraph{On the gap between the idealized view and reality}

\begin{figure}%
    \begin{center}
        \begin{tikzpicture}

            \draw (-.73,-2) -- (4.9,-2) -- (4.9,3) -- (-.73,3) -- (-.73,-2);
		    \node at (2.25,-2.5) {$x_1$};
		    \node at (-1.5,0.75) {$x_2$};
            
            \draw[very thick, color=magenta100,domain=-1.25:1.25, variable=\y] plot ({2*\y*\y*\y*\y},{\y});
            \draw[very thick, color=darkorange100,domain=-1.82:2.82, variable=\y] plot ({-0.5*\y*\y+0.5*\y+1.875},{\y});
            \draw[very thick, color=green100,domain=-1.7:1.2, variable=\y] plot ({-0.4*pow(\y+0.25,6)+3},{\y});
            \node[circle,fill,inner sep=2pt,blue100,label={[right]:\textcolor{blue100}{$s$}}] (dot) at (0.5,0) {};
            \node at (3.5,1.5) {\textcolor{magenta100}{$\deg_{x_2}(p_1) {=} 4$}};
            \node at (1.5,2.5) {\textcolor{darkorange100}{$\deg_{x_2}(p_3) {=} 2$}};
            \node at (2.2,-1.72) {\textcolor{green100}{$\deg_{x_2}(p_2) {=} 6$}};

            \draw[dotted, thick, black] (0.2,-0.6) -- (0.2,-1.3);
            \draw[dotted, thick, red100] (0.4,-1.3) -- (0.42,-1.6);
            \draw[dotted, thick, black] (0.2,0.6) -- (0.2,1.15);
            \draw[dotted, thick, black] (0.4,0.7) -- (0.4,2.3);
        \end{tikzpicture}
    \end{center}
    \caption{Redundancies in the \textsc{Lowest degree barriers} heuristic}
    \label{fig:obs_imperfection}
\end{figure}
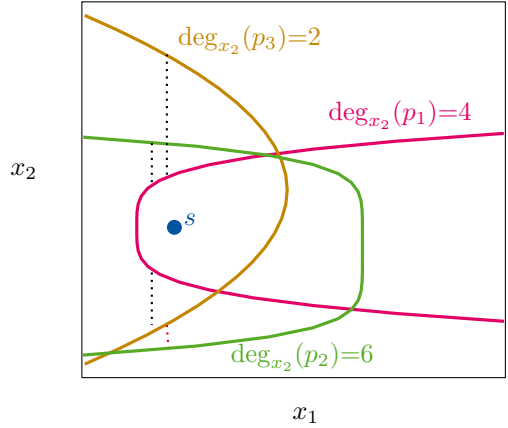

For the \textsc{lowest degree barriers} heuristic, this might lead to the computation of a redundant set of resultants regarding the minimal requirement on the indexed root ordering, as shown in \Cref{fig:obs_imperfection}, where the relation between the first roots of $p_2$ and $p_3$ (depicted in red) is superfluous but its corresponding resultant is added because all roots are considered individually and not their connection via the defining polynomials.

\subsection{Comparison with the refinement-based approach}

The refinement-based approach to single cell construction \cite{brown2015onecell} saves resultants compared to full CAD projection in the same way as our levelwise variant, by exploiting the transitivity of the induced ordering on real root functions. Complexity-wise, the approaches are the same, as both add up to two resultants per polynomial on a level.

However, the influence on the ordering of constraints on the refinement-based variant means the quality of the constructed cell varies. In the worst case, the resulting ordering corresponds to the \textsc{chain} heuristic: in \Cref{fig:obs_chain}, this is achieved for the example when the polynomials are merged in the ordering $p_4, p_3, p_2, p_1$. Then, the upper cell boundary is updated in every step to a lower boundary. In the best case, the \textsc{biggest cell} heuristic is achieved: for our example in \Cref{fig:obs_biggestcell}, this is when the polynomial $p_1$ defining the sector's boundary is merged first.

For the section case, it should also be noted that the levelwise approach can always apply the equational constraints rule as the cell description is known before the projection. The refinement-based approach only starts applying the equational constraints rule when the cell collapses to a section, until then it adds discriminants of all polynomials, which may not actually be needed. The illustrating examples may be used to explain this observation. Consider \Cref{fig:obs_chain} but with the sample $s$ moved to the upper root of $p_1$. When merging polynomials in the ordering $p_4, p_3, p_2, p_1$, the polynomials $p_4, p_3, p_2$ are merged as in the sector case until the cell collapses to a section when merging $p_1$. When merging $p_1$ first, then the section case is identified directly and the reduced projection applied when merging $p_4, p_3, p_2$, meaning the discriminants and coefficients for those polynomials need not be added.

\subsection{Potential for non-connected ``cell'' descriptions}
Recall that we aim to compute a single cell on which the input polynomials are sign-invariant, but McCallum's CAD projection theory uses the stronger property of order-invariance.  It has been observed before that this can allow for small optimizations at the top layer.  For example, when using equational constraint projection we must take discriminants if the projection is in a middle layer (to ensure order-invariance is provided which is a hypothesis of the next lifting) but can avoid these at the top layer as we need never lift over that \cite{EBD20}.

The visualization of our proof rules in \Cref{fig:rules_overview} alerted us to another such optimization, which if enacted has some strange consequences.  Note that a cell being connected at level $i$ is required for it to be order-invariant, but not sign-invariant.  Thus, we need not ensure connectivity at the top level, i.e. we could avoid taking resultants of the upper and lower bound polynomials to satisfy \Cref{def:map:connectedness}.  Without connectedness, it would not be accurate to describe what we are constructing as a cell, rather it is a semi-algebraic set (see also \Cref{fig:map:connectedness}). However, it is still describing a portion of space on whose points the respective polynomials are all sign-invariant.  Thus, in the context of the MCSAT search which we discuss in the next section, the set is still describing a portion of space on whose points the constraints are all unsatisfiable for the same reason and so its negation is still a valid explanation clause to further that search. 

We note that the resultants saved by this optimization may still have to be computed later, if the cell is used in further propagation.  However, this will not always be the case and so often this optimization may save computation.  Discovery of this optimization illustrates the advantages of the proof system presentation used in this paper.

\subsection{Factorization of polynomials and cell size}
To ensure the output of CAD is correct we must compute a \emph{square-free basis} of the current set of polynomials $P$ (i.e. a set of square-free polynomials without common factors which define the same varieties as $P$) before the application of a projection operator.

The approach of the rules presented above is to fully factorize each polynomial, resulting in what is called the \emph{finest} square-free basis.  This is a fairly standard choice made in CAD implementations as the effort of computing a full factorization pays off compared to the heavy resultant, discriminant, and real root isolation computations, which are all simpler for smaller polynomials.

When building a full CAD the choice of a square-free basis does not affect the decomposition computed, just the time taken to compute it.  However, for the single-cell construction, this makes a difference also in the size of the resulting cell, specifically the cell can be larger if we factor.  This can easily be observed by considering \Cref{ex:motivation}.  Here polynomial $p_1 \cdot p_2 \cdot p_3$ is already square-free. Using this directly in the one-cell construction algorithm without factorization as a whole would simply result in computing the discriminant of the polynomial (and some coefficients): the discriminant must have as factors all the cross resultants of $p_1, p_2, p_3$ by definition. So in this case, no improvement over a full projection is achieved, and we would find the smaller cell from \Cref{fig:recapproach}.  However, if we factor to consider the set $\{p_1, p_2, p_3\}$ instead then we can obtain the larger cell from \Cref{fig:recapproach}.

\paragraph{The limits of factorization}

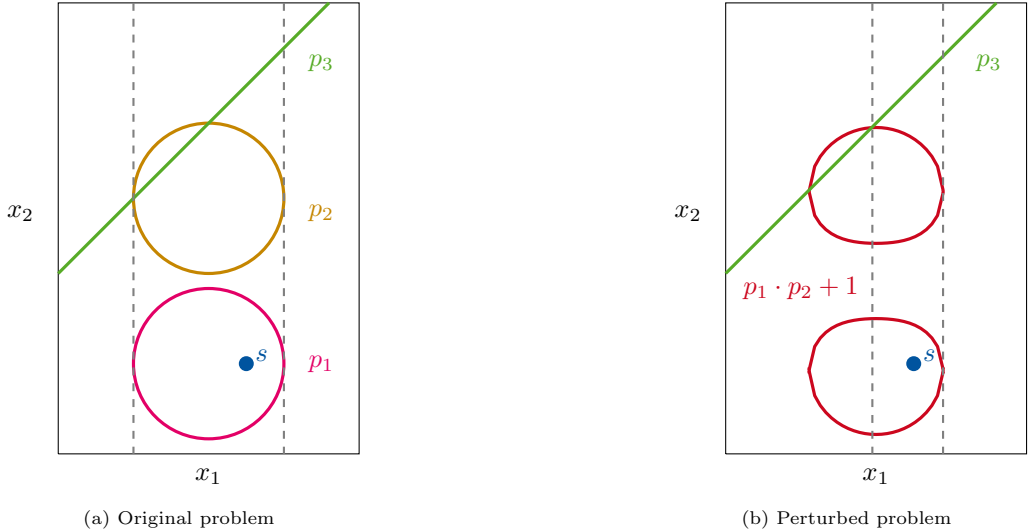
\begin{figure}
    \begin{subfigure}[b]{0.5\textwidth}
        \center
        \begin{tikzpicture}

            \draw (-2,-.2) -- (2,-.2) -- (2,5.8) -- (-2,5.8) -- (-2,-.2);
		    \node at (0,-.5) {$x_1$};
		    \node at (-2.5,3) {$x_2$};
            
            \draw[very thick, magenta100] (0,1) ellipse (1cm and 1cm);
            \draw[very thick, darkorange100] (0,3.2) ellipse (1cm and 1cm);
            \draw[very thick, color=green100,domain=-2:1.6] plot (\x,{\x+4.2});
            \node[circle,fill,inner sep=2pt,blue100,label={[right]:\textcolor{blue100}{$s$}}] (dot) at (0.5,1) {};

            \node[color=magenta100] at (1.5,1) {$p_1$};
            \node[color=darkorange100] at (1.5,3) {$p_2$};
            \node[color=green100] at (1.5,5) {$p_3$};

            \draw[dashed, thick, gray] (-1,5.8) -- (-1,-.2);
            \draw[dashed, thick, gray] (1,5.8) -- (1,-.2);

        \end{tikzpicture}

        \caption{Original problem}
        \label{fig:irredcirclesa}
    \end{subfigure}
    \begin{subfigure}[b]{0.5\textwidth}
        \center
        \begin{tikzpicture}
            \draw (-2,-.2) -- (2,-.2) -- (2,5.8) -- (-2,5.8) -- (-2,-.2);
		    \node at (0,-.5) {$x_1$};
		    \node at (-2.5,3) {$x_2$};
            
            \draw[very thick, color=red100,domain=-0.89072:0.89072] plot (\x,{0.1*(21-sqrt(-100*\x*\x-20*sqrt(96-121*\x*\x)+221))});
            \draw[very thick, color=red100,domain=-0.89072:0.89072] plot (\x,{0.1*(21+sqrt(-100*\x*\x-20*sqrt(96-121*\x*\x)+221))});
            \draw[very thick, color=red100,domain=-0.89072:0.89072] plot (\x,{0.1*(21-sqrt(-100*\x*\x+20*sqrt(96-121*\x*\x)+221))});
            \draw[very thick, color=red100,domain=-0.89072:0.89072] plot (\x,{0.1*(21+sqrt(-100*\x*\x+20*sqrt(96-121*\x*\x)+221))});
            \draw[very thick, color=red100] (-0.89072, 3.28) -- (-0.89072, 3.31);
            \draw[very thick, color=red100] (0.89072, 3.27) -- (0.89072, 3.31);
            \draw[very thick, color=red100] (-0.89072, 0.87) -- (-0.89072, 0.95);
            \draw[very thick, color=red100] (0.89072, 0.87) -- (0.89072, 0.95);
            
            \draw[very thick, color=green100,domain=-2:1.6] plot (\x,{\x+4.2});
            \node[circle,fill,inner sep=2pt,blue100,label={[right]:\textcolor{blue100}{$s$}}] (dot) at (0.5,1) {};

            \node[color=red100] at (-1,2) {$p_1 \cdot p_2 + 1$};
            \node[color=green100] at (1.5,5) {$p_3$};

            \draw[dashed, thick, gray] (-0.05,5.8) -- (-0.05,-.2);
            \draw[dashed, thick, gray] (0.89,5.8) -- (0.89,-.2);

        \end{tikzpicture}

        \caption{Perturbed problem}
        \label{fig:irredcirclesb}
    \end{subfigure}
    \caption{By perturbing a problem making a reducible polynomial irreducible, the problem gets harder.}
    \label{fig:irredcircles}
\end{figure}

We note that this described gain in size of the computed cell from factorization does not mean we build optimal cells for all problems where there is a geometric separation.  \Cref{fig:irredcirclesa} shows an example with two circles and a linear polynomial for which the constructed cell we ideally build is the inside of the circle defined by $p_1$. If we were originally presented with $p_1 \cdot p_2$, then factorization would allow a one-cell algorithm to ignore the intersections of $p_2$ and $p_3$ to construct the entire inner circle.  But consider the similar example from \Cref{fig:irredcirclesb}, in which we have perturbed the problem to consider the irreducible polynomial $p_1 \cdot p_2 + 1$. Graphically, the two problems seem to be similar, and the delineation of the roots are identical.  But in the second case the single cell construction cannot treat the two ovals separately: we must consider the irrelevant intersection and thus build a smaller than ideal cell.

This shows some limits of the single cell construction by way of a well observed truth in computer algebra: problems which appear similar to human beings (i.e. when viewed geometrically) are not always equally hard algebraically. 

\section{Experimental evaluation}
\label{sec:experiments}
The presented proof rules are, due to their generality, potentially applicable with only small additions to a variety of problems and algorithms related to non-linear arithmetic, such as quantifier elimination by CAD \cite{Collins1975}, various CAD optimizations e.g. \cite{mccallum1999equational, CH91}, non-uniformly cylindrical decompositions for quantifier elimination \cite{Brown2015, Brown2017}, the generation of explanations when using cylindrical algebraic coverings for a traditional SMT solver \cite{abraham2020covering}, and the generation of explanations in MCSAT \cite{jovanovic2012solving,jovanovic2013design,brown2015onecell}. As the latter was the main motivation for our work, the evaluation of this paper will focus on generating theory explanations in MCSAT for non-linear arithmetic.

\subsection{Generating explanations for MCSAT}
\label{sec:mcsat_explanations}

Recall the description of MCSAT in \Cref{ssec:mcsat}.  
We are interested in when MCSAT resolves theory conflicts. I.e. when there is a set of constraints $C$ in real variables $x_1,\ldots,x_n,x_{n+1}$ that should be satisfied according to the Boolean model and an assignment $s: \{ x_1, \ldots, x_n \} \to \reals$ such that $s$ cannot be extended to a value for $x_{n+1}$ satisfying $C$. The task then is to exclude a cell around $s$ that generalizes this conflict, i.e. a region cell where the reason for unsatisfiability of $C$ is invariant.

This reason of unsatisfiability is maintained when all input polynomials are sign-invariant on the generalized cell. To achieve this, we could do a full McCallum projection step, obtaining a set of properties of one level below allowing to construct a cell around $s$.

However, this is already too strong, as we need the set of input polynomials $P = \{ p \mid (p \sim 0) \in C \} \subset \rationals[x_1, \ldots, x_n,x_{n+1}]$ to be delineable over a cell \emph{containing the current sample $s \in \reals^n$} for maintaining the desired property. We achieve this by determining the indexed root expression of the real roots of the set $\{ p \in \factors{P} | \level{p} = n+1 \}$ over $s$ in $x_{n+1}$ and ordering them such that $\xi_1(s) \leq \xi_2(s) \leq \ldots \leq \xi_k(s)$. Finally, we ensure that all lower level factors $\{ p \in \factors{P} | \level{p} < n+1 \}$ are sign-invariant, check that each polynomial is not nullified (if not, we stop), make each polynomial individually delineable and add the resultants of the pair of polynomials $(\xi_j.p, \xi_{j+1}.p)$ for $j \in [1..k-1]$. Thus, the input of the presented one-cell algorithm is given as

\begin{align*}
    Q &=  \{ \sgninv{p} \mid p \in \factors{P},\ \level{p}<n+1 \} \\
    &\quad \cup \{ \del{p} \mid p \in \factors{P},\ \level{p}=n+1 \} \\
    &\quad \cup \{ \ordinv{\res{x_{n+1}}{\xi_j.p}{\xi_{j+1}.p}} \mid j \in [k-1] \}.
\end{align*}
This approach is similar to the chain heuristic for indexed root orderings presented in \Cref{def:heuristics_preceq_chain}.

Note that this approach could be embedded nicely into our system as a proof rule, taking over some optimizations. Furthermore, there are alternative approaches for elimination in the first levels, i.e. by computing a covering of unsatisfying intervals of input constraints. These possibilities are part of our plans for future work.

Further, note that in MCSAT, conflicts might also depend on previously computed cells which are expressed by conjunctions of \emph{extended constraints} where a variable is compared with an indexed root expression; these constraints can also occur in the input. To handle an extended constraint $x_{n+1} \sim \iroot{x_{i+1}}{p}{j}$ with $p \in \rationals[x_1,\ldots,x_{n+1}]$, we simply add $p$ to the set $P$ of input polynomials.

\subsection{Implementation}

For the evaluation of the presented algorithm, we employ the \textsc{SMT-RAT} \cite{smtrat,corzilius2015smt} solver, which provides an \emph{MCSAT} engine allowing the combination of multiple explanation backends. Several incomplete and complete methods are combined in the sense that these backends are called sequentially until one returns an explanation.

Currently available backends are the \textit{Fourier-Motzkin variable elimination (FM)} \cite{jovanovic2013design}, \textit{interval constraint propagation (ICP)} \cite{kremercylindrical}, \textit{virtual substitution (VS)} \cite{abrahamembedding}, the complete model-based CAD cell construction algorithm from \textit{NLSAT} \cite{jovanovic2012solving} using Collin's projection operator as well as the refinement-based single cell construction algorithm \cite{brown2015onecell}. Furthermore, SMT-RAT employs a fully dynamic activity-based variable ordering heuristic for scheduling theory variable assignments and Boolean decisions \cite{nalbach2019variable}.

All variants of the presented levelwise one-cell construction algorithm are implemented as backends in \textsc{SMT-RAT}. This is a preliminary version not fully exploiting the power of all proof rules, in particular, it is not checked whether a property is already implied by some properties in the projection.

\noindent For the evaluation, we compare the following solver variants:

\begin{description}
    \item[NL] The model-based projection using Collin's operator from \emph{NLSAT} \cite{jovanovic2012solving} as a complete explanation backend.  I.e. to use when one of the following variants which are all based on McCallum projection hits a nullification which they cannot handle (see \Cref{def:nullification}).
    \item[OC-*] The refinement-based one-cell construction algorithm \cite{brown2015onecell}. We use the same MCSAT embedding as described above. Furthermore, the refinement-based method is able to return an explanation when a polynomial is nullified in the sector case in some special cases which our levelwise approach cannot handle yet; for better comparability, these special cases are excluded from the following tests. In case of failure, the complete explanation from \emph{NLSAT} is called. To further specify the algorithm's behaviour and make it reproducible, we implemented heuristics for the order of merging of initial polynomials. These specify the \textbf{*} in the variant name as follows.
    \begin{description}
    	\item[ASC] The merge-operation is called on the initial polynomials in \textit{ascending} order by their total degree.
    	\item[DSC] The merge-operation is called on the initial polynomials in \textit{descending} order by their total degree.
    \end{description} 
    \item[LW-*-*] The new levelwise one-cell construction algorithm with different heuristics applied in the section and sector case. In case of failure, the complete explanation from \emph{NLSAT} is called. The first \textbf{*} in the variant name dictates the employed heuristic for the section case and the second for the sector case. Possible substitutions for the stars are as follows.
    \begin{description}
        \item[EQ] \textsc{equational constraint} heuristic is applied (only for section case).
        \item[BC] \textsc{biggest cell} heuristic is applied (only for sector case).
        \item[CH] \textsc{chain} heuristic is applied. 
        \item[LDB] \textsc{lowest degree barriers} heuristic is applied.
    \end{description} 
    \item[\textit{[solver]}+] with \textit{[solver]} being one of the solver variants above, uses the FM-, ICP- and VS-based backends serially, in this order, before resorting to \textit{[solver]}. Furthermore, we apply general preprocessing to the input before calling the main solver \cite{corzilius2015smt}.
\end{description} 

All variants are executed on the \textsc{SMT-LIB} benchmark library \cite{barrett2010smt} for quantifier-free non-linear real arithmetic, abbreviated as \textsc{QF\_NRA}.
This set contains 11552 problem instances.

The machine used for testing has four 2.1 GHz AMD Opteron CPUs with 12 cores each.
In the created test series, each instance was solved with 15 minutes timeout and 6 GB of memory.

For reproducibility, the implementation which generated the following results is available at \url{https://doi.org/10.5281/zenodo.5764569}.

\subsection{Results}

\paragraph{Examined solvers}

First of all, we observed that \texttt{OC-ASC} solves as many instances as \texttt{OC-DSC} but needs slightly less time; thus, we omit \texttt{OC-DSC}. Furthermore, we observe that \texttt{LW-EQ-CH} and \texttt{LW-EQ-LDB} solve more instances than \texttt{LW-CH-CH} and \texttt{LW-LDB-LDB}. In our basic implementation, saving leading coefficients and discriminants pays off compared to the other heuristics. Thus, for further examination, we focus on the \texttt{LW-EQ-*} variants which always use the equational constraints projection in the section case. A brief summary of solved instances of all solvers can be seen in \Cref{tbl:results:summary}. Furthermore, from now on, \texttt{VB-LW} (respectively \texttt{VB-LW+}) is the virtual best of the \texttt{LW-EQ-*} (respectively \texttt{LW-EQ-*+}) solvers. \texttt{VB} and \texttt{VB+} are the corresponding virtual bests with respect to \texttt{LW-EQ-*} and \texttt{OC-ASC}.

\begin{table}
	\center
	\begin{tabular*}{0.8\linewidth}{@{\extracolsep{\fill}}|c||c|c c|c c|}
		\hline
		{} &  solved &  sat &  unsat &  with single cell &  without single cell \\
		\hline
		\hline
		NL         &    9057 &        4518 &          4539 &                         &                         \\
		OC-ASC     &    9316 &        4605 &          4711 &                     4194 &                        5122 \\
		OC-DSC     &    9316 &        4605 &          4711 &                     4194 &                        5122 \\
		LW-EQ-BC   &    9308 &        4598 &          4710 &                     4186 &                        5122 \\
		LW-EQ-CH   &    9311 &        4602 &          4709 &                     4189 &                        5122 \\
		LW-EQ-LDB  &    9304 &        4595 &          4709 &                     4182 &                        5122 \\
		VB-LW      &    9403 &        4650 &          4753 &                     4281 &                        5122 \\
		VB         &    \textbf{9435} &        \textbf{4671} &          \textbf{4764} &                     \textbf{4313} &                        5122 \\

		LW-CH-CH   &    9305 &        4600 &          4705 &                     4183 &                        5122 \\
		LW-LDB-LDB &    9296 &        4585 &          4711 &                     4174 &                        5122 \\
		\hline
		NL+        &    9535 &        4699 &          4836 &                         &                         \\
		OC-ASC+    &    9797 &        4788 &          5009 &                      615 &                        9182 \\
		LW-EQ-BC+  &    9795 &        4786 &          5009 &                      613 &                        9182 \\
		LW-EQ-CH+  &    9797 &        4787 &          5010 &                      615 &                        9182 \\
		LW-EQ-LDB+ &    9795 &        4786 &          5009 &                      613 &                        9182 \\
		VB-LW+     &    9797 &        4787 &          5010 &                      615 &                        9182 \\
		VB+        &    \textbf{9800} &        \textbf{4790} &          \textbf{5010} &                      \textbf{618} &                        9182 \\
		\hline
		\hline
		Total      &   11552* &        5069* &          5379* &                         &                            \\
		\hline
	\end{tabular*}

	\textit{* For 1104 instances, it is not known whether they are satisfiable or unsatisfiable.}
		
	\caption{Details on the instances solved by each solver: the number solved (first column), the number of satisfiable and unsatisfiable solved instances (the second and third columns) and the number of solved instances where the solver made at least one call to the single-cell construction and the number that did not require any such call to solve (the final two columns). Note that the last column is, for the first block of solvers, the number of instances that may be solved with Boolean reasoning and construction of sample points alone, i.e. without any theory calls.  For the second block of solvers this also includes the instances that are solved using the additional incomplete theory backends. \label{tbl:results:summary}}
\end{table}

\paragraph{General observations}

Before we compare the different approaches, we make some general comments on the results based on exemplary solvers in \Cref{tbl:res:details}.

As already observed in \Cref{tbl:results:summary} and confirmed by \Cref{tbl:res:details}, large parts of the benchmark set are relatively easy. Around half of the benchmarks do not involve a single explanation call; and when enabling the additional incomplete backends, $79\%$ of the benchmarks can be solved without a single call to the single cell construction. Considering the total number of explanation calls made, only $3.14\%$ of them use the single cell construction for the \texttt{VB-LW+} solver, meaning that even for the problems where single cell is needed, it is only needed rarely.  

The fail rate of the levelwise backend (the cases where a nullification occurs) is smaller on the \texttt{VB-LW+} solver ($6\%$) than on the \texttt{VB-LW} solver ($17.98\%$). That means, that nullifications are more probable on the simple parts of the problem.

It should also be noted that \texttt{VB-LW+} needs significantly less explanation calls than \texttt{VB-LW}; which means, that the incomplete backends have explanations of higher quality.

To summarize, we can only make meaningful statements on our heuristics based on the solver variants without the additional backends, as otherwise, there are too few calls to the single cell construction in solved instances. The possible reasons for this are twofold. On the one hand, our procedure and its implementation may not yet be suitable to solve the harder instances in the benchmark set. On the other hand, the benchmark set might not contain enough interesting or diverse benchmarks for an evaluation. 

\begin{table}
	\begin{tabular*}{\linewidth}{@{\extracolsep{\fill}}|l||c c c c|}
		\hline
		{} &        LW-EQ-BC &      LW-EQ-BC+ &               VB-LW &            VB-LW+ \\
		\hline
		\hline
		solved instances                     &            9308 &           9795 &             9403 &           9797 \\
		mean running time (s)               &            4.36 &           5.22 &             5.56 &           5.26 \\
		sum calls to explanation         &           83628 &          58826 &           107964 &          58953 \\
		\quad of them: sum calls to SC                  &  83628 (100.0\%) &   1740 (2.96\%) &  107964 (100.0\%) &   1849 (3.14\%) \\
		\quad \quad of them: sum calls to NL               &  15394 (18.41\%) &     117 (0.2\%) &   19412 (17.98\%) &     119 (0.2\%) \\
		\hline
		solved instances with single cell             &            4186 &            613 &             4281 &            615 \\
		mean running time (s)      &            9.13 &          20.34 &             11.7 &          21.81 \\
		sum calls to explanation  &           83628 &          10363 &           107964 &          10490 \\
		\quad of them: sum calls to SC          &  83628 (100.0\%) &  1740 (16.79\%) &  107964 (100.0\%) &  1849 (17.63\%) \\
		\quad \quad of them: sum calls to NL          &  15394 (18.41\%) &    117 (1.13\%) &   19412 (17.98\%) &    119 (1.13\%) \\
		\hline
		\end{tabular*}

		\caption{Comparison of four exemplary solvers: firstly, the subset of all solved instances and secondly the subset of all solved instances where at least one call to the single cell construction was made are considered. For each, the statistics describing the total number of instances, the mean running time, the sum of constructed cells as well as the sum of attempts to generate a cell using single cell construction or NLSAT backend are given. For the last two, these sums are included in the previous one; and their share of the total number of constructed cells is given. Note that the explanation calls involving single cell are the ones not being solved by the additional backends and the calls involving NLSAT are the ones where the single cell construction failed.}
	\label{tbl:res:details}
\end{table}

\paragraph{Overall results}

\begin{figure}
	\center

	\begin{subfigure}[t]{0.49\textwidth}
		\includegraphics{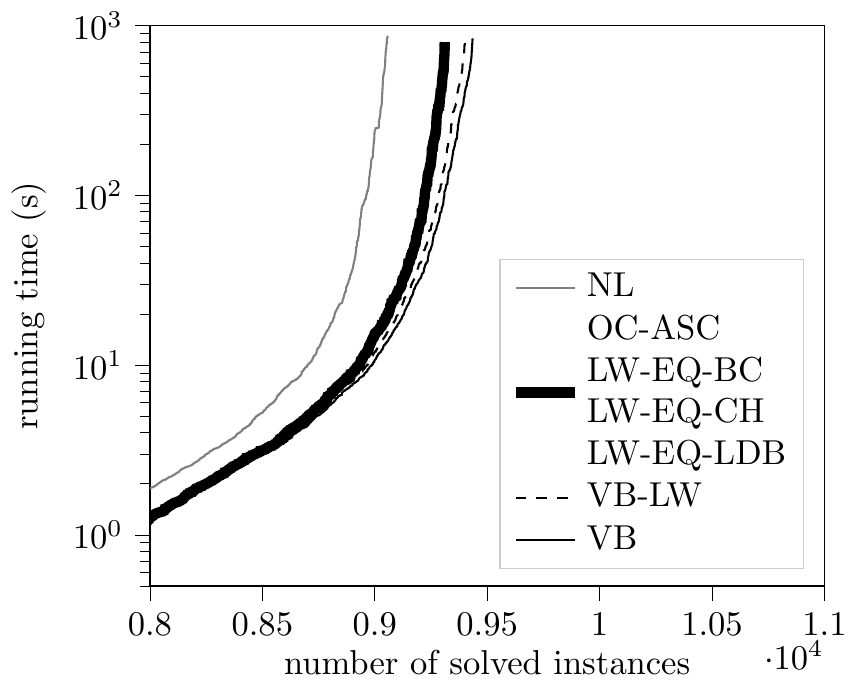}
		\caption{Comparison of solvers without additional backends.}
		\label{fig:res:performanceprofile:pure}
	\end{subfigure}
	\begin{subfigure}[t]{0.49\textwidth}
		\includegraphics{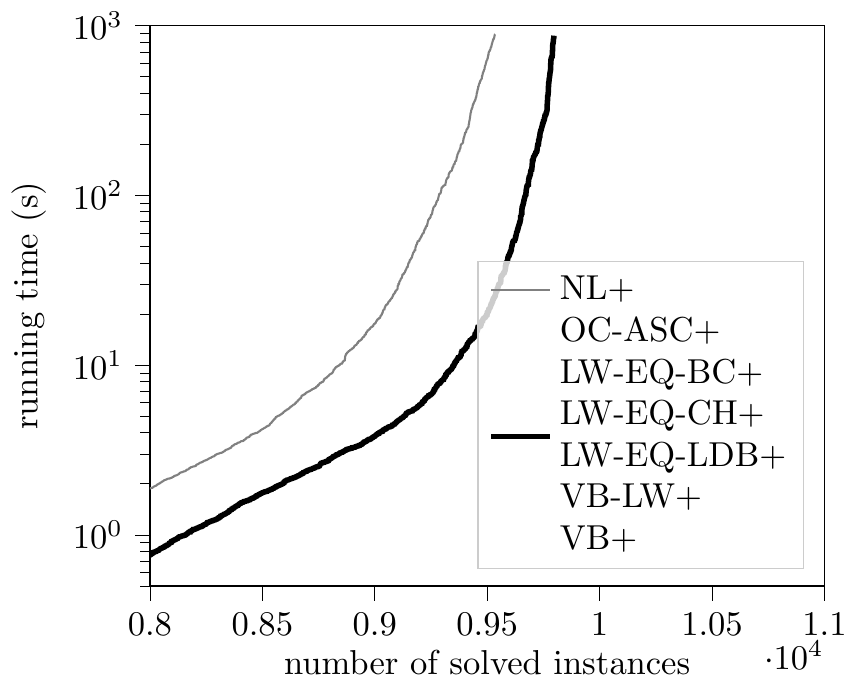}
		\caption{Comparison of solvers with additional backends.}
		\label{fig:res:performanceprofile:additional}
	\end{subfigure}

	\caption{Performance profile (running times).}
	\label{fig:res:performanceprofile}
\end{figure}

All solvers are depicted in the performance profile in \Cref{fig:res:performanceprofile}.

First of all, the \texttt{NL} and \texttt{NL+} solvers perform significantly worse than the single cell variants, which justifies the investigation of this new levelwise cell construction approach.

The refinement-based approach \texttt{OC-ASC} as well as \texttt{LW-EQ-*} perform similarly. Among the levelwise variants, the \texttt{LW-EQ-CH} solves the most; however, these differences are not significant and depend on the implementation and heuristics chosen in the MCSAT solver. Considering \texttt{OC-ASC+} and \texttt{LW-EQ-*+}, these differences vanish even more when combining them with incomplete methods handling simple sub-problems. These results are summarized in \Cref{tbl:results:summary}.

\paragraph{Virtual best and orthogonality of heuristics}

\begin{figure}
	\centering
	\begin{tikzpicture}
		\draw (0,1.73205) ellipse (2cm and 2cm);
		\draw (1,0) ellipse (2cm and 2cm);
		\draw (-1,0) ellipse (2cm and 2cm);
		\node at (0,4) {\textsc{LW-EQ-BC}};
		\node at (2.3,-2.3) {\textsc{LW-EQ-CH}};
		\node at (-2.3,-2.3) {\textsc{LW-EQ-LDB}};
		\node at (0,2.7) {27}; %
		\node at (1.8,-.5) {30}; %
		\node at (-1.8,-.5) {40}; %
		\node at (1.25,1.25) {42}; %
		\node at (0,-1) {25}; %
		\node at (-1.25,1.25) {25}; %
		\node at (0,.65) {9214}; %

	\end{tikzpicture}

	\caption{Commonly solved instances.}
	\label{fig:results:orthogonal}
\end{figure}
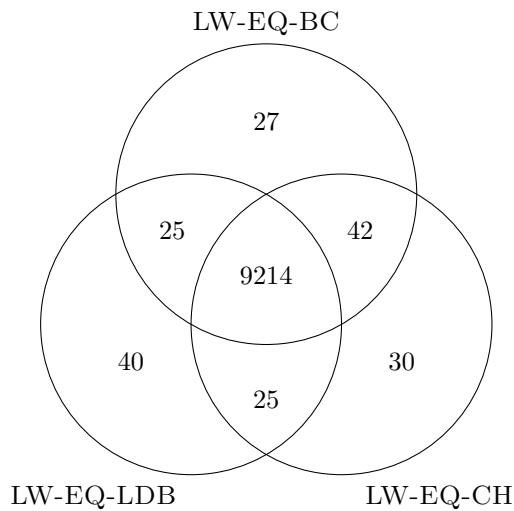

\texttt{VB-LW} performs significantly better than the \texttt{LW-EQ-*} solvers. This means that although the number of solved instances is similar for all heuristics, each heuristic solves instances that the others do not solve.  Thus these heuristics are \emph{orthogonal} to some degree. We depict the number of instances solved by differing combinations of solvers in \Cref{fig:results:orthogonal}. Note that the same holds for \texttt{VB}, meaning that \texttt{OC-ASC} is orthogonal to \texttt{LW-EQ-*} as well.

Note that the \texttt{VB-LW+} solver does not solve significantly more instances than any of the \texttt{LW-EQ-*+} solvers. That means the differences of the heuristics only become noticeable in the simple parts of the instances. An explanation could be that harder parts of instances require heavy resultant computations which could quickly shift the instance to unsolvable within the timeout.

\begin{figure}[t]
	\begin{subfigure}[t]{0.49\textwidth}
		\centering
		\includegraphics{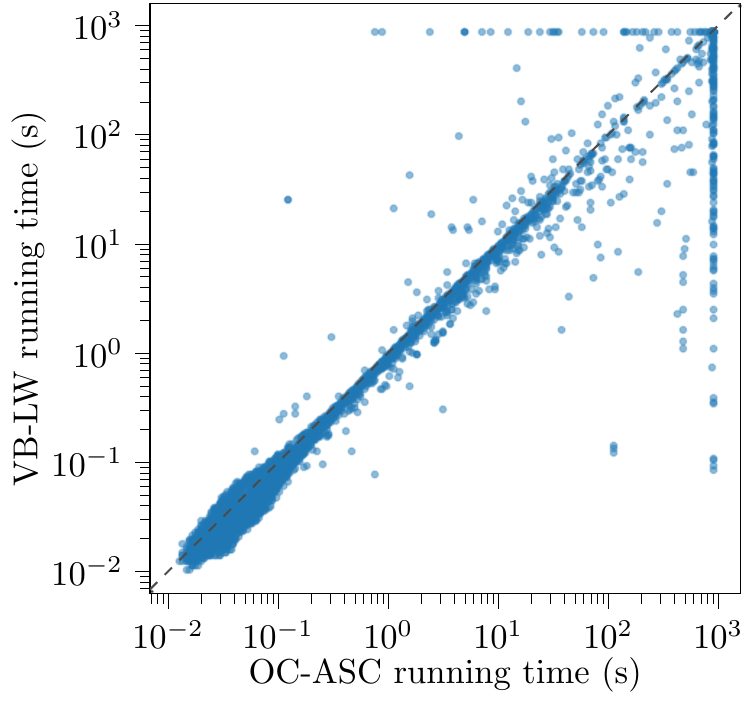}
	
		\caption{Running times on all instances.}
		\label{fig:res:scatter_runtime}
	\end{subfigure}\hfill
	\begin{subfigure}[t]{0.49\textwidth}
		\centering
		\includegraphics{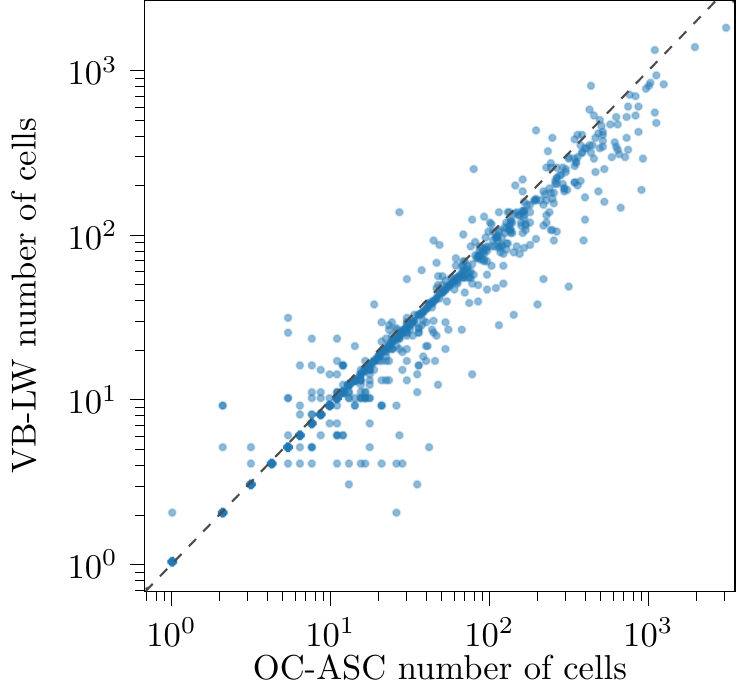}
	
		\caption{Number of constructed cells on instances solved by both solvers.}
		\label{fig:res:scatter_cells}
	\end{subfigure}
	\caption{Scatter plots comparing \texttt{VB-LW} and \texttt{OC-ASC} on the individual instances.}
	\label{fig:res:scatter}
\end{figure}

Now focusing on the ``pure'' solvers without additional backends, we observe that on simple instances the refinement based approach \texttt{OC-ASC} and the virtual best of the levelwise approaches \texttt{VB-LW} behave similar on simple instances while they are more orthogonal on harder instances, as indicated by \Cref{fig:res:scatter_runtime}. 

The SMT-LIB benchmark set is split up into families which each have a similar structure. We note that all variants \texttt{LW-EQ-*} and \texttt{OC-ASC} solve roughly the same amount of benchmarks in each family individually, as seen in \Cref{tbl:res:families}, while the virtual best solvers do solve more. That means that the orthogonality of the variants is split up over the various benchmark families.

\begin{table}
	\begin{tabular*}{\linewidth}{@{\extracolsep{\fill}}|l||c c c c c c c c c c c c|}
		\hline
		{} &  1&2&3&4&5&6&7&8&9&20&11&12 \\
		\hline
		\hline
		OC-ASC    &                  25 &                   0 &                                             0 &                           86 &           9 &            2 &                 26 &     6 &    2180 &        8 &         6913 &     61 \\
		LW-EQ-BC  &                  22 &                   0 &                                             0 &                           89 &           9 &            2 &                 22 &     6 &    2170 &        8 &         6918 &     62 \\
		LW-EQ-CH  &                  25 &                   0 &                                             0 &                           88 &           9 &            3 &                 22 &     5 &    2174 &        8 &         6914 &     63 \\
		LW-EQ-LDB &                  22 &                   0 &                                             0 &                           88 &           9 &            2 &                 22 &     7 &    2170 &        8 &         6913 &     63 \\
		VB-LW     &                  26 &                   0 &                                             0 &                           90 &           9 &            4 &                 25 &     7 &    2249 &        8 &         6918 &     67 \\
		VB        &                  27 &                   0 &                                             0 &                           94 &           9 &            4 &                 28 &     7 &    2267 &        8 &         6923 &     68 \\
		\hline\hline
		Total     &                 405 &                   9 &                                            69 &                          135 &          63 &          821 &                 61 &    20 &    2752 &       45 &         7006 &    166 \\
		\hline
		\end{tabular*}

		\caption{Solved instances by solver and family. SMT-LIB benchmark families: (1) 20161105-Sturm-MBO (2)  20161105-Sturm-MGC   (3) 20170501-Heizmann-UltimateInvariantSynthesis (4)   20180501-Economics-Mulligan   (5) 2019-ezsmt   (6) LassoRanker   (7) UltimateAutomizer   (8) hong  (9) hycomp   (10) kissing  (11) meti-tarski  (12) zankl.}
	\label{tbl:res:families}
\end{table}

\paragraph{Number of constructed cells}

\begin{figure}[t]
	\center

	\begin{subfigure}[t]{0.49\textwidth}
		\includegraphics{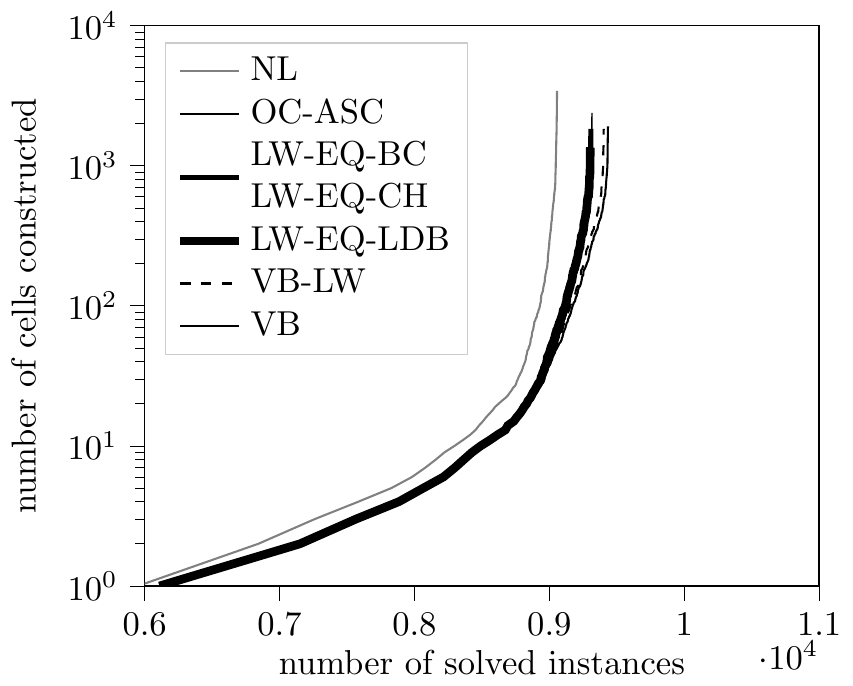}
		\caption{Comparison of solvers without additional backends.}
		\label{fig:res:performanceprofile_cells:pure}
	\end{subfigure}
	\begin{subfigure}[t]{0.49\textwidth}
		\includegraphics{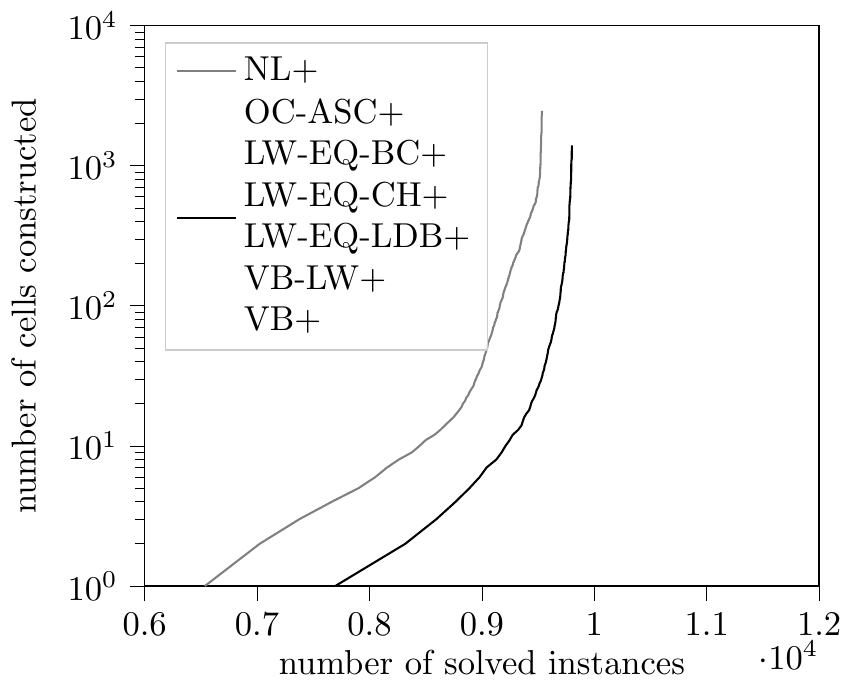}
		\caption{Comparison of solvers with additional backends.}
		\label{fig:res:performanceprofile_cells:additional}
	\end{subfigure}

	\caption{Performance profile (number of cells).}
	\label{fig:res:performanceprofile_cells}
\end{figure}

\Cref{fig:res:performanceprofile_cells} depicts a performance profile where we consider the number of constructed cells (that is, the number of times the explanation function is called by MCSAT) instead of the running time. We can clearly see that the solvers that solve more instances tend to compute fewer cells during a run. That is, the quality of the explanations is better. Again, considering the refinement based approach and the levelwise approaches, differences are only significant without the other backends. In particular, the virtual best solver clearly solves more instances with fewer cells. \Cref{fig:res:scatter_cells} makes this even more clear by comparing the number of cells constructed by \texttt{VB-LW} and \texttt{OC-ASC} for every instance.

\paragraph{Comparison of heuristics}

For further comparison of the three heuristics, we collected more statistics for each instance: the dimension of constructed cells (i.e. the number of levels considered during its construction), the maximum degree in the main variable of the polynomials occurring in the computation, and the size of the computed projection (i.e. number of resultants, discriminants and coefficients). A summary of these statistics is shown in \Cref{tbl:results:comparison}.

\begin{table}[t]
	\center
	\begin{tabular*}{0.95\linewidth}{@{\extracolsep{\fill}}|c||c c c c c c|}
		\hline
		{} &  \shortstack{mean \#cells \\ per inst.} &  \shortstack{mean dim. \\ of cells} &  \shortstack{mean max \\ deg. per inst.} &  \shortstack{mean \#res. \\ per inst.} &  \shortstack{mean \#disc. \\ per inst.} &  \shortstack{mean \#coeff. \\ per inst.} \\
		\hline
		\hline
		LW-EQ-BC  &                               4.38 &                                 8.59 &                          \textbf{6.51} &                                    8.95 &                                      13.94 &                                     13.62 \\
LW-EQ-CH  &                               4.39 &                                 8.52 &                          6.52 &                                    9.07 &                                      14.04 &                                     13.80 \\
LW-EQ-LDB &                               4.39 &                                 \textbf{8.48} &                          6.57 &                                    9.07 &                                      14.05 &                                     13.88 \\
VB-LW     &                               \textbf{4.32} &                                 8.49 &                          6.52 &                                    \textbf{8.77} &                                      \textbf{13.75} &                                     \textbf{13.48} \\
		\hline
	\end{tabular*}
		
	\caption{Statistics on the $3226$ instances solved by all three solvers with at least one call to the single cell construction and without a call to the NLSAT explanation (single cell construction always succeeds): mean number of cells per instance, mean dimension of constructed cells, mean maximum degree of polynomials occurring in an instance, mean number of resultants/discriminants/coefficients in the projection per instance.}
	\label{tbl:results:comparison}
\end{table}

First note that the virtual best has the lowest value or close to the lowest value in all categories, which means that they might indicate the performance of a solver to some degree. However, interpretation needs to be careful, as the differences are relatively small, confirming the previous observations. Further, note that we do consider fewer instances as in previous analysis. However, we do have two minor observations to make here.

Regarding the heuristic \texttt{LW-EQ-LDB}: its idea was to minimize the degrees of the polynomials in the projection, however, the average maximum degree is slightly higher than for the other two heuristics. \texttt{LW-EQ-BC} needs slightly less projection steps, but is similar to the other solvers in terms of number of cells created. This means that the size of the cells are similar, although again, the intention was to produce the biggest possible cells. To emphasize, in the analysis in \Cref{fig:res:performanceprofile_cells} which is based on all solved instances (including the ones where the NLSAT backend was used as fallback), it needs more cells than \texttt{LW-EQ-LDB}.

To summarize, this simple analysis can not confirm that the ideas behind the different heuristics take effect on the benchmarks or prove different behaviour on all benchmarks. However, as stated above, on individual benchmarks, they do behave differently, as shown by the performance of the virtual best solver.

\section{Conclusions and future work}
\label{sec:conclusion}
\subsection{Future work}
\label{sec:futurework}

The formulation of the presented proof system allows heuristic to influence the shape of the constructed cells. The experimental evaluation shows potential for further development of those. Furthermore, for some applications such as incremental linearization \cite{cimatti2017invariant}, under-approximations of the constructed cell might be beneficial if they can be computed more efficiently; more concretely, the resultant computations get trivial if the lower and upper bounds are replaced by one or more linear polynomials, resulting in boxes or polyhedra.

The proof system could be applied in the future in contexts other than MCSAT, such as the \emph{cylindrical algebraic coverings method} \cite{abraham2020covering} or quantifier elimination algorithms such as \emph{NuCAD} \cite{Brown2015, Brown2017}.

By relying on the theory of McCallum projection, the presented proof system is incomplete. Recently, there has been progress on the Lazard projection operator \cite{Lazard1994, MH16, MPP19, BM20, NDS19, NDS20}, which is complete while maintaining the advantages of the McCallum operator.  Thus, it is promising to extend our framework to exploit the Lazard projection.

As indicated in \Cref{fig:map:sgninvord:2,fig:map:irordering:2}, the notion of delineability is stronger than we need for single cell construction. We will investigate the role of the leading coefficients and the zeros of a resultant.

Finally, the presented set of inference rules allows producing fine-grained proof graphs, which could enable the certification and external automated verification of results.

\subsection{Conclusion}

We introduced the new concept of levelwise single cell construction, motivated by maintaining the savings of the existing refinement based approach of Brown \cite{brown2015onecell} while allowing for more flexibility and new optimizations.

The theoretical part of this paper consists of a proof system in order to enable fine-grained projections based on a given sample.  To demonstrate the possibilities of the proof system, we gave a simple algorithm which builds upon this proof system as well as several heuristics for the application of the given rules. Finally, we gave a qualitative evaluation as well as some notable observations.

We evaluated our algorithm by an implementation applied to explanation generation in MCSAT. We showed that our basic heuristics yield different performances for different instances indicating that there is room for further algorithmic development, as well as a more elaborated implementation of the proof rules.

Importantly, our proof system allows for a wide range of improvements through experimentation with heuristics, approximations and on theoretical matters as well as extensions to other algorithms than the single cell construction.

\section*{Acknowledgements}

Jasper Nalbach was supported by the DFG RTG 2236 \textit{UnRAVeL}. James Davenport and Matthew England were supported by the EPSRC DEWCAD Project, \textit{Pushing Back the Doubly-Exponential Wall of Cylindrical Algebraic Decomposition} (grant references EP/T015748/1 and EP/T015713/1).

\bibliographystyle{plainurl}
\bibliography{literature}

\appendix

\section{Correctness of the proof system}
\label{sec:proofsystemcorrectness}

Note that throughout this section, when proving the correctness of a mapping, we implicitly use the assumptions from the mapping's definition in the proofs.

\defmapdel*

\begin{lemma}
    \Vref{def:map:del} is correct.  
\end{lemma}

\begin{proof}
    Follows immediately from \cite[Theorem 3.1]{brown2001improved} and \cite[Theorem 2]{mccallum1998improved}.
\end{proof}
\defmapnonnull*

\begin{lemma}
    \Vref{def:map:nonnull} is correct.
\end{lemma}

\begin{proof}
    The first rule follows from the definition of the discriminant by the Sylvester matrix. 

    For the second rule, observe that $p$ is nullified on any point $r \in R$ if and only if $c_j(r) = 0$ for all $j \in [m]$. Then, the statement follows immediately.
    See also \cite[Lemma 5]{brown2015onecell}.
\end{proof}
\defmapinvtrivial*

\begin{lemma}
    \Vref{def:map:invtrivial} is correct.
\end{lemma}

\begin{proof}
    Trivial.
\end{proof}
\defmapinvreducible*

\begin{lemma}
    \Vref{def:map:invreducible} is correct.
\end{lemma}

\begin{proof}
    Note that if $p$ is reducible, then $p \notin \factors{p}$. Both statements follow by \cite[Lemma 3.2.2]{mccallum1985improved}.
\end{proof}
\defmapordinv*

\begin{lemma}
    \Vref{def:map:ordinv} is correct.
\end{lemma}

\begin{proof}
    In the first case, from $p(s)\neq0$ and the sign-invariance of $p$ on $R$ follows $p(r) \neq 0$ for all $r \in R$. Thus, $p$ is order-invariant on $R$ as an immediate consequence of the definition of order-invariance.

    In the second case, from $p(s)=0$ and the sign-invariance of $p$ on $R$ follows $p(r) = 0$ for all $r \in R$. As $p$ is analytically delineable on $\proj{R}{[i-1]}$, $p$ is order-invariant in each $p$-section on $\proj{R}{[i-1]}$, $R$ is a $p$-section on $\proj{R}{[i-1]}$, thus the order-invariance of $p$ on $R$ follows immediately.
\end{proof}
\defmapnozero*

\begin{lemma}
    \Vref{def:map:nozero} is correct.
\end{lemma}

\begin{proof}
    By assumption $p$ is (analytically) delineable on a connected superset $R' \supseteq \proj{R}{[i-1]}$, thus by \Cref{def:delineability} the number of roots of $p$ is constant over $R'$. As $\realRoots{p(s,x_{i+1})} = \emptyset$ and $s \in \proj{R}{[i-1]} \subset R'$, $p$ does not have any roots over $R'$, thus $p$ is sign-invariant on $R' \times \reals \supset R$.
\end{proof}
\defmapsubmanifold*

\begin{lemma}
    \Vref{def:map:submanifold} is correct.
\end{lemma}

\begin{proof}
    As the endpoints of $\Isymb$ are described by analytic functions, the statement follows immediately from \cite[Theorem 2.2.3 and Theorem 2.2.4]{mccallum1985improved}.
\end{proof}
\defmapeqproj*

\begin{lemma}
    \Vref{def:map:eqproj} is correct.
\end{lemma}

\begin{proof}
    As $\representation{\Isymb,s}$ holds on $R$, it holds $b.p(r)=0$ for all $r \in R$.

    In the first case, if $p=b.p$, $p$ is trivially sign-invariant on $R$. 
    
    In the second case, if $p \neq b.p$, by the order-invariance of $\res{x_i}{b.p}{p}$ on $\proj{R}{[i-1]}$, \cite[Theorem 2.2]{mccallum1999equational} yields that $p$ is sign-invariant on $R$.
\end{proof}

\defmapirordering*

\begin{lemma}
    \Vref{def:map:irordering} is correct.
\end{lemma}

\begin{proof}
    First consider $\xi \preceq \xi'$. Either $\xi.p = \xi'.p$, then $\theta_{\xi,s} = \theta_{\xi',s}$ by definition of delineability (\Cref{def:delineability}), or $\res{x_{i+1}}{\xi.p}{\xi'.p}$ is order-invariant on $R$, then either $\theta_{\xi,s} = \theta_{\xi',s}$ or $\theta_{\xi,s} < \theta_{\xi',s}$ on $R$ holds by \Cref{lemma:p:resultant}.

    Now consider $\xi \preceq^t \xi'$ (where $\preceq^t$ is the transitive closure of $\preceq$), then there exist $\xi_1, \ldots, \xi_k$ such that $\xi = \xi_1 \preceq \ldots \preceq \xi_k = \xi'$.  We show that either $\theta_{\xi,s} = \theta_{\xi',s}$ or $\theta_{\xi,s} < \theta_{\xi',s}$ on $R$ holds by induction. The base case for $k=2$ is proven by the preceding paragraph.
    Assume that the statement holds for $\xi_1$ and $\xi_j$ with $j<k$ (induction hypothesis), i.e. either $\theta_{\xi_1,s} = \theta_{\xi_j,s}$ or $\theta_{\xi_1,s} < \theta_{\xi_j,s}$ on $R$. Observe that $\theta_{\xi_j,s} = \theta_{\xi_{j+1},s}$ or $\theta_{\xi_j,s} < \theta_{\xi_{j+1},s}$ on $R$ by the preceding paragraph. It follows by transitivity that $\theta_{\xi_1,s} = \theta_{\xi_{j+1},s}$ or $\theta_{\xi_1,s} < \theta_{\xi_{j+1},s}$ on $R$.
\end{proof}

\newcommand{\ra}{r}
\newcommand{\rb}{r'}
\newcommand{\rc}{r''}

\newcommand{\pa}{p}
\newcommand{\pb}{q}

\newcommand{\sa}{R}

\newcommand{\fa}{f}
\newcommand{\fb}{g}

\begin{definition}[Degree invariance]
    Let $i \in \integers$, $R \subseteq \reals^i$, and $p \in \rationals[x_1,\ldots,x_{i+1}]$ of level $i+1$. Then $p$ is called \emph{degree-invariant} on $R$ if and only if $\vdeg{x_{i+1}}{p(r,x_{i+1})} = \vdeg{x_{i+1}}{p(r',x_{i+1})}$ for all $r,r' \in R$.
\end{definition}

\begin{theorem}
    \label{lemma:p:resultant}
    Let $i \in \naturals$, $R \subseteq \reals^{i}$ be a connected analytic submanifold and $p_1,p_2 \in \rationals[x_1,\ldots,x_{i+1}]$ irreducible and coprime such that $\level{p_1}=\level{p_2}={i+1}$, $\theta_1, \theta_2: R \to \reals$ be real root functions of $p_1$ and $p_2$ respectively, and ${\sim} \in \{ =,<,> \}$. 
    
    If $p_1,p_2$ are not nullified on any point in $R$ and $\res{x_{i+1}}{p_1}{p_2}$ is order-invariant on $R$ and $p_1$ and $p_2$ are analytically delineable on $R$, then $\theta_1(r) \sim \theta_2(r) \iff \theta_1(r') \sim \theta_2(r')$ for all $r,r' \in R$.
\end{theorem}

\begin{proof}
    The hypotheses state that $\theta_1$ and $\theta_2$ are continuous and analytic real root functions over $R$.  If the two functions are identical over $R$, then the theorem holds.  So assume they are not, i.e. assume there is a point $s\in R$ at which the two functions have different values.  We will show that the two functions differ everywhere in $R$ which, because they are continuous, proves the theorem.

    Let $s \in R$, $\alpha \in \reals$ such that $\alpha < r'$ for every $r' \in \reals$ such that $p_1(s,r')=0$ or $p_2(s,r')=0$ (we can choose such an $\alpha$ as $p_1$ and $p_2$ are delineable on $R$), and choose $S \subseteq R$ such that $s \in S$, and $\alpha > r'$ for every $r \in S$ and every $r' \in \reals$ such that $p_1(r,r')=0$ or $p_2(r,r')=0$.

    Let $p_1^* = p_1(x_1,\ldots,x_{i},x_{i+1}-\alpha)$ and $p_2^* = p_2(x_1,\ldots,x_{i},x_{i+1}-\alpha)$. The mapping that shifts $x_{i+1}$ by $\alpha$ is a homeomorphism that maps the roots of $p_1$ and $p_2$ to the roots of $p_1^*$ and $p_2^*$, respectively.  So we can study the roots of $p_1$ and $p_2$ by understanding the roots of $p_1^*$ and $p_2^*$.  But for $p_1^*$ and $p_2^*$ we are guaranteed that $p_1^*(r,0) \neq 0$ and $p_2^*(r,0) \neq 0$ for all $r \in S$, i.e. the constant coefficients with respect to $x_{i+1}$ of $p_1^*$ and $p_2^*$ do not vanish on $S$. Further, by \cite{GKZ94}, the resultant is invariant under this transformation.

    Let $\overline{p_1} = x_{i+1}^{d_1} p_1^*(x_1,\ldots,x_{i},1/x_{i+1})$ and $\overline{p_2} = x_{i+1}^{d_2} p_2^*(x_1,\ldots,x_{i},1/x_{i+1})$, where the $d_1 = \vdeg{x_{i+1}}{p_1}$ and $d_2 = \vdeg{x_{i+1}}{p_2}$.
    
    Moreover, since neither $p_1$ nor $p_2$ is nullified in $S$,     $\overline{p_1}$ and $\overline{p_2}$ are non-nullified in $S$.
    The transformation from $p_1$ and $p_2$ to $\overline{p_1}$ and $\overline{p_2}$ defines an analytic homeomorphism from $S \times (\reals \setminus \{ 0 \})$ to $S \times (\reals \setminus \{ \alpha \})$ that maps the zeros of $\overline{p_1}$ and $\overline{p_2}$ off the $x_{i+1} = 0$ hyperplane to the zeros of $p_1$ and $p_2$ (which do not cross the $x_{i+1} = \alpha$ hyperplane). Specifically, the homeomorphism is given by the mapping $(x,y) \mapsto (x,1/y - \alpha)$.
    Moreover, the leading coefficients of $\overline{p_1}$ and $\overline{p_2}$, which are the same as the constant coefficients (in $x_{i+1}$) of $p_1^*$ and $p_2^*$, are non-zero throughout $S$. This means that $\overline{p_1}$ and $\overline{p_2}$ are not nullified anywhere in $S$ and their leading coefficients (in $x_{i+1}$) are non-vanishing on $S$, and thus $\overline{p_1}$ and $\overline{p_2}$ are degree-invariant on $S$.
    We note that $\overline{p_1}$ and $\overline{p_2}$ are delineable on $S$ (as $p_1$ and $p_2$ are delineable on $S$ and the homeomorphism); in particular, for every root function $\theta$ of $p_1$ or $p_2$, there exists a corresponding root function $\overline{\theta}$ of $\overline{p_1}$ or $\overline{p_2}$, respectively.
    Further, the resultant is unchanged by this transformation, once again by \cite{GKZ94}, so $\res{x_{i+1}}{\overline{p_1}}{\overline{p_2}}$ is order-invariant in $S$.

    Now, let $\overline{\theta_1}$ and $\overline{\theta_2}$ be the root functions of $\overline{p_1}$ and $\overline{p_2}$ corresponding to $\theta_1$ and $\theta_2$. By \Cref{theorem:resultantresult}, $\overline{\theta_1}(s) \neq \overline{\theta_2}(s)$ implies $\overline{\theta_1}(r) \neq \overline{\theta_2}(r)$ for all $r \in S$; and thus, $\theta_1(r) \neq \theta_2(r)$ for all $r \in S$.

    Since we can set $s$ to any value in $R$, it follows that $\theta_1(r) \neq \theta_2(r)$ for all $r \in R$.
\end{proof}

\begin{theorem}
    \label{theorem:resultantresult}  
    Let $p_1, p_2 \in \reals[x_1,\ldots,x_i]$ of level $i$. Let $R$ be a connected analytic submanifold in $\realring^{i-1}$ on which $p_1$ and $p_2$ are degree-invariant and non-nullified,
    and on which $p_r = \mbox{res}_{x_i}(p_1,p_2)$ is order-invariant.  Let $\theta_1$ be a real root function of $p_1$ and $\theta_2$ be a real root function of $p_2$, both defined on $R$.  If there is a point $\ra\in R$ at which the values of $\theta_1$ and $\theta_2$ differ, then the values of $\theta_1$ and $\theta_2$ differ throughout $R$.
\end{theorem}

\begin{proof}
    We prove this theorem by contradiction.  Suppose there are points at
    which $\theta_1$ and $\theta_2$ have the same value.  Choose such a point
    $\rb\in R$ and path $\pi:[0,1]\to R$ such that
    $\pi(0) = \ra$, $\pi(1) = \rb$ and for all $x\in(0,1)$ we have
    $\theta_1(\pi(x)) \neq \theta_2(\pi(x))$.  Note that we are
    guaranteed to be able to find such $\rb$ and $\pi$ by the continuity
    of $\theta_1$ and $\theta_2$.

    Let $U$ be a
    neighbourhood of $\rb$ and $V$ a neighbourhood of the origin, both
    within $\realring^{i-1}$.  Denote by $d$ the dimension of $R$.
    We now refer to the concept of a \emph{coordinate system}, as
    described in \cite{mccallum1985improved}.
    By Theorem 2.2.1 of \cite{mccallum1985improved}, there is a coordinate
    system $\phi$ mapping $U$ to $V$ such that
    $R \cap U =
    \{ x \in U \ |\ \phi_{d+1}(x) = 0, \ldots , \phi_{i-1}(x) = 0\}$.  In
    other words, the image $S$ of $R \cap U$ is the set of points in $V$
    with zeros for the last $i - 1 - d$ coordinates.%
    We view $p_1$ and
    $p_2$ in this new coordinate system as
    $\overline{p_1} = p_1(\phi^{-1}(x_1,\ldots,x_{i-1}),x_i)$
    and
    $\overline{p_2} = p_2(\phi^{-1}(x_1,\ldots,x_{i-1}),x_i)$,
    respectively.
    Note that for any $a \in U$, $p_1$ evaluated at $a$ gives the same
    univariate polynomial in $x_i$ as $\overline{p_1}$ evaluated at
    $\phi(a)$ (as is also true for $p_2$). In particular, since $p_1$ and $p_2$
    are degree-invariant in $R$, $\overline{p_1}$ and
    $\overline{p_2}$ are degree-invariant in $S$, moreover, the degrees of
    $p_1$ and $\overline{p_1}$ are the same, as are the degrees of
    $p_2$ and $\overline{p_2}$.
    Because $\phi^{-1}$ only acts on the coefficients of $p_1$ and $p_2$
    (as polynomials in $x_i$),
    and because the degrees remain unchanged, the resultant construction
    commutes with 
    respect to $\phi^{-1}$.  In other words,
    recalling that $p_r = \mbox{res}_{x_i}(p_1,p_2)$, we have
    $p_r\circ \phi^{-1} = \mbox{res}_{x_i}(\overline{p_1},\overline{p_2})$.
    Moreover, by the ``Remark'' on p.45 of \cite{mccallum1985improved},
    the order of $p_r$ at point $a$ is the same as the order of
    $p_r\circ \phi^{-1}$ at $\phi(a)$, for any $a \in U$.  This means that $p_r$ is
    order-invariant in $R \cap U$ if and only if $p_r\circ \phi^{-1}$ is
    order-invariant in $S$.  We will show
    that the order-invariance of
    $\mbox{res}_{x_i}(\overline{p_1},\overline{p_2})$ in $S$ contradicts
    the assumption that $\theta_1(\rb) = \theta_2(\rb)$.

    For convenience, let
    $f = \theta_1 \circ \phi^{-1}$
    and
    $g = \theta_2 \circ \phi^{-1}$, noting that $f$ and $g$ are themselves
    analytic functions and thus proper real-root functions for
    $\overline{p_1}$ and $\overline{p_2}$, and that they agree in value at
    $\phi(\rb)$.  The function $f$ defines a root of $\overline{p_1}$ on
    $S$, not on all of $V$. Define $f^*(y_1,\ldots,y_{i-1}) = f(y_1,\ldots,y_d,0,\ldots,0)$, which
    extends the domain of $f$ to all of $V$, and define $g^*$
    analogously.
    Note that $f^*$ and $g^*$ are \emph{analytic} functions of $V$.
    [This is actually the key to this proof.  Had we tried in the same way
    to extend 
    $\theta_1$ and $\theta_2$ from functions of $R$ to functions of
    $U$, we would not have been guaranteed to get analytic functions.]
    By division we get
    $\overline{p_1} = (x - f^*)q_1 + \overline{p_1}(f^*)$.  The remainder,
    $\overline{p_1}(f^*)$, is zero throughout $S$ because $f$, which $f^*$
    extends, is a root function of $\overline{p_1}$ on $S$.  However,
    since the value of $f^*$ is independent of $y_{d+1},\ldots,y_{i-1}$,
    the remainder $\overline{p_1}(f^*)$ is actually zero throughout $V$.  Thus,
    $\overline{p_1}$ factors as $\overline{p_1} = (x - f^*)q_1$ over $V$.
    By the same logic, we get the factorization
    $\overline{p_2} = (x - f^*)q_2$.  Note that the coefficients of $q_1$
    are all polynomials in $f^*$, so all coefficients are analytic
    functions of $V$.

    It is shown in \cite{BCL:83}, Theorem 1, Chapter \emph{Computing in
    Algebraic Extensions}, that
    $res(A,B) = a_m^n \prod_{i=1}^m B(\ra_i)$, where $m$ is the degree
    of $A$, 
    $a_m$ is the
    leading coefficient of $A$, the $\ra_i$s are the $m$ complex roots
    of $A$, and $n$ is the degree of $B$.  It easily follows that

    $\mbox{res}(A,BC) = \mbox{res}(A,B) \cdot \mbox{res}(A,C)$,
    from which we get:
    $$\mbox{res}_{x_i}((x_i - f^*)q_1,(x_i - g^*)q_2) =
    \mbox{res}_{x_i}(x_i - f^*,x_i - g^*) \cdot 
    \mbox{res}_{x_i}(q_1,x_i - g^*) \cdot 
    \mbox{res}_{x_i}(x_i - f^*,q_2) \cdot 
    \mbox{res}_{x_i}(q_1,q_2).$$
    For any analytic function, if there are points of
    order $k$ and also of order $k + j$, the set of points of order at
    least $k+j$ is a closed subset of the set of all points of order at
    least $k$. 
    This means that if the order 
    of a product of analytic functions is constant in $S$ then the orders of each
    of the individual factors must be constant as well.  Thus,
    since the order of $\mbox{res}_{x_i}(\overline{p_1},\overline{p_2})$
    is constant in $S$, the order
    of $\mbox{res}_{x_i}(x_i - f^*,x_i - g^*)$, which we note is
    $f^* - g^*$, is constant as well.
    There are points in $R$ arbitrarily close to $\rb$ at which
    $\theta_1$ and $\theta_2$ are unequal.  Let $\rc$ be such a point,
    sufficiently close such that $\phi(\rc)\in S$.
    Then at $\phi(\rc)$, the functions $f^*$ and $g^*$ are unequal.
    Thus the order of $\mbox{res}_{x_i}(x_i - f^*,x_i - g^*)$ is
    zero at $\phi(\rc)$, which means its order is zero everywhere in $S$.
    So $f^* \neq g^*$ for every point in $S$.  However, since
    $\theta_1$ and $\theta_2$ are equal at $\rb$, $f^*$ and $g^*$ are
    equal at $\phi(\rb)$, which is a contradiction.

\end{proof}
\defmapsgninvord*

\begin{lemma}
    \Vref{def:map:sgninvord} is correct.
\end{lemma}

\begin{proof}
    As $p$ is delineable on $\proj{R}{[i-1]}$, the variety of $p$ on $\proj{R}{[i-1]}$ is described by the real root functions $\theta_{\xi,s},\ \xi \in \irexpr{p}{s}$.
    As $\representation{\Isymb,s}$ holds on $R$, it follows that if $l \neq -\infty$, then $\theta_{l,s}$ exists on $\proj{R}{[i-1]}$, and if $u \neq \infty$, then $\theta_{u,s}$ exist on $\proj{R}{[i-1]}$.

    As $s \in \proj{R}{[i-1]}$, $\preceq$ matches $s$, and either $\xi \preceq^t l$ or $u \preceq^t \xi$ for all $\xi \in \irexpr{p}{s}$, it holds $\xi(s) \leq l(s)$ or $u(s) \leq \xi(s)$ for all $\xi \in \irexpr{p}{s}$. To prove sign-invariance of $p$ on $R$, we thus only need to show that either $\theta_{\xi,s} < \theta_{l,s}$ on $R$ (if $\xi(s)<l(s)$), $\theta_{u,s} < \theta_{\xi,s}$ on $R$ (if $u(s)<\xi(s)$), or $\theta_{\xi,s} = \theta_{b,s}$ on $R$ (if $\xi(s)=b(s)$) for all $\xi \in \irexpr{p}{s}$.
    
    This statement holds as $\preceq$ matches $s$, $\xi \preceq^t l$ or $u \preceq^t \xi$ for all $\xi \in \irexpr{p}{s}$, and $\irordering{\preceq,s}$ holds on $\proj{R}{[i-1]}$.
\end{proof}
\defmapconnectedness*

\begin{lemma}
    \Vref{def:map:connectedness} is correct.
\end{lemma}

\begin{proof}
    We only consider the case $\Isymb=(\textit{sector},l,u), l \neq -\infty, u \neq \infty, l.p \neq u.p$, as the other cases are similar but simpler.

    As $\representation{\Isymb,s}$ holds, $\theta_{l,s}$ and $\theta_{u,s}$ are well-defined on $\proj{R}{[i-1]}$.
    As $s \in R$, $\preceq$ matches $s$, $l \preceq^t u$, and $\irordering{\preceq,s}$ holds on $\proj{R}{[i-1]}$, it either holds $\theta_{l,s} < \theta_{u,s}$ on $\proj{R}{[i-1]}$ or $\theta_{l,s} = \theta_{u,s}$ on $\proj{R}{[i-1]}$. The property $\representation{\Isymb,s}$ holds on $R$ and $\proj{R}{[i-1]}$ is connected, thus $R$ is connected (noting that empty sets are connected).
\end{proof}
\defmapsample*

\begin{lemma}
    \Vref{def:map:sample} is correct.
\end{lemma}

\begin{proof}
    Follows immediately from the definitions.
\end{proof}
\defmaprepr*

\begin{lemma}
    \Vref{def:map:repr} is correct.
\end{lemma}

\begin{proof}
    The graph of any delineable polynomial $p$ over a connected subset $R' \supseteq R$ is described by real root functions $\theta_1 < \ldots < \theta_k$ on $R'$. It follows that for any indexed root expression $\xi$ with $\xi.p = p$ and  $\xi.k \leq k$ it holds $\dom{\theta_{\xi,s}} \supseteq R'$ and $\theta_{\xi,s} = \theta_{\xi.k} = \xi$ on $R'$.

    Now, the result follows immediately from the definition of $\irint{\Isymb}$ and the previous statement applied on the cell's boundaries.
\end{proof}

\end{document}